\documentclass[preprint]{JHEP3}
\pdfoutput=1
\JHEPspecialurl{http://jhep.sissa.it/JOURNAL/JHEP3.tar.gz}
\usepackage{epsfig,multicol,bbm}
\usepackage{citesort}
\newcommand\fverb{\setbox\fverbbox=\hbox\bgroup\verb}
\newcommand\fverbdo{\egroup\medskip\noindent%
\fbox{\unhbox\fverbbox}\ }
\newcommand\fverbit{\egroup\item[\fbox{\unhbox\fverbbox}]}
\newbox\fverbbox

\usepackage{placeins}

\newcommand{\lsim}{\raisebox{-0.13cm}{~\shortstack{$<$ \\[-0.07cm] $\sim$}}~} 
\newcommand{\gsim}{\raisebox{-0.13cm}{~\shortstack{$>$ \\[-0.07cm] $\sim$}}~} 
\newcommand{\beq}{\begin{eqnarray}} 
\newcommand{\eeq}{\end{eqnarray}} 
\newcommand{\tb}{\tan \beta}
\newcommand{\h}{h}

\preprint{KCL-PH-TH 15-08 \\ LPT-Orsay 15-15}

\title{Fully covering the MSSM Higgs sector at the LHC}

\author{A. Djouadi$^1$, L. Maiani$^{2}$, A. Polosa$^2$, J. Quevillon$^3$ and V. Riquer$^2$\\
$^1$ Laboratoire de Physique Th\'eorique d'Orsay, Universitî\'e  Paris XI and CNRS, \\ F--91405 Orsay, France.\\
$^2$ Department of Physics and INFN, Universit\`a di Roma Sapienza, Pizzale Aldo Moro 5,\\ I--00185 Roma, Italia.\\
$^3$ Theoretical Particle Physics and Cosmology Group, King's College London, \\ WC2R 2LS London, UK.}

\abstract{
In the context of the Minimal Supersymmetric extension of the Standard Model (MSSM), 
we reanalyze the search for the heavier CP--even $H$ and CP--odd $A$ neutral Higgs 
bosons at the LHC in their production in the gluon--fusion mechanism and their decays 
into gauge and lighter $h$ bosons and into top quark pairs. We show that only when 
considering these processes, that one can fully cover the entire parameter space of the 
Higgs sector of the model. Indeed, they are sensitive to the low $\tan\beta$ and high 
Higgs mass ranges,
complementing the traditional searches for high mass resonances decaying into $\tau$--lepton 
pairs which are instead sensitive to the large and moderate $\tb$ regions. The complementarity 
of the various channels in the probing of the complete $[\tb, M_A]$ MSSM parameter space 
at the previous and upcoming phases of the LHC is illustrated in a recently proposed 
simple and model independent approach for the Higgs sector, the $h$MSSM, that we also refine 
in this paper.
}


\begin{document} 

\renewcommand{\thefootnote}{\arabic{footnote}}
\setcounter{footnote}{0}

\section{Introduction}

The probing of the electroweak symmetry breaking mechanism and the search for possible
extensions of the Standard Model (SM) of particle physics has become the main mission of 
the CERN Large Hadron Collider (LHC). Among these extensions, Supersymmetry (SUSY) \cite{SUSY} 
is considered as the most appealing one as it addresses several shortcomings of the SM, 
including the problem of the large hierarchy between the Planck and electroweak scales. 
While the search for SUSY was unsuccessful at the first LHC run, the increase of the center 
of mass energy of the machine from 8 TeV  to the 14 TeV level will significantly improve 
the sensitivity to the new particles that are predicted by the weak scale theory. These 
consists not only of the superpartners of the known fermions and gauge bosons but, also, 
of the additional Higgs bosons beyond the state with a mass of 125 GeV that 
has been observed by the ATLAS and CMS collaborations in the first LHC phase \cite{LHC-Higgs}. 

As a matter of fact, in low--energy SUSY scenarios, at least two Higgs doublet fields 
$H_u$ and $H_d$ are required to break the electroweak symmetry and to generate the 
isospin--up and down type fermion and the $W/Z$ boson masses. In the simplest scenario, 
the Minimal Supersymmetric Standard Model (MSSM), the spectrum consists of five states 
\cite{Review0,Review1,Review}: two charged $H^\pm$, a CP-odd $A$ and two CP-even Higgs particles $h$ and $H$, with $h$ being the state observed at the LHC while $H$ is heavier as present 
LHC data is strongly indicating \cite{ATLAS+CMS-h}. 

The phenomenology of the Higgs sector
is described entirely by two input parameters, one Higgs mass that is usually taken to be 
that of the pseudoscalar $A$ boson $M_A$ and the ratio $\tan\beta$ of the vacuum expectation 
values of the two doublet fields, which is generally assumed to lie in the range $1 
\lsim \tb \lsim m_t/m_b \approx 60$. This is the case at tree--level where, for instance,
the lightest $h$ boson mass is an output and is predicted to be $M_h\! \lsim \! M_Z 
|\cos2\beta|$, i.e. $M_h\! \leq \! M_Z$ at high $\tb$ for which $|\cos2\beta| \! \simeq \! 1$.
However, this relation is violated since important radiative corrections,  
that introduce a dependence on many SUSY parameters, occur in this context 
\cite{CR-1loop,CR-eff,CR-2loop}. It has been recently shown that,  to a good approximation, 
the MSSM Higgs sector can be again parametrised using the two basic inputs $\tb$ and $M_A$,
provided that the crucial LHC information $M_h \simeq 125$ GeV is used 
\cite{habemus,Rome,Orsay}. 

It is known that two efficient channels can be used to directly search for the heavier 
MSSM Higgs particles at the LHC  and probe part of the $[\tan\beta,  M_A]$ parameter 
space\footnote{Of course, there are also indirect limits on the MSSM parameter space, in 
particular from the measurement of the couplings of the observed $h$ particle at the LHC; 
see for instance Refs.~\cite{habemus,Hcoup1}. These limits are nevertheless slightly model
dependent as, for instance, they can be affected by SUSY particle  contributions
to the $h$ production and decay rates. These indirect limits, that exclude low values of 
the $A,H,H^\pm$ masses, are complementary to those from the direct Higgs searches on which we will focus in this paper.}. 
The first one is the search for light charged Higgs bosons that would emerge from the decays 
of the copiously produced top quarks and would decay almost exclusively into a $\tau$
lepton and its associated neutrino for $\tb \gsim 1$. For almost all values of 
$\tan\beta$, the latest ATLAS \cite{ATLASH+} and CMS  \cite{CMSH+} results now 
practically rule out the mass range $M_{H^\pm} \lsim 160$ GeV, which approximately 
corresponds to $M_A \lsim 140$ GeV in the MSSM.  The second efficient channel is the search 
for high mass resonances decaying into $\tau$--lepton pairs, which would be the signature
of the production of the heavy neutral $H/A$ states and their decay into $\tau$ leptons. 
The rates for this channel can be very large at high $\tan\beta$ values, as a consequence 
of the strong enhancement of the $H/A$ couplings  to bottom quarks and $\tau$--leptons. 
This process is particularly favored as, for a heavy enough $A$ boson, one has 
the mass degeneracy relation $M_H\! \approx\! M_A$ that in practice leads to search
for a single resonance and  allows to  combine the rates for $A$ and $H$ production. 
The most recent ATLAS \cite{tau-ATLAS} and CMS \cite{tau-CMS} results with the data 
collected at the first LHC phase, exclude at the 95\% confidence level (CL) a significant
portion of the $[\tb, M_A]$ plane for sufficiently high $\tb$ values. 

Except in the narrow mass range $M_Z \lsim M_A \lsim 140$ GeV, where the lower value corresponds 
to the exclusion limit from negative Higgs searches at the LEP collider \cite{PDG,LEP} and the upper one is due to the present limit from charged Higgs boson searches at the LHC (which can be  straightforwardly interpreted in the $[\tb, M_{H^\pm}]$ parameter space as the $H^\pm$
properties depend only on these two parameters in the low mass range), the low $\tan\beta$ region
of the MSSM has not been considered so far by the experimental collaborations. The reason is that in the benchmark scenarios  that are used to interpret the various experimental limits on 
the cross sections times branching ratios in the context of the MSSM \cite{benchmark-old,benchmarks}, the SUSY--breaking scale is usually set to relatively low values, $M_S \approx 1$ TeV, that do 
not allow for a heavy enough $h$ state at too low $\tb$. Indeed, the radiative corrections to the mass $M_h$ depend logarithmically on the scale $M_S$ and, for instance, one
cannot obtain a value $M_h \approx 125$ GeV for $\tb \lsim 3$--5 in the MSSM, even if one favorably tunes the other SUSY parameters that enter the loop radiative corrections, in particular the stop mixing parameter $X_t$  which also plays an important role in this context.
This is the case of the so--called maximal mixing  or $M_h^{\rm max}$ scenario which is
defined such that the value of $M_h$ is maximized, i.e. for a stop mixing parameter $X_t 
\simeq \sqrt 6 M_S$ in the dimensional reduction scheme \cite{benchmark-old}. The situation
is even worse for different values of the $X_t$ parameter. 

In fact, in most of the $[\tb, M_{A}]$ parameter space, the measured value $M_h \approx 125$
GeV, which should be now considered as an essential information on the model, is not satisfied in the $M_h^{\rm max}$ benchmark scenario with $M_S=1$ TeV nor in the alternative benchmark scenarios that are presently used to interpret the experimental searches in the context of the MSSM.  If one allows for an uncertainty of a say 3 GeV in the determination of 
the $h$ mass in the MSSM, from unknown higher order contributions for instance \cite{DeltaH},
the situation is acceptable if the $h$ mass is confined in the range $122\;{\rm GeV} \lsim 
\! M_h \! \lsim 128 \; {\rm GeV}$. Nevertheless, it remains annoying that for each point of 
the $[\tb, M_A]$ parameter space, one has a  different $M_h$ value in these benchmark scenarios.  

A straightforward and easily implementable solution to this problem has been proposed in    
Refs.~\cite{habemus,Rome,Orsay}: if the experimental constraint $M_h \approx 125$ GeV is 
enforced, one 
in fact removes the dependence of the Higgs sector on the dominant radiative correction 
and, hence, on the additional SUSY  parameters, in particular $M_S$ and $X_t$.  
One can again parametrise the MSSM Higgs sector using only the two basic inputs $\tb$ and 
$M_A$, exactly like it was the case  at tree--level. The masses of the heavier $H$ and $H^\pm$
states as well as the mixing angle $\alpha$ in the CP--even sector are given by very simple
expressions in terms of $\tb$ and $M_A$ with the constraint $M_h=125$ GeV. It was shown 
that this approximation is very good in most of the MSSM parameter space that is currently accessible at the LHC, even when subleading radiative corrections are also considered \cite{habemus}. 

In this minimal and almost model independent approach, called the $h$MSSM in 
Ref.~\cite{habemus},  one has access to the entire $[\tb, M_A]$ parameter space without 
being in conflict with the LHC data, as the information $M_h=125$ GeV is  taken into account from the very beginning (this is not always the case for the Higgs couplings which conflict 
with the measured ones at low $M_A$). In particular, the low $\tb$ region can naturally 
be accessed, but at the expense of assuming a very high SUSY scale $M_S$. The reason is that 
at $\tb$ values too close to unity, the tree--level $h$ mass becomes very small, $M_h \! 
\approx \! M_Z|\cos2\beta| \! \to  \! 0$. To increase $M_h$  to $\approx 125$ GeV, 
the radiative corrections that grow logarithmically with $M_S$ 
need to be maximized and hence, a very large scale, $M_S \gsim {\cal O} (100)$ TeV for 
$\tb \lsim 2$, is required.  

The low $\tb$ region can be  directly probed by the search for the heavier $H/A$ 
(and eventually $H^\pm$) states and for relatively low Higgs masses, $M_H \! \approx \! M_A
\!  \lsim \! 350$ GeV, two ways have been suggested.  First, one can use the same 
constraint discussed above from the search of resonances decaying into $\tau$--lepton pairs 
\cite{Orsay}. Indeed, the rates for $A/H$ production are appreciable at low $\tb$
as the dominant process,  the gluon--fusion mechanism, is now primarily mediated by loops of 
top quarks that have significant couplings to the $H/A$ bosons; at the same time, the decay 
of at least the $A$ boson into $\tau\tau$ pairs has a still appreciable rate. The second
way is to reinterpret the existing ATLAS and CMS exclusion limits from the search for 
a heavy SM--like Higgs boson decaying into a pair of massive gauge bosons \cite{CMS-WW,CMS-ZZ} 
in the context of the MSSM.  At low $\tb$ and not too large $M_H$ values for which we are not yet in the decoupling regime with a vanishing $H$ coupling to massive gauge bosons,  the rates for the decays $H \to \! VV$ with $V\!=\!W,Z$, as well as for $gg\! \to \! H$ production,  
are still significant. In addition, searches for the  
resonant $hh$ \cite{ATLAS-hh,CMS-hh} and $hZ$ \cite{CMS-hZ} topologies have been performed 
at the LHC with the available 25 fb$^{-1}$ data at $\sqrt s\!=\!7$+8 TeV, and one can reinterpret them in the context of the MSSM where the production cross section for $gg\! \to 
\! H/A$ and the branching ratios for the decay modes  $H\! \to \! hh$ and $A \! \to \! Zh$ 
below the $t\bar t $ threshold can be substantial; see Refs.~\cite{Orsay,Heavy}. 

The two types of searches mentioned above, with results that were preliminary and obtained
with a subset of the LHC data collected at $\sqrt s=7$+8 TeV, have been used in Ref.~\cite{Orsay} to set constraints on the [$\tb, M_A]$~plane; excluded regions have been delineated
using some approximations and extrapolations. In the present paper, we update this discussion first by using the latest ATLAS and CMS results, especially the final $H/A \to \tau^+\tau^-$ 
and $t \to b H^+ \to b \tau \nu$ analyses \cite{ATLASH+,CMSH+,tau-ATLAS,tau-CMS} as well
as heavy SM Higgs searches in the $H\to WW,ZZ$ channels \cite{CMS-WW,CMS-ZZ}, with the 
full set of 25 fb$^{-1}$ data collected in the first LHC phase. In addition, constraints from more
appropriate analyses in the  $A\! \to \! hZ$ and $H\! \to  \! hh$ topologies where the 
resonant case has now been considered \cite{ATLAS-hh,CMS-hh,CMS-hZ} will be included. 
We will then extrapolate these results to estimate the sensitivity of the 14 TeV LHC run, 
with at least an order of magnitude higher integrated luminosity than the one accumulated so far.

Above the $t\bar t$ threshold, i.e. for $M_{A,H} \gsim 350$ GeV, the previously discussed 
search channels will have little relevance at low $\tb$ values as, because their couplings 
to $b$ quarks and $\tau$ leptons are not enhanced anymore, the heavier $H$ and $A$ bosons 
will dominantly decay into $t\bar t$ pairs, the top--quark Yukawa coupling $\propto m_t/\tb$
becoming then large. As already mentioned, the main Higgs production channel will be the 
gluon-fusion process $gg\to H/A$ in which the top quark loop generates the dominant
contribution. We will see that the production times the decay rates in the processes $gg \to 
H/A \to t\bar t$ are indeed substantial in a large part of the MSSM parameter space. 
We perform a naive estimate of the sensitivity that can be achieved in the search for 
$t\bar t$ resonances, a sensitivity that could allow to probe a significant part 
of the low $\tb$ region of the MSSM, complementing the searches for $\tau^+\tau^-$ 
resonances that are instead sensitive to the high $\tb$ region.  
 
The main message of the present paper is that combining the searches for heavy resonances 
decaying into $\tau^+ \tau^-$ and $t\bar t$ pairs, and including the $H\to WW,ZZ,hh$ and $A 
\to hZ$ channels at $M_A \lsim 350$ GeV, one can possibly probe the \underline{entire} 
$[\tb, M_A]$ MSSM plane (and not only the high--$\tb$ region as is presently the case) 
up to large pseudoscalar Higgs masses; $M_A$ values close to 1 TeV could be 
reached for any $\tb$ with $\approx 3000$ fb$^{-1}$ data at the LHC with $\sqrt s=14$ TeV. 
This can be done in a model independent way and without relying on any 
additional theoretical assumption or indirect experimental constraint.

The paper is organized as follows. We first summarise our simple
parametrisation of the MSSM Higgs sector, further discussing and refining the $\h$MSSM 
approach. In section 3, we discuss the heavier Higgs production and decay rates 
focusing on low $\tb$ and summarise the possible impact of superparticles. 
In section 4, the probing of the $[\tb, M_A]$ MSSM parameter space
is discussed when all the search channels, including a projection for the the $H/A \! 
\to \! t\bar t$ channel, are combined first at $\sqrt s=7$+8 
TeV with  25 fb$^{-1}$ data and then at $\sqrt s\!=\!14$ TeV and higher 
luminosities.  A brief conclusion is given in a final section.

\section{The hMSSM approach}

In this section, we briefly describe the $h$MSSM introduced in Ref.~\cite{habemus}, an approach that allows to parameterize the CP--conserving MSSM Higgs sector in a simple (as only two 
inputs are needed) and ``model independent" (in the sense that we do not consider or fix any other SUSY parameter)  way, using the information that the lightest $h$ boson has a 
mass $M_h \approx 125$ GeV. The approach is based on several assumptions that we first summarize.
 
The first basic assumption of the $h$MSSM is that in the basis $(H_d,H_u)$ of the two 
MSSM Higgs doublet fields that break electroweak symmetry, the CP--even $h$ and $H$  
mass matrix  can be simply written in terms of the $Z$ and $A$ boson masses and the 
angle $\beta$ as 
\beq
M_{\Phi}^2=
\left(
\begin{array}{cc}
  M_Z^2 \cos^2\beta + M_A^2 \sin^2\beta & -(M_Z^2+M_A^2)\sin\beta \cos\beta \\
 -(M_Z^2+M_A^2) \sin\beta \cos\beta & M_Z^2 \sin^2\beta + M_A^2 \cos^2\beta \\
\end{array}
\right) \!+\! 
\left(
\begin{array}{cc}
 \Delta {\cal M}_{11}^2 &  \Delta {\cal M}_{12}^2 \\
 \Delta {\cal M}_{12}^2 &\Delta {\cal M}_{22}^2 \\
\end{array}
\right)
\label{mass-matrix}
\eeq
in which the radiative corrections are introduced through a $2\times 2$ general matrix  
$\Delta {\cal M}_{ij}^2$. This is the usual starting point of the analyses of the neutral
MSSM Higgs sector \cite{Review1} and the  
calculation of the Higgs 
masses and couplings including radiative corrections and in which the SUSY scale, 
taken to be the geometric average of the two stop masses, $M_S= \sqrt {m_{\tilde t_1} 
m_{\tilde t_2}}$, can be as high as a few TeV. However, if $M_S$ is orders of magnitude 
higher than the TeV scale, the evolution  from this high scale down to the electroweak 
scale might mix the quartic couplings of the MSSM Higgs sector in a non trivial way, such 
that the structure of the mass matrix at the low energy scale is different from the one given in 
eq.~(\ref{mass-matrix}). 

In the $h$MSSM, we assume that the form of the CP--even 
Higgs mass matrix is as given above even at the very high SUSY scales, $M_S \gsim 
{\cal O}(100$ TeV) that, as it will be seen later, are needed to consider $\tb$ values 
close to unity\footnote{The validity of this approximation is currently studied by the LHC 
Higgs cross section working group \cite{Pietro}. Preliminary results in an effective two 
Higgs doublet 
model that is RG improved to resum the large logarithms 
involving $M_S$, have been given in Ref.~\cite{Lee+Wagner1,Cheungetal} and a more refined 
analysis is under way  \cite{Pietro,Lee+Wagner2}.}. 

A second basic assumption of the $h$MSSM is that in the $2\times 2$ matrix above 
for the radiative corrections,  only the $\Delta{\cal M}^{2}_{22}$ entry  is relevant, 
$\Delta{\cal M
}^{2}_{22} \gg \Delta{\cal M}^{2}_{11}, \Delta{\cal M}^{2}_{12}$. In this case,  one can simply trade $\Delta {\cal M}^{2}_{22}$ for the by now known mass value $M_h\!=\! 125$ GeV  using 
\begin{eqnarray}
\Delta {\cal M}^{2}_{22}= \frac{M_{h}^2(M_{A}^2  + M_{Z}^2 -M_{h}^2) - M_{A}^2 M_{Z}^2 
\cos^2 2\beta } { M_{Z}^2 \cos^{2}{\beta}  +M_{A}^2 \sin^{2}{\beta} -M_{h}^2}
\label{dM22}
\end{eqnarray}
This assumption is valid in most cases as the by far dominant radiative correction from 
the stop--top sector that is quartic in the top quark mass, enters only in this entry
\cite{CR-1loop,CR-eff}: 
\begin{eqnarray}
\Delta {\cal M}_{22}^2 \sim  \frac{3 v^2\sin^2\beta}{8\pi^2} \lambda_t^4 \left[ \log 
\frac{M_S^2}{\overline{m}_t^2} + \frac{X_t A_t}{\,M_S^2} \left( 1 -\frac{X_t A_t}{12\,M_S^2} \right) \right]
\label{epsilon}
\end{eqnarray}
which depends on, besides $M_S$, the stop mixing parameter given by $X_t\!=\! A_t
\!-\! \mu/\tb$ with $A_t$ the stop trilinear coupling and $\mu$ the higgsino mass parameter.
$\lambda_t\!= \! \sqrt 2 \overline{m}_t/(v \sin\beta)$ is the top Yukawa coupling with $v$ 
the standard vacuum expectation value $v \approx 246$ GeV, and  $\overline{m}_t$ the running 
${\rm \overline{MS}}$ top quark mass to account for 
the leading two--loop corrections in a renormalisation group (RG) improved approach. 

The maximal value of the 
$h$ mass, $M_h^{\rm max}$ is given in the approximation above by 
\begin{eqnarray}
M_h^2 \to  M_Z^2 \cos^2 2 \beta + \Delta {\cal M}_{22}^2
\end{eqnarray}
and is obtained for the following choice of parameters \cite{benchmark-old}: a decoupling
regime with a heavy pseudoscalar $A$ boson,  large enough $\tb$ values that allow to
maximize the tree--level term $M_Z^2 \cos^22\beta \to M_Z^2$,  heavy stop squarks with a
sufficiently large $M_S$ value to enhance the logarithmic correction $\log (M_S^2/
\overline{m}_t^2)$ and, finally, a stop mixing parameter such that $X_t=\sqrt{6}M_S$, the 
so--called maximal mixing scenario that maximizes the stop loops and hence $M_h$. 
If the SUSY parameters are optimized as above, the maximal $M_h$ value can 
reach the level of $M_h^{\rm max} \approx 130$ GeV for $M_S$ of the order of the TeV 
scale, a range that is in general assumed in order to avoid a too large fine--tuning in 
the model. However, if $\tb$ is small, the tree--level contribution  $M_Z^2\cos^22\beta$ 
to the $h$ mass squared  becomes small as $|\cos2\beta| \! \to \! 0$, thus requiring a
substantial correction $\Delta {\cal M}_{22}^2$  to obtain a sufficiently large $M_h$. 
To achieve this, eq.~(\ref{epsilon}) shows that one has to substantially increase $M_S$. 

In Ref.~\cite{habemus}, the approximation $ \Delta{\cal M}^{2}_{11}, \Delta{\cal M}^{2}_{12} 
\ll  \Delta{\cal M}^{2}_{22}$  has been checked in various scenarios and found to be rather
good if $M_S$  much larger than the other soft--SUSY breaking parameters that enter 
the subleading radiative corrections, such as the higgsino mass $\mu$ 
and the sbottom trilinear coupling $A_b$ or more generally the sbottom mixing parameter 
$X_b\!= \!A_b \!-\!\mu \tan\beta$. 
This assumption should be particularly justified at low and moderate $\tan\beta$ values 
where first, the bottom--Yukawa coupling $\lambda_b = \sqrt 2 \bar m_b/ (v \cos\beta)$ is 
not strongly enhanced. In the approach of Ref.~\cite{Review} to parameterize the correction matrix 
of eq.~(\ref{mass-matrix}) including the dominant corrections from the stop  
and sbottom sectors  (and which has been used in Ref.~\cite{habemus} to check this second 
$h$MSSM  assumption),  the entries $\Delta{\cal M}^{2}_{11}$ and $\Delta{\cal M}^{2}_{12}$ 
of the mass matrix when $\lambda_b$ is set to zero are simply given at lowest order by
\cite{Review1,CR-eff}
\beq
\Delta {\cal M}_{11}^2 \simeq  - \frac{v^2 \sin^2\beta}{32 \pi^2} \lambda_t^4 \times 
\frac{\mu^2   X_t^2}{M_S^4} \ , \ \ 
\Delta {\cal M}_{12}^2 \simeq  - \frac{v^2 \sin^2\beta}{32 \pi^2} \lambda_t^4 \times 
\frac{ \mu X_t}{M_S^2}  \left( 6- X_t A_t/M_S^2 \right) 
\label{Mii-approx}
\eeq
They are proportional to $\mu/M_S$ and hence, are small if $|\mu| \lsim M_S$. Note that from 
the expressions above one can see that the two entries $\Delta{\cal M}^{2}_{11}$ and 
$\Delta{\cal M}^{2}_{12}$ are small not only for $M_S \! \gg \! |\mu|$, but also when 
stop mixing is small, $M_S \! \gg \! X_t $. For moderate $\tb$  (and also at large 
$\tb$ if the sbottom corrections can still be neglected),  one has $A_t \approx X_t$ and the off--diagonal entry is further suppressed for maximal  $X_t=\sqrt 6 M_S$.
Thus, the approximation of retaining only the entry $\Delta {\cal 
M}_{22}^2$ for the radiative corrections should be good at least at low $\tb$ where a very 
high SUSY scale is required to obtain a heavy enough $h$ state, suggesting that one naturally 
has $M_S \gg |\mu|$ and eventually also $M_S \gg X_t$.  
 
In this $h$MSSM approach the mass of the neutral CP even $H$ particle and the mixing angle 
$\alpha$ that diagonalises the $h,H$ states, will be given by the extremely simple expressions
\begin{eqnarray}
M_{H}^2 &= &\frac{(M_{A}^2+M_{Z}^2-M_{h}^2)(M_{Z}^2 \cos^{2}{\beta}+M_{A}^2
\sin^{2}{\beta}) - M_{A}^2 M_{Z}^2 \cos^{2}{2\beta} } {M_{Z}^2 \cos^{2}{\beta}+M_{A}^2
\sin^{2}{\beta} - M_{h}^2} \nonumber \\
\ \ \  \alpha &= &-\arctan\left(\frac{ (M_{Z}^2+M_{A}^2) \cos{\beta} \sin{\beta}} {M_{Z}^2
\cos^{2}{\beta}+M_{A}^2 \sin^{2}{\beta} - M_{h}^2}\right)
\label{hMSSM-output}
\end{eqnarray}
in terms of the inputs $M_A$, $\tb$ and the mass of the lightest $h$ state 
$M_h\!=\! 125$ GeV. 

The mass of the charged Higgs boson is simply given by the tree--level relation 
\begin{eqnarray}
M_{H^\pm}^2 = 
 { M_A^2 + M_W^2}
\label{eq:MH+}
\end{eqnarray}
as the SUSY radiative corrections in this particular case are known to be very small in 
general. According to Ref.~\cite{H+mass} where a detailed analysis of the radiative 
corrections has been recently performed, the leading one--loop correction to $M_{H^+}^2$ 
reads when expanding in powers of the SUSY scale as it is justified at low $\tb$ 
\begin{eqnarray}
\Delta M_{H^\pm}^2 = - \frac{3 \alpha}{16 \pi \sin^2\theta_W}  \frac{m_t^4}{M_W^2 \sin^2\beta} 
\frac{\mu^2}{M_S^2} +{\cal O} (1/M_S^4) 
\label{eq:MH+RC}
\end{eqnarray}
and is therefore very small for $M_S \gg |\mu|$. In fact, even for $M_S \approx |\mu|$
one obtains $\Delta M_{H^\pm}^2 \! \approx \! - 10^3~(250)~{\rm GeV}^2$ for $\tan\beta \approx 1
(\tb \gg 1)$ and, hence, a relative correction $|\Delta M_{H^\pm}/ M_{H^\pm}|$ that is only
about 5\% (1\%) for $M_{H^\pm} \approx 100$ GeV and negligibly small for higher $H^\pm$ 
masses. Hence, retaining only the tree--level relation eq.~(\ref{eq:MH+}) as done in the 
$h$MSSM should be a very good approximation in this case.

A third assumption of the $h$MSSM is that all couplings of the Higgs particles to fermions 
and gauge bosons are given in terms of $\tan \beta$ and the mixing angle $\alpha$ only and,
hence, the entire phenomenology of the Higgs particles is determined when the two inputs 
$\tb$ and $M_A$ are fixed. This means that possible corrections 
not incorporated in the mixing angle $\alpha$, such as direct vertex corrections, are assumed
to have a small impact\footnote{The direct corrections can be particularly important at high 
$\tb$ in the bottom--quark Yukawa coupling, as is the case of the so--called $\Delta_b$ correction which can 
be large if $ |\mu| \tan\beta \gg M_S$. We will show in the next section that even in this 
case, the impact of this direct correction is limited in LHC phenomenology.}. In particular, 
the couplings of the neutral Higgs bosons, collectively denoted by $\Phi$,  to 
up and down-type fermions and to massive gauge bosons (including the coupling of two 
Higgs and one gauge bosons)  when normalized to the SM--Higgs couplings, 
are simply given by:  
\beq 
\begin{tabular}{ccccc} 
$\  \Phi  \ $ & $ g_{\Phi \bar{u}u} $    &   $ g_{\Phi \bar{d} d} $  &
$g_{ \Phi VV} $  & $g_{ \Phi A Z} / g_{ \Phi H^+ W^-} $ \\ 
$h$ & $ ~~\cos\alpha/ \sin\beta ~~$  &  
$~~-\sin\alpha/ \cos\beta  ~~$  & $~~ \sin(\beta-\alpha) ~~$ & $~~~ \propto 
\cos(\beta-\alpha) ~~~$ \\ 
$H$ & $\sin\alpha/ \sin\beta$  & $\cos\alpha / \cos\beta$ & 
$\cos(\beta-\alpha)$ &  $ \propto \sin(\beta-\alpha) $ \\ 
$A$  & \ $\; {\rm cot}\beta \; $\ & \ $ \; \tb \; $ \   & \ $ \; 0 \; $ & 
$ \propto 0/1$ 
\end{tabular}
\label{Hcoup}
\eeq

The trilinear self-couplings among the Higgs bosons are also given in terms of $\beta$ and 
$\alpha$. This is clearly the case at tree--level but, to a good approximation, it
remains true when radiative corrections are incorporated. Indeed, besides the corrections that 
affect the angle $\alpha$ as discussed above, the trilinear couplings receive direct corrections whose 
dominant component turns out to be  simply the one that appears in the correction 
matrix $\Delta {\cal M}^{2}$ and hence, the correction $\Delta {\cal M}^{2}_{22}$ of 
eq.~(\ref{dM22}) \cite{Deltah3}. Thus, the trilinear MSSM Higgs couplings are also fixed in terms of $M_A, \tb$ and $M_h$ to a good approximation.  In units of $\lambda_0=-iM_Z^2/v$, 
the $Hhh$ and $hhh$ self--couplings,  which are the only ones that will matter for LHC 
phenomenology, will be then given by
\begin{eqnarray}
\lambda_{hhh} &=& 3\cos2\alpha \sin(\beta+\alpha) + 3 
\frac{\Delta {\cal M}^{2}_{22} }{M_Z^2} \frac{ \cos\alpha}{\sin\beta} \cos^2 \alpha
\nonumber \\[-3mm] 
\lambda_{Hhh} &=& 2\sin2\alpha \sin(\beta+\alpha) -\cos 2\alpha \cos(\beta+\alpha)  
+ 3 \frac{ \Delta {\cal M}^{2}_{22} }{M_Z^2} \frac{ \sin\alpha}{\sin\beta} \cos^2 \alpha
\label{eq:Hhh}
\end{eqnarray}
We note that at least for the $hhh$ self--coupling,  one should incorporate the 
radiative corrections in  the same approximation that has been used for the Higgs masses.  
This would be the only way to achieve a consistent decoupling limit and to make that the 
$\lambda_{hhh}$ self--coupling indeed reaches the SM value in this limit, 
$\lambda_{hhh}\!=\! 3M_h^2/M_Z^2$ for $\alpha\! = \! \beta \! -\! \frac{\pi}{2}$. For 
the sake of consistency, one should 
include the radiative corrections to the other self--couplings in the same approximation 
as for  $\lambda_{hhh}$. This then fully justifies the choice that we adopt in the $h$MSSM
and the expression of eq.~(\ref{eq:Hhh}) for the Higgs self--couplings. 

From the discussion above, one can conclude that the $h$MSSM approach has two very 
interesting aspects: its economy as only two input parameters are needed to describe 
the entire MSSM Higgs sector and its simplicity, as the Higgs masses and couplings are 
given by the very simple relations eqs.~(\ref{hMSSM-output})--(\ref{eq:Hhh}). This would 
allow to considerably simplify phenomenological analyses of the MSSM Higgs sector which, 
because of the large number of SUSY parameters to be taken into account, rely  up to now 
either on large scans of the parameter space or resort to benchmark scenarios in 
which most of these parameters are fixed. Nevertheless, the most interesting aspect of 
the $h$MSSM is that it easily allows  to describe scenarios with large values 
of the SUSY scale, $M_S \gg 1$ TeV, but weak--scale masses for the extended Higgs sector. 

Because of the large log($M_S/m_h)$ that occur, the high SUSY scale scenarios are 
notoriously difficult to analyze and,  before resuming the large logarithms,  the MSSM 
Higgs spectrum could not be calculated in a reliable way. Until very recently, this was
the case of the numerical tools that deal with the MSSM, such as the renormalisation 
group  program {\tt Suspect} \cite{suspect} or the program {\tt FeynHiggs} \cite{feynhiggs} 
that is more specialized on the MSSM Higgs sector, which were not reliable at too high $M_S$ 
outside the decoupling regime. A new version of {\tt FeynHiggs} in which some partial
resummation of the large logarithmic terms is performed has become available and allows to 
address low $\tb$ values in a somewhat reliable 
way\footnote{This new version of the program {\tt FeynHiggs} has been used to perform 
the following comparison \cite{Pietro}: for a $(\tb, M_A)$ set and given other  
MSSM inputs (those of the $M_h^{\rm max}$ scenario for instance), one calculates $\alpha, 
M_H$ and $M_h$ using {\tt FeynHiggs} and with the specific value obtained for $M_h$
(which is in general  $M_h \neq 125$ GeV), one recalculates the $h$MSSM values of $\alpha$ 
and $M_H$. The relative difference between {\tt FeynHiggs} and the $h$MSSM was found to 
be extremely small in the entire $[\tb, M_A]$ plane. Even at very low $M_A$ and $\tb$ values, 
the outputs for $M_H$  and $\alpha$ differ by less than 1\%. This proves once more that the 
second assumption of the $h$MSSM, i.e. that one can consider only the $\Delta M_{22}$ radiative
correction, is fully justified. We thank Pietro Slavich, who originally performed this 
comparison, for a discussion on this issue.}
(albeit with $M_h$ values still below $M_h \approx 125$ GeV at $\tb \lsim 2$). 
The $h$MSSM approach is currently being implemented 
in an updated version of the program {\tt Suspect}. 

An immediate advantage of the $h$MSSM  is that it re-opens the possibility 
of studying the MSSM low $\tb$ region \cite{Orsay}, which was for a long time 
overlooked. Indeed, as only SUSY scales of the order of $M_S\approx 1$ TeV were 
assumed in the analyses performed in the past, one always had a too light $h$ boson 
with a  mass below the limit $M_h \gsim 114$  GeV derived from the negative searches 
of a SM--like $h$ boson at the LEP2 collider \cite{PDG,LEP}. For a scale $M_S=1$ TeV, 
the possibilities $\tb\! \lsim\! 3$ and $\tb\! \lsim\! 10$ were excluded for, respectively 
the maximal--mixing scenario $X_t \! = \! \sqrt 6 M_S$ and the no--mixing scenario $X_t\!=\!0$. 
The situation became worse with the observation of the $h$ state at the LHC and the determination that its mass is $M_h\! \approx \! 125\;$GeV, i.e. well beyond the LEP limit.  
In fact, for $M_S\!  \approx\! 1$ TeV, this relatively large $M_h$ value cannot be reached   
in a large part of the [$\tb, M_A]$ parameter space that is being explored at the LHC
in the search for the additional Higgs bosons. 

Nonetheless, fixing the SUSY scale at $M_S \approx 1$ TeV is a very strong theoretical assumption and is currently challenged not only by the measured $M_h$ \cite{high-MS}
but also by direct sparticles searches at the LHC \cite{Craig}, especially in constrained 
MSSM scenarios. In the search for the MSSM Higgs bosons at
the LHC, one would like to avoid any such assumption and interpret the experimental results, 
for instance imposing the relevant experimental constraints in the absence of any evidence,
in a way that is as model-independent as possible.  The $h$MSSM approach, as no assumption 
on the SUSY scale nor on any other SUSY parameter (except eventually that they should be
smaller than $M_S$) is made, is more suitable in this respect. In fact, one is considering
simply in this case a two--Higgs doublet model of type II \cite{Lisbon} where the 
MSSM relations eqs.~(\ref{hMSSM-output}--\ref{eq:MH+}) are enforced; the superparticles
are assumed to be too heavy to have an impact on the Higgs sector (as it will be shown to
be generally the case in the next section). The only price to pay is that when the very low 
$\tb$ region is addressed, one is implicitly considering a very large SUSY--breaking scale,
making the MSSM a very unnatural and fine-tuned scenario.

\begin{figure}[!t]
\vspace*{-2.2cm}
\begin{center} 
\mbox{\hspace*{-2cm}
\includegraphics[scale=0.9]{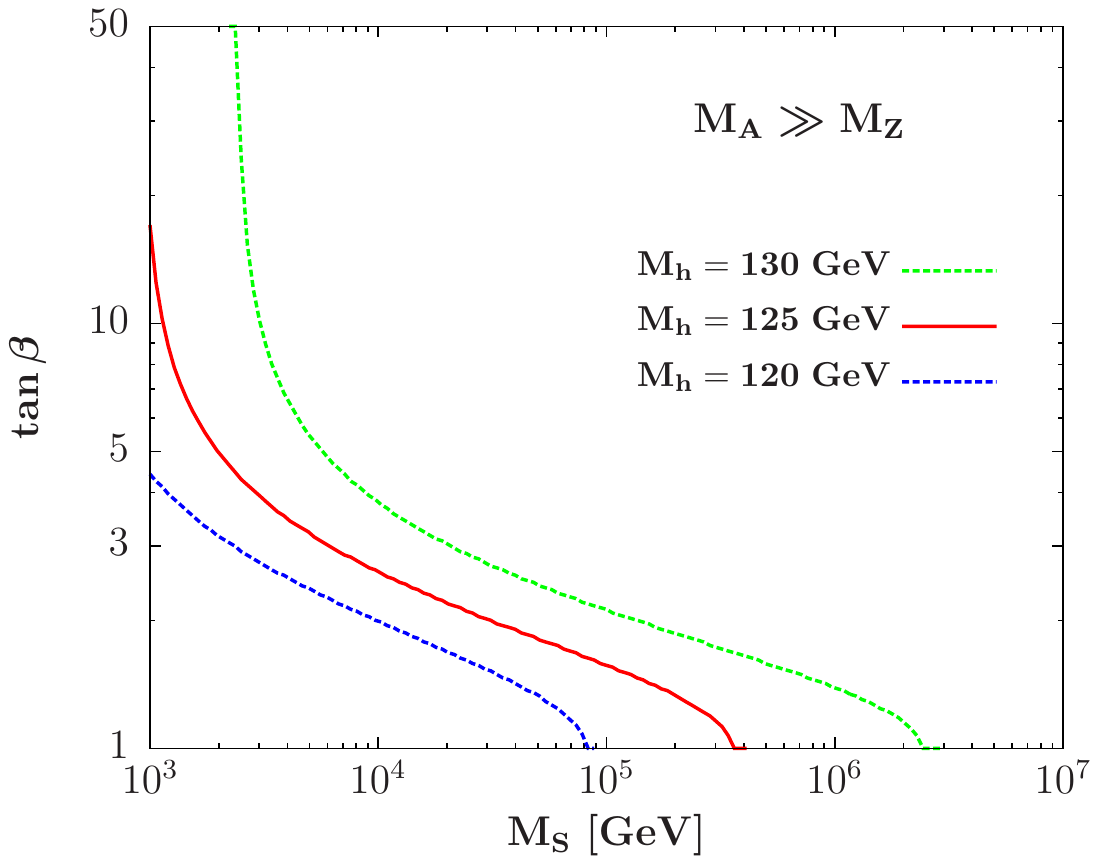}
}
\vspace*{-17cm}
\caption{Contour plots in the $[\tan\beta,M_{S}]$ plane where the there values $M_h=120, 
125$ and 130 GeV are obtained; the decoupling limit and maximal stop mixing are assumed. 
The SM inputs are $m_t=173$ GeV and $\alpha_S=0.118$ for the top quark mass and strong 
coupling constant
\cite{PDG}.}
\label{contour-MS}
\end{center}
\vspace*{-8mm}
\end{figure}

To illustrate this feature, we display in Fig.~\ref{contour-MS} contours in the $[\tb, M_S]$
plane in which one obtains the value $M_h\!=\!125$ GeV for the $h$ mass, as well as $M_h=
120$ and 130 GeV. The latter examples are when one  assumes that a possible mass shift of 
$\Delta M_h\!=\!5$ GeV  is missing from unaccounted for subleading corrections (e.g. the
contributions of the charginos and neutralinos that we have ignored here) or unknown higher
order terms (a theoretical uncertainty of $\approx 3$ GeV in the determination of $M_h$ 
is usually assumed \cite{DeltaH}). The limit $M_A\! \approx \!M_S$ and maximal stop mixing 
$X_t\!=\!\sqrt 6 M_S$ are assumed. 

The figure has been in fact obtained from an analysis of the split--SUSY scenario
where the large logarithms have been indeed resummed \cite{bds}. As can be seen, at 
high $\tb$, $M_S$ values in the vicinity of the TeV scale can be accommodate while in the low 
$\tb$ region, extremely large values of the SUSY scale $M_S$ are necessary to obtain  $M_h=125$ GeV. 

This is particularly the case for $\tb$ close to unity where a value $M_S 
\approx 400$ TeV is required for $\tb =1$. The situation becomes even worse for the more natural
small mixing situation $X_t\ll M_S$ and outside the decoupling regime when the tree--level
$h$ mass is reduced by Higgs mixing. In both cases, huge $M_S$ values will 
be needed for $\tb \approx 1$ to reach $M_h=125$ GeV. For $\tb\! \simeq \! 2$, 
the situation is less dramatic as in the configuration of Fig.~\ref{contour-MS}, only $M_S\!
\approx\! 20$ TeV is needed to reach $M_h\!=\!125$ GeV (or the target value $M_h \!
\approx \! 120$ GeV if uncertainties are included). We thus expect  that our $h$MSSM
approach should be valid down to $\tb$ values as low as $\tb \approx 2$, but in our analysis 
we will extend the validity of the approach to $\tb$ values close to unity.
 
\begin{figure}[!h]
\vspace*{-2cm}
\begin{center} 
\mbox{\hspace*{-4cm}
\includegraphics[scale=0.7]{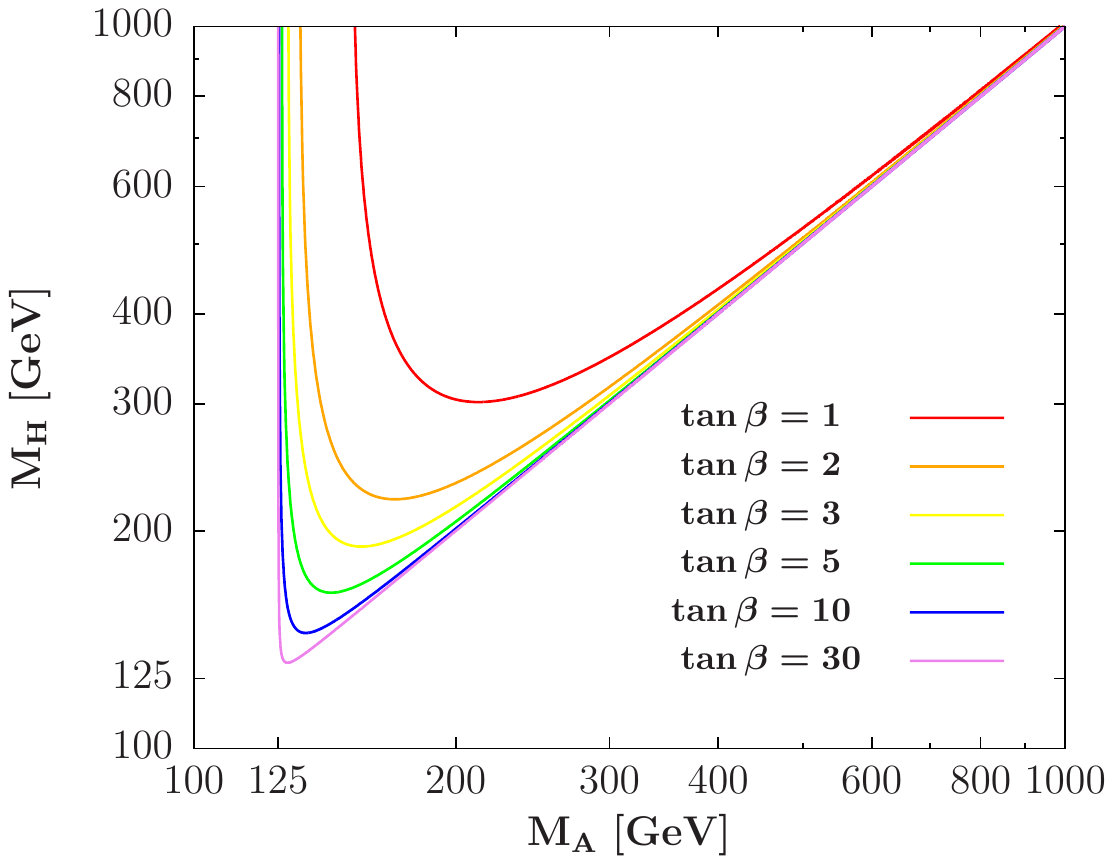}\hspace*{-7cm}
\includegraphics[scale=0.7]{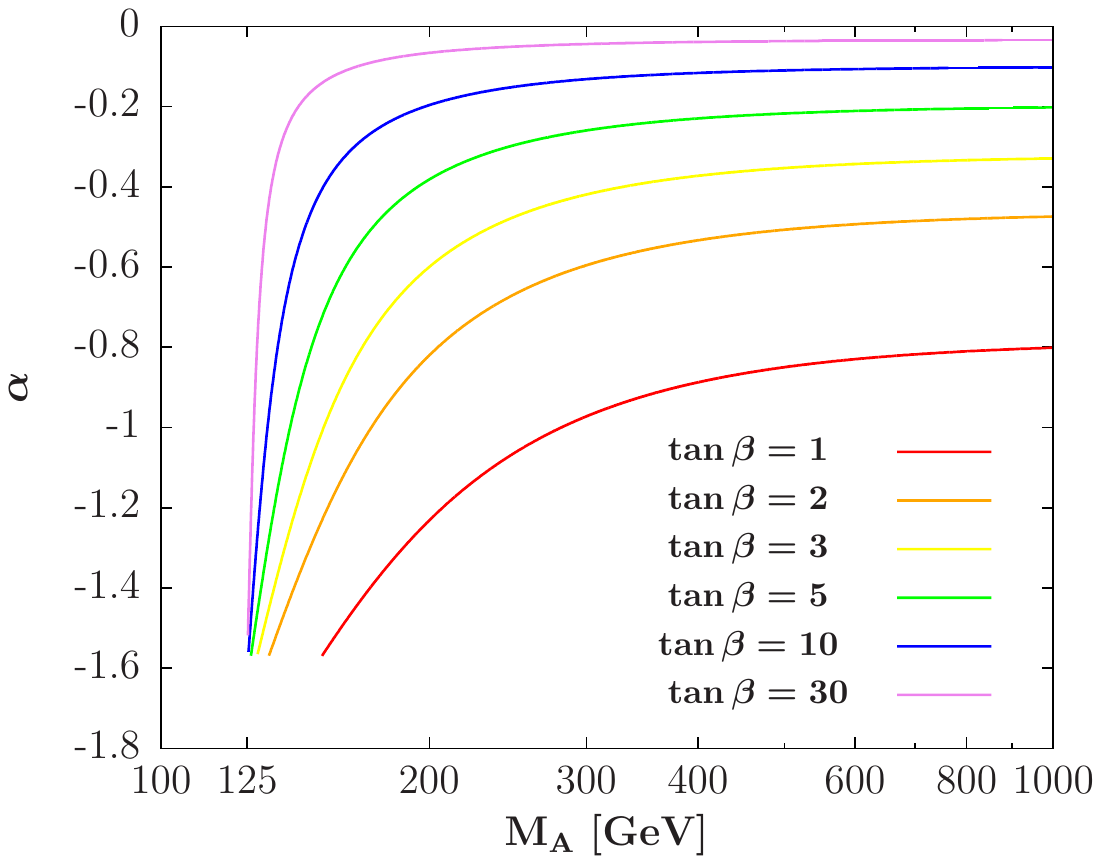} 
}
\vspace*{-12.9cm}
\caption{The CP--even $H$ boson mass (left) and the mixing angle $\alpha$ (right) 
in the $h$MSSM as a function of $M_A$ for representative values of $\tan\beta$;
the $h$ mass is fixed to the value $M_h=125$ GeV.}
\label{MH-alpha} 
\end{center}
\vspace*{-5mm}
\end{figure}

Let us now illustrate the values that one obtains for the two outputs of the $h$MSSM,  
the CP--even $H$ mass $M_H$ and the mixing angle $\alpha$; 
the charged Higgs mass is simply given by eq.~(\ref{eq:MH+}). These are shown in 
Fig.~\ref{MH-alpha} as a function of $M_A$ for several representative $\tb$ values, 
from unity to $\tb\!=\!30$. One can see that at sufficiently high $\tb$ values, $\tb \gsim 10$, 
$M_H$ becomes very close to $M_A$ and the angle $\alpha$ close to $\beta-\frac12 \pi$,
as soon as the pseudoscalar mass becomes larger than $M_A \gsim 200$ GeV.
This is a reflection of the well known fact that the decoupling limit, in which the $A$ and 
$H$ states are degenerate in mass and have the same couplings to fermions and vanishing couplings to gauge bosons, is attained very quickly at high $\tb$. Hence, the $h$MSSM 
approach should be a good approximation as it describes correctly this decoupling regime. 
In turn, at low $\tb$, the mass difference $M_H\!-\!M_A$ can be large and the angle $\alpha$ significantly different from $\beta- \frac12 \pi$ even for $M_A \! \approx \! 400$ GeV
meaning that the decoupling limit is reached slowly in this case.  (For $M_A\! \approx \! M_h$
we are close to the regime in which the $h$MSSM is not valid and one gets $M_H \! \to \! 
\infty$ and $\alpha -\pi/2$; this feature  will be discussed shortly).

\begin{figure}[!h]
\vspace*{-1.7cm}
\begin{center} 
\mbox{\hspace*{-4.cm}
\includegraphics[scale=0.7]{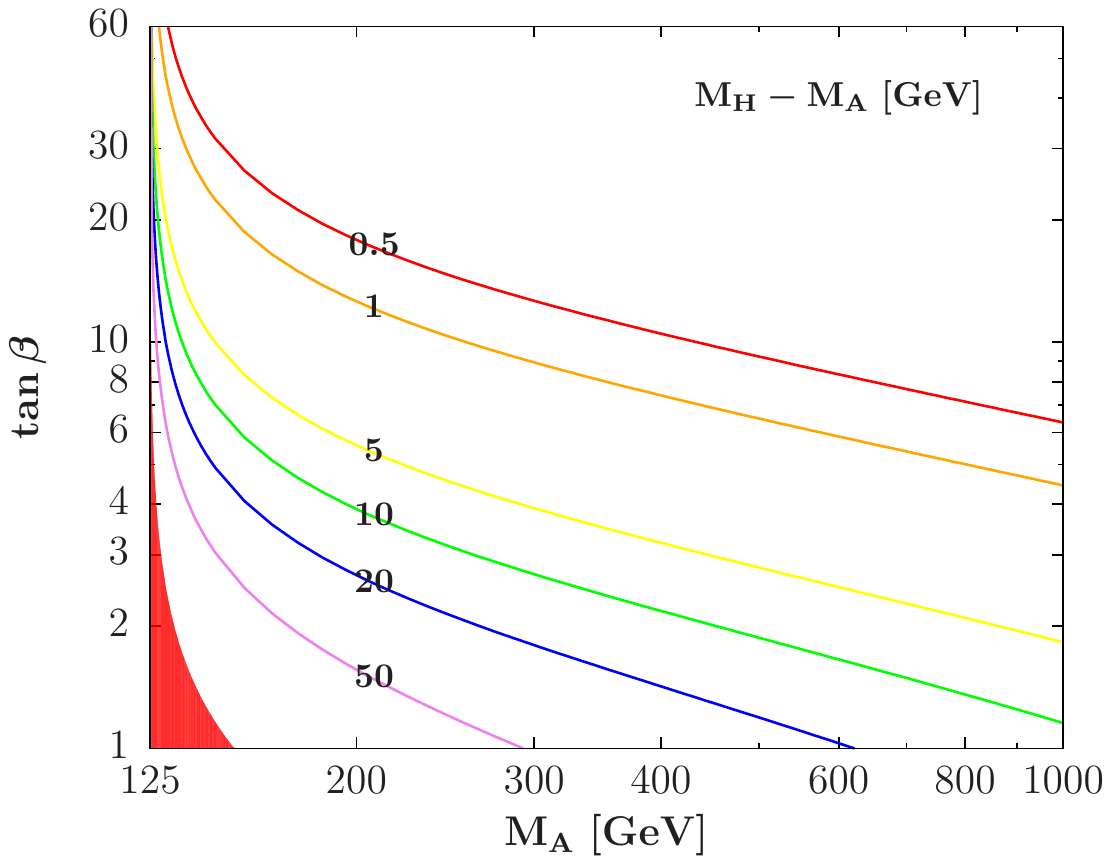}\hspace*{-7cm} 
\includegraphics[scale=0.7]{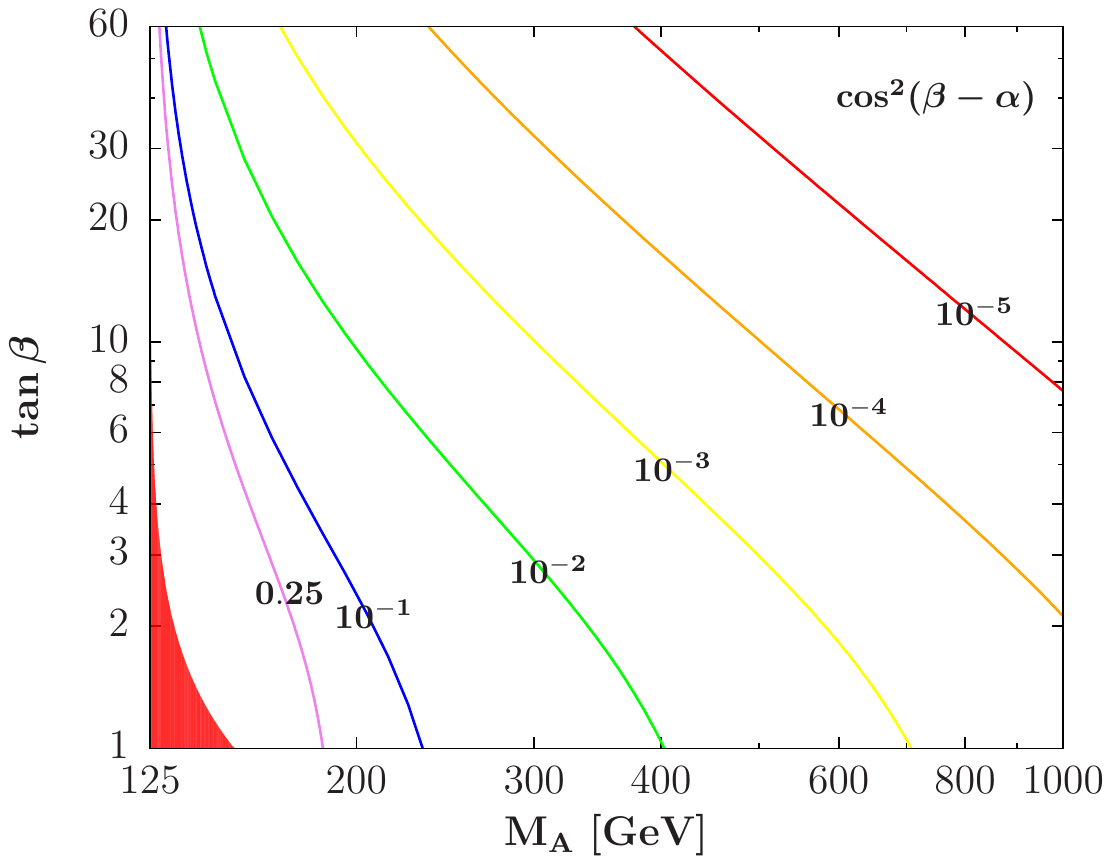}
}
\vspace*{-13.1cm}
\caption{Contours for the mass difference $M_H-M_A$ (left) and the coupling squared 
$\cos^2(\beta-\alpha)$ (right) in the [$\tb, M_A$] plane. The $h$ mass is fixed to the 
value $M_h=125$ GeV and  the red area at low $\tb$ and $M_A$ is the one where the $h$MSSM 
is ill defined. }
\label{differ} 
\end{center}
\vspace*{-5mm}
\end{figure}

This statement is made more explicit in Fig.~\ref{differ} where contours for the
heavier Higgs mass difference $M_H-M_A$ and the square of the reduced $H$ coupling  
to massive gauge bosons, $g_{HVV}^2\!=\!\cos^2(\beta-\alpha)$, which is a very good 
measure of the distance to the decoupling limit.
At $M_A \approx 500$ GeV, the difference $M_H-M_A$ is less than 1 GeV for $\tb \gsim 10$
while it is about 10 GeV for $\tan\beta \approx 2$. However, even in this case,
the mass difference represents about  2\% of the $A/H$ masses and, hence in view of 
the experimental resolution, one can still consider that the two states $A$ and $H$ 
are degenerate in mass.     

There is one problem with the $h$MSSM approach at very low $\tb$, though. If $\tb \approx 1$, 
the denominator of eq.~(\ref{dM22}) that expresses the dominant radiative correction 
$\Delta M_{22}^2$ in terms of the measured mass $M_h$ becomes close to $M_Z^2+ M_A^2- 2M_h^2$ 
and, at low $M_A$, it approaches zero rendering  eq.~(\ref{dM22}) ill defined. 
For $\tb \simeq 1$, the pseudoscalar mass range $M_A \lsim 160$ GeV is inaccessible, but the 
forbidden range shrinks to $M_A \lsim M_h$ for $\tb \gsim 3$. As a lower bound $M_A \! \gsim \! 
M_Z$ has  been set in the model--independent searches of the MSSM Higgs bosons 
at the LEP2 collider \cite{LEP}, this area in which the $h$MSSM is not defined is therefore 
rather small. We will show later that this  forbidden area is entirely 
excluded by the present LHC MSSM Higgs searches, in
particular by $H^\pm$ and $A$ searches as in both cases, the constraints can be interpreted  only in terms of $\tb$ and $M_A$ and hence without using 
the $h$MSSM relations of eq.~(\ref{hMSSM-output}).

In fact, this ``theoretically forbidden" 
hMSSM area is also excluded by the measurement of the observed Higgs boson production 
and decay rates at the LHC. Indeed, for these low $M_A$ and $\tb$ values, we are very far from 
the decoupling limit in which the couplings of the $h$ boson are close to their  SM--like  values,
as the LHC Higgs data in the various channels seem to strongly indicate \cite{ATLAS+CMS-h}.
However, we will refrain from using this argument to exclude the low $M_A, \tb$  possibility, as 
we will prefer to perform the direct Higgs searches in a model independent manner, without relying 
on any indirect constraint.

\FloatBarrier


\section{MSSM Higgs production and decays at the LHC}

We come now to the discussion of the decays and the production at the LHC of the 
heavier $A,H$ and $H^\pm$ particles in the $h$MSSM. We will be mostly interested  
in the low $\tb$ region, but we will first summarize the main features at 
high and moderate $\tb$. 

\subsection{Neutral Higgs decays}

At high $\tb$ values, say $\tb \gsim 10$, the decay pattern of the heavier neutral $H/A$
bosons is extremely simple \cite{Review,hdecay} as a  result of the strong enhancement of the 
couplings to down--type quarks and charged leptons that are proportional to $\tb$, not 
only for the 
$A$ state but also for the $H$ boson. Indeed, as in the decoupling limit $M_{A} \gg M_{Z}$ 
on has $\alpha \rightarrow \beta - \frac12 \pi$, the $Hb\bar b$ and $H\tau \tau$ couplings   normalised to the SM Higgs coupling take the limit
\begin{equation}
g_{Hdd} \equiv \cos\alpha / \cos\beta \stackrel{\small M_A \gg M_Z} \longrightarrow \tb  \equiv 
 g_{Add}
\end{equation}

The neutral $\Phi=A/H$ states will decay almost exclusively into  $\tau^+\tau^-$ and  
$b\bar b$ pairs,  with branching ratios of BR$(\Phi \to \tau \tau) \approx  10\%$ 
and BR$(\Phi \to b \bar b) \approx 90\%$. This is a simple consequence of the fact 
that the partial widths are proportional to  respectively $(m_\tau \tb)^2$ and $3 
(\overline{m}_b \tb)^2$ when the color factor is included; $m_\tau=1.78$ GeV and 
the $\overline{\rm MS}$ bottom quark mass defined at the scale of the Higgs mass 
is $\overline{m}_b \approx 3$ GeV, implying  
thus,   
\begin{equation}
{\rm BR} (\Phi \to \tau\tau) \approx 1- {\rm BR} (\Phi \to b\bar b) \approx  
m_\tau^2/( m_\tau^2+ 3 \overline{m}_b^2) \approx 0.1 \, . 
\end{equation}

At high $\tb$, all other decay channels of the $H/A$ states are  strongly 
suppressed. This is particularly the case of the decays into top quark pairs, despite 
of the large value $m_t\! \gg \! m_b$, as the Higgs coupling to up--type quarks are 
inversely proportional to $\tb$, 
\begin{equation}
g_{Huu} \equiv  \sin\alpha / \sin\beta \stackrel{\small M_A \gg M_Z} \longrightarrow 1/\tb 
\equiv  g_{Auu}
\end{equation}
rendering very small the $\Phi\!=\!H,A$ partial widths, given in the Born approximation by 
\begin{equation}
\Gamma(\Phi \rightarrow t \bar{t} ) = 3 G_F m_t^2/ (4\sqrt{2} \pi ) \times 
\, g_{\Phi t t}^2 \, M_{\Phi} \, \beta^{p}_t
\label{width-tt}
\end{equation}
where $\beta_t=(1-4m_t^2/M_{\Phi}^2)^{1/2}$ and  $p =3\,(1)$ for the CP--even
(CP--odd) Higgs boson.

This is also the case of Higgs decays involving gauge and Higgs particles in the final
state. In particular, one should have in principle also the decay modes $H\to VV$ with 
$V=W,Z$ and $H\to hh$ in the case of the CP--even and $A\to Zh$ in the case of the 
CP--odd Higgs bosons. However, the partial decay widths of the $H$ particle into massive 
gauge bosons $\Gamma(H\to VV)$ are proportional to the square of the reduced coupling
\begin{equation}
g_{HVV} \equiv  \cos (\beta-\alpha)  \stackrel{\small M_A \gg M_Z} \longrightarrow 0 \equiv  
g_{AVV}             
\end{equation}
which becomes zero in the decoupling limit as is the case for the pseudoscalar $A$ boson, 
that has no tree-level couplings to $VV$ states as a result of CP--invariance. 
For the latter state, the possibility $A \!\to \! hZ$  for $M_A 
\! \ge \!  M_h\! +\! M_Z \! \approx \! 220$ GeV, i.e. near the decoupling limit, will have a suppressed rate as the coupling $g_{AhZ} = g_{HVV}$ tends to zero at large $\tb$. Indeed, 
an expansion in terms of $1/M_A^2$ gives \cite{Orsay}
\beq
g_{HVV} = g_{AhZ}  \stackrel{\small M_A \gg M_Z} \longrightarrow  M_Z^2/(2M_A^2)
\times \sin4 \beta  - \Delta M_{22}^2/ (2M_A^2) \times  \sin 2 \beta \label{eq:HVV}
\eeq
and, at high $\tb$, both $\sin 4\beta$ and $\sin2\beta$ are proportional to
$\cot \beta$  so that the limit $g_{HVV} \to 0$ is reached faster in this case. 
The same is true for the decay $H\to hh$ when $M_H \!  \ge \! 2M_h$ as the trilinear 
Higgs coupling of eq.~(\ref{eq:Hhh}) for $M_H \gsim 2M_h$ reaches the limit
\beq
g_{Hhh}   \stackrel{\small M_A \gg M_Z} \longrightarrow - 3  
\Delta M_{22}^2 /( 2M_Z^2)  \times \sin 2 \beta \label{eq:HhhD}
\eeq
and is thus strongly suppressed at high $\tb$  that implies $\sin 2\beta \! = \! 2\tb
/(1\! + \! \tan^2\beta)\!   \to\! 0$. 

The situation is drastically different at low values of $\tb$ when the heavy Higgs 
states are kinematically allowed to decay into top quark pairs, $M_H\! \approx \! M_A  
\! \gsim \! 2m_t$ \cite{Review}. Indeed,  $H/A \to t\bar t$ become 
by far dominant $g_{\Phi tt} \propto m_t/\tb$ is so strong that it leaves no chance 
to the other possible channels. This is clearly the case for the $H/A \to  b\bar b, 
\tau^+\tau^-$ rates which become negligibly small as the 
couplings $g_{\Phi dd}$ are not enhanced anymore and $m_t \gg m_b, m_\tau$. 
This is also the case of the decays $A \to hZ$ and  $H \to VV$ at large  $M_\Phi$  
since the couplings approach zero in this case. For the $H\to hh$ decay, there 
is still a component of the $g_{Hhh}$ coupling of eq.~(\ref{eq:HhhD}), the one $\propto 
\sin2\beta \approx 1$ for $\tb \approx 1$, that is non--zero in the decoupling limit. 
However, besides  the fact that the $H tt$ coupling is larger than the $Hhh$ coupling, 
the partial decay width for the process $H\to hh$ decreases as $1/M_H$ 
\beq 
\Gamma(H \rightarrow hh ) =  (G_F M_Z^4)/ (16\sqrt{2} \pi M_H) \times 
\lambda_{Hhh}^2 \left(1- 4M_h^2/M_H^2\right)^{1/2}
\eeq
while $\Gamma(H\! \to\! tt) \propto M_H$ and, hence, largely dominates at 
high $M_H$ (slightly above $M_H \! \approx \! 2m_t$ however, BR$(H \! \to \! hh )$ 
stays significant as $\Gamma(H \! \to \! t\bar t)$ is suppressed by $\beta_t^3$ near
threshold).

The situation is opposite when the decays $H\to VV$ are considered. Indeed, because
of the contributions of the longitudinal components of the $V=W,Z$ bosons that grow
with the energy scale, the partial decay widths increases as $M_H^3$ and not as $M_H$
\begin{eqnarray}
\Gamma(H \rightarrow VV ) \! = \!  \frac{G_F \delta_V}{16\sqrt{2} \pi}
g_{HVV}^2 M_{\Phi}^3  \beta_V \left(1- \frac{4M_V^2}{M_H^2}+\frac{12 M_V^4}{M_H^4} 
\right) 
 \stackrel{\small M_A \gg M_Z} \longrightarrow \frac{G_F \delta_V}{16\sqrt{2} \pi}
g_{HVV}^2 M_{H}^3~~
\end{eqnarray}
where again $\beta_V=(1-4M_V^2/M_{H}^2)^{1/2}$ and $\delta_V=2(1)$ for $V=W(Z)$. At
low $\tb$ and high $M_H$ values, one of the components of $g_{HVV}$ given in 
eq.~(\ref{eq:HVV}) (the one $\propto \sin 4\beta$) vanishes while the other component 
tends to $g_{HVV}   \to  - \frac12  \Delta M_{22}^2 /M_A^2$. Because of the enhancement of the 
decay rate by $M_H^3$, one would have then a partial width $\Gamma(H\to VV)$ that is 
suppressed by a power $1/M_H$ only and hence, does not become completely negligible 
compared to $H\to t\bar t$ even at very high $M_H$. For instance, at $M_H \approx 500$ 
GeV, the branching ratios for the decays $H \to WW$ and $H \to ZZ$ are still at the
2\% and 1\% level respectively. This is appreciable and, at least in the case of the
$ZZ$ decay, it is of the same order of magnitude as the branching fraction of the 
observed 125 GeV $h$ boson, with the advantage that the $ZZ$ pair has a much larger 
invariant mass with a significantly smaller background (which compensates for the 
smaller Higgs production  cross section as will be seen later). 

If the $H/A$ states have masses below the $2m_t$ kinematical threshold, the two--body 
$H/A \to \bar t t$ decays are not allowed anymore. Off--shell three--body decays $A/H \to 
\bar t t^* \to \bar t bW$ are possible, but the rates are suppressed by an additional electroweak 
factor and the virtuality of one of the top quarks \cite{H3body}.  The gauge and Higgs decays of 
the $H/A$ states would then become significant at low $\tb$ values. In the mass range 
$M_h\!+\!M_Z \! \lsim \! M_A \! \lsim \! 2m_t$, the decay $A \to hZ$  will be dominant: 
the reduced coupling $g_{AhZ} = g_{HVV} \propto  M_h^2/M_A^2$ is only moderately suppressed and 
the full $AhZ$ coupling is  still substantial compared to the tiny $Ab\bar b$ coupling. 
Likewise, for $2M_h \lsim M_H \lsim 2m_t$, the decay mode $H\to hh$ is dominant 
as the coupling $g_{Hhh}$ at low $\tb$ will stay appreciable. 
The two--body decays into massive gauge bosons $H\to WW$ and $ZZ$ are also significant 
below the $2m_t$ threshold.

The bosonic decays will also be non--negligible at intermediate values of $\tb$, $\tb \approx 
\sqrt { \overline{m}_t/\overline{m}_b} \approx 5$--10, when the $A/H$ couplings to top 
quarks are suppressed while those to bottom quarks are not yet strongly enhanced. However, 
below the $2m_t$ threshold, when the Higgs couplings to the bosonic states are not 
too suppressed and the only competition will be due to the $\Phi \to 
b\bar b$ decays that is only slightly enhanced, the rates for the $H\to WW,ZZ,hh$ and 
$A \to Zh$ channels will be smaller than at low $\tb$ values.      

\begin{figure}[!h]
\vspace*{-1.5cm}
\mbox{
\hspace*{-7mm}
\includegraphics[scale=0.28]{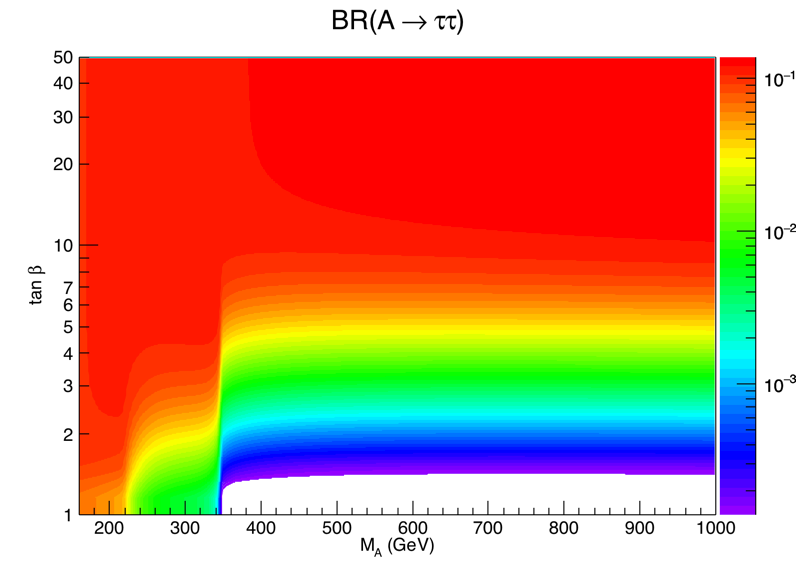} 
\hspace*{-2mm} 
\includegraphics[scale=0.28]{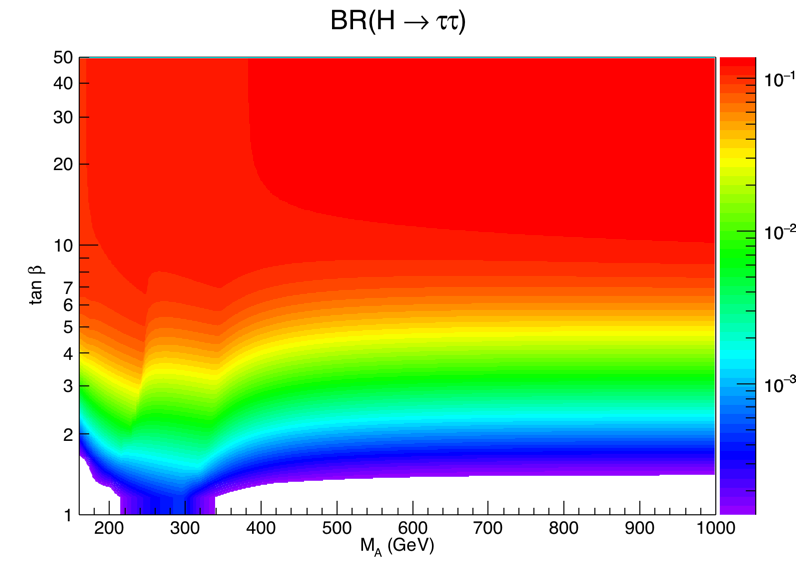} \\
}\\[1mm]
\mbox{
\hspace*{-7mm}
\includegraphics[scale=0.28]{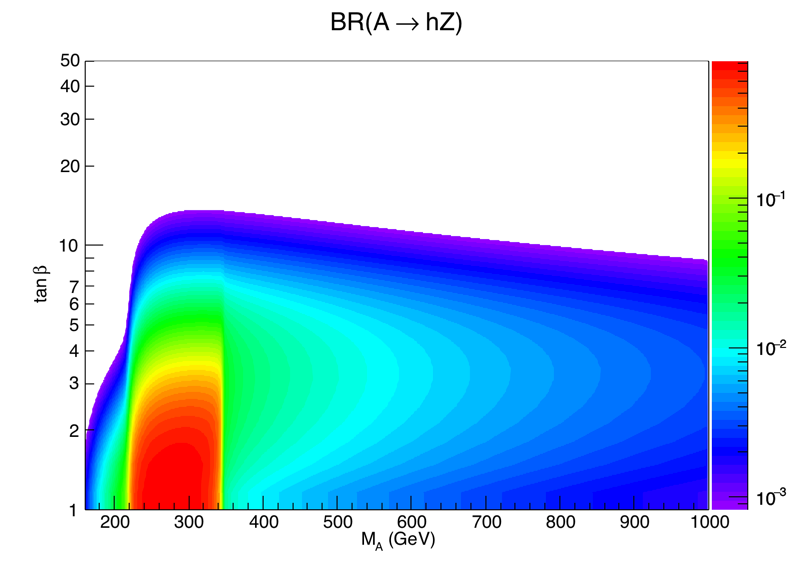}  
\hspace*{-2mm}
\includegraphics[scale=0.28]{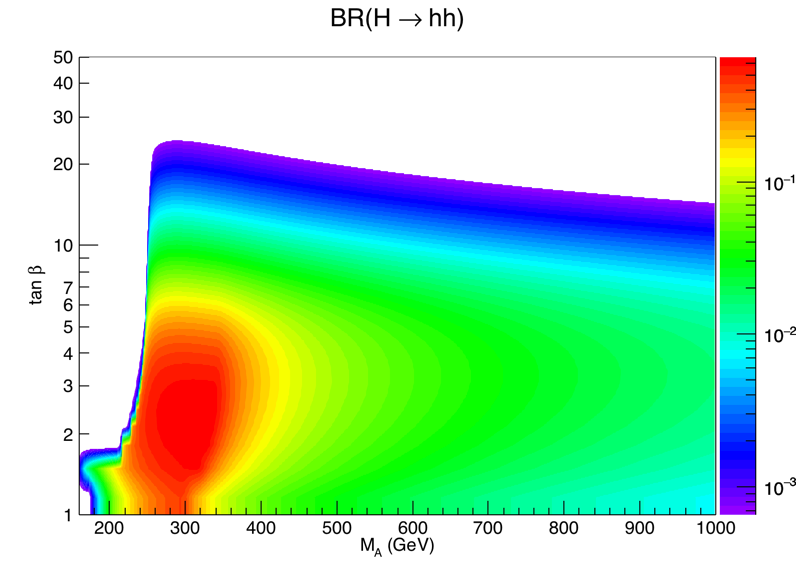} 
} \\[1mm]
\mbox{
\hspace*{-7mm}
\includegraphics[scale=0.28]{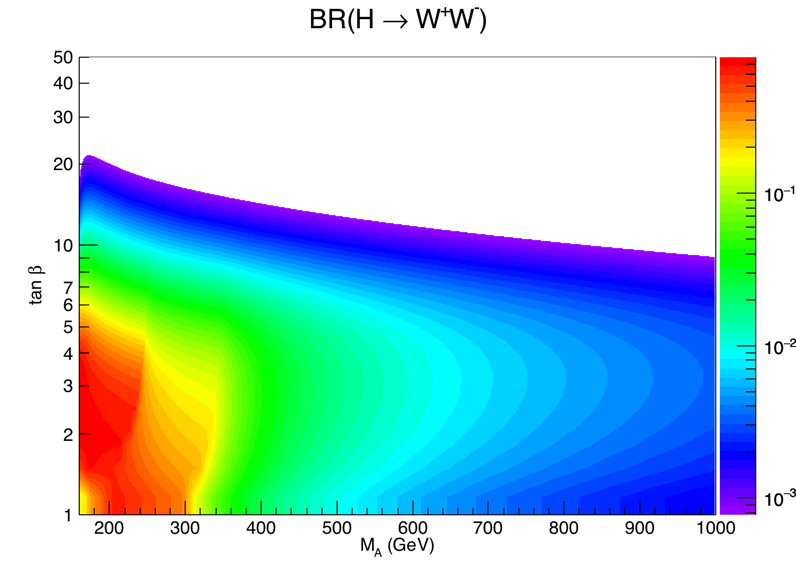} 
\hspace*{-2mm}
\includegraphics[scale=0.28]{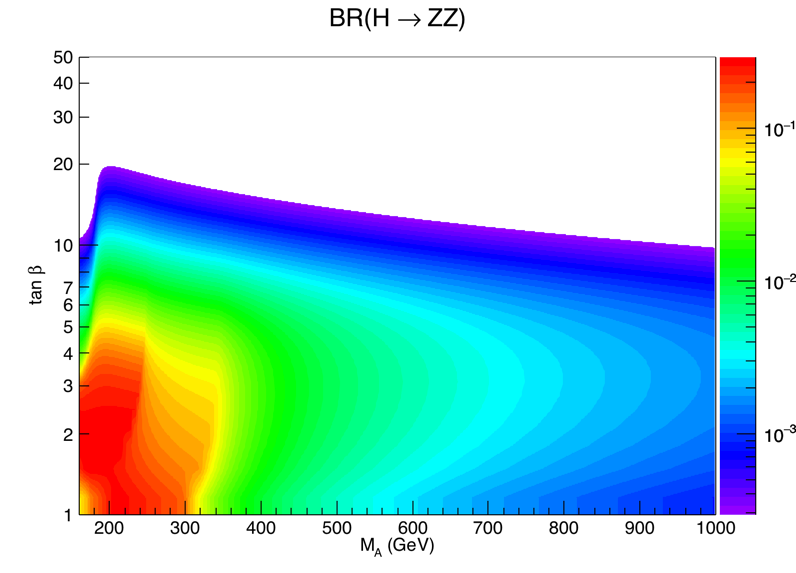} 
}\\[1mm]
\mbox{
\hspace*{-7mm} 
\includegraphics[scale=0.28]{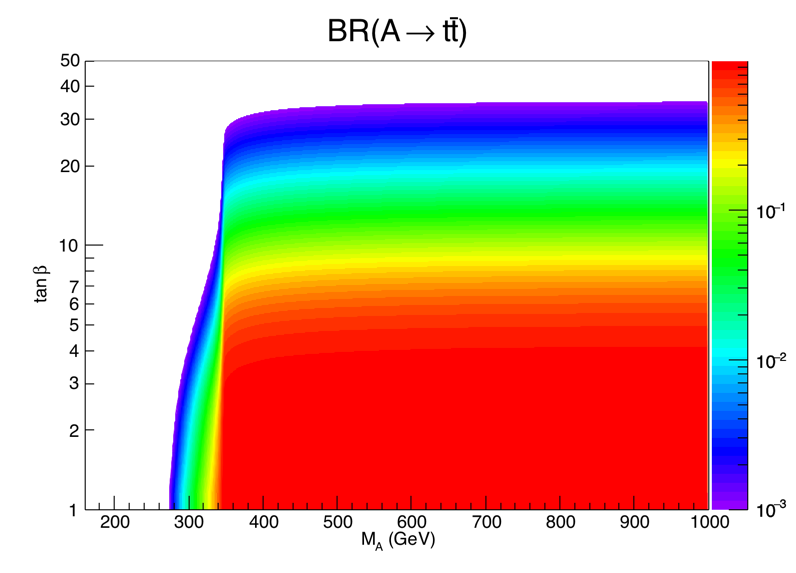}
 \hspace*{-2mm} 
\includegraphics[scale=0.28]{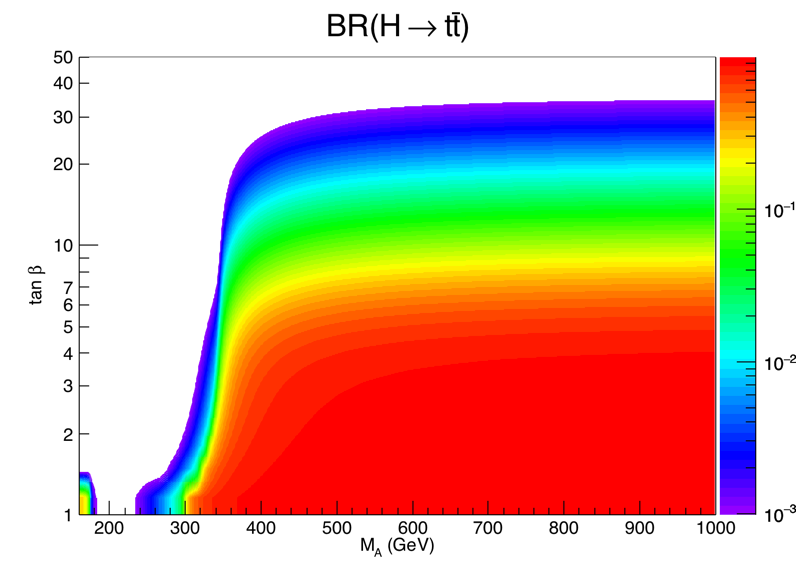}
}
\vspace*{-7mm}
\caption{The branching ratios of the neutral Higgs bosons in the 
$[\tb, M_A]$ parameter space of the $h$MSSM with the constraint $M_h=125$ GeV. Shown 
are the rates for the decays  $A/H\to \tau \tau$ (top), $A\to hZ$  and $H\to hh,WW,ZZ$ 
(middle) and $A/H\to t\bar t$ (bottom).}
\label{fig:Brs}
\end{figure}

The branching fractions for the various Higgs decays discussed above are displayed 
in Fig.~\ref{fig:Brs} in the $[\tb, M_A]$ plane assuming the $h$MSSM with $M_h=125$
GeV. The Fortran program {\tt HDECAY} \cite{hdecay} where the $h$MSSM relations were
implemented has been used. The color code is such that the red  area is when the 
considered decay rates are large, while the blue area is when they are 
small (for alternative and more easily readable figures for the branching ratios,
as well as for the production cross sections, see for instance 
Refs.~\cite{habemus,Orsay}). The white 
areas are when the decay rates are very small, below the minimal value of the scale 
in the color axis. As can be seen, the $H/A\to \tau \tau$ decays are important at high 
$\tb$ values. The branching ratios for the decays $H/A\to  b\bar b$ follow that
of $\tau \tau$ final states when  a factor 9 is included and are, hence, largely 
dominant. The decays $H/A \to t\bar t$ are by far leading at low $\tb$ for 
$M_{A,H} \gsim 350$ GeV (one notices that at least for $A$, they are also significant slightly 
below the $t\bar t$ threshold). The bosonic decays $H\to WW,ZZ,hh$ and also
$A \to hZ$ have reasonable rates 
only for $\tb \lsim 3$ and below the $2m_t$ threshold. 

\begin{figure}[!h]
\mbox{
\hspace*{-6mm}
\includegraphics[scale=0.28]{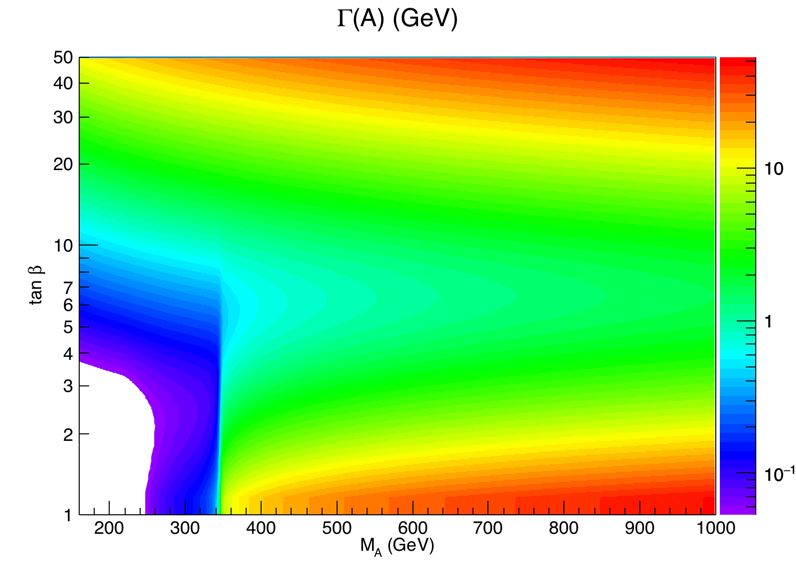} 
\hspace*{-4mm}
\includegraphics[scale=0.28]{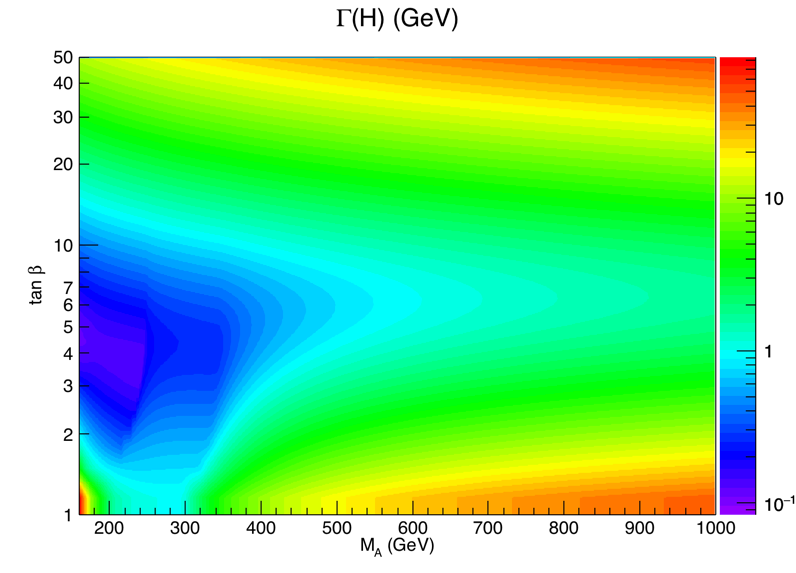}
}
\vspace*{-7mm}
\caption{The total decay widths of the neutral $A$ (left) and $H$ (right)
bosons in the $[\tb, M_A]$ parameter space of the $h$MSSM with the constraint $M_h=125$ GeV.}
\label{fig:Gam}
\vspace*{-3mm}
\end{figure}

A final word should be devoted to the total Higgs decay widths, which are displayed for 
the $A$ and $H$ particles in Fig.~\ref{fig:Gam}, again in the $[\tb, M_A]$ $\h$MSSM plane. 
In the low and high $\tb$ regimes, 
one can consider only the dominant fermionic decays of the $\Phi$ states and, to a good approximation, the total decay widths  read (up to phase space factors)
\beq
\Gamma_H^{\rm tot} \approx \Gamma_A^{\rm tot} (\approx \Gamma_{H^\pm}^{\rm tot})  
\approx G_F M_{\Phi} / (4\sqrt{2} \pi) 
 \, [(m_\tau^2 + 3 \overline{m}_b^2)\tan^2 \beta + 3m_t^2 \cot^2\beta]
\label{gamtot}
\eeq
For $\tb \! \approx \! 1$ and $\tb  \! \approx  \! 60$, one obtains a total decay width that is approximately $\Gamma_\Phi^{\rm tot}  \! \approx  \! 5\% M_\Phi$ and, compared to
the Higgs mass, it is not very large. Hence, to a good approximation the $A/H$ states 
can be considered as narrow resonances in most cases. 

Note that for the $H$ state, as the branching ratios and total decay width are
shown as a function of $M_A$, some peculiar features can be observed. These are explained  
by the fact that there is a large  splitting between $M_H$ and $M_A$ at low $\tb$ and $M_A$
values which lead to, for instance, the opening of the $H\to t\bar t$ mode already at $M_A \lsim  
200$ GeV and, hence, suppressed $H\to WW,ZZ$ decays but a large total decay width for
$\tb \approx 1$.


\FloatBarrier

\subsection{Neutral Higgs production}

Let us turn now to the production of the neutral MSSM $\Phi=H/A$  bosons at the 
LHC. Also in this case,  the cross sections crucially depend on the considered $\tb$ 
regime and in most cases, the two processes that play a leading role are the 
gluon--fusion mechanism $gg\!  \to\!  \Phi$ which is initiated by a heavy quark loop
\cite{ggH} and the associated Higgs production with $b$--quarks, $gg/q\bar q \! \to \! 
b\bar b \! 
+\! \Phi$, which at high energies and if no--additional $b$--quark is considered in 
the final state, is equivalent to the fusion process  $b \bar b \to \Phi$ \cite{bbH}.
All other processes, in particular vector boson fusion and associated production with 
a massive gauge boson for the CP--even $H$ state, $qq \to Hqq$ and $q \bar q \to HV$, 
and associated production with top--quark pairs for both the $H$ and $A$ states, $pp 
\to \Phi t \bar t$, have much smaller rates as the couplings $g_{HVV}$ and $g_{\Phi tt}$ 
are suppressed and/or the available phase space is not favorable. 

At leading order in perturbation theory, the partonic cross sections for the $b\bar b \to 
\Phi$ and $gg \to \Phi$ processes can be written in terms of the partonic c.m. energy 
$\hat s$ and $M_\Phi$ as \cite{Review0}
\beq
\hat \sigma(b \bar b \to \Phi) &=& \frac{\pi}{12} \; g_{\Phi bb}^2 \: 
\delta ( \hat s -M_\Phi^2) \\
\hat \sigma(gg\to \Phi) &=& \frac{G_F \alpha_s^2}{288 \sqrt 2 \pi} \,
M_\Phi^2 \delta (\hat s- M_\Phi^2)  \, \bigg| \frac34 \sum_Q 
g_{\Phi QQ} A_{1/2}^\Phi (\tau_Q)\bigg|^2
\eeq
In the case of the $gg\to \Phi$ process, the quarks $Q$ running in the loop should be
taken to be the heavy bottom and top quarks with Higgs couplings given in eq.~(\ref{Hcoup})
and masses incorporated into the reduced variables $\tau_Q= M_\Phi^2/4m_Q^2$.  The 
form factors for spin--$\frac{1}{2}$ fermion loops in the case of a CP--even $H$ and a 
CP--odd $A$ bosons are given by
\begin{eqnarray}
A_{1/2}^H(\tau) & = & 2 [\tau +(\tau -1)f(\tau)]\, \tau^{-2}  \nonumber \\   
A_{1/2}^A(\tau) & = & 2\tau^{-1}f(\tau)
\label{eq:A1/2}
\end{eqnarray}
where the function  $f(\tau)$ above and below the $\tau=1$ kinematical threshold is defined as
\begin{eqnarray}
f(\tau)=\left\{
\begin{array}{ll}  \displaystyle
\arcsin^2\sqrt{\tau} & \tau\leq 1 \\
\displaystyle -\frac{1}{4}\left[ \log\frac{1+\sqrt{1-\tau^{-1}}}
{1-\sqrt{1-\tau^{-1}}}-i\pi \right]^2 \hspace{0.5cm} & \tau>1
\end{array} \right.
\label{eq:ftau}
\end{eqnarray}
While the amplitudes are real for $M_\Phi \le 2m_Q$, they develop an 
imaginary part above the kinematical threshold. At very low Higgs masses, compared 
to the internal quark mass, the amplitudes for a scalar and a pseudoscalar states
reach constant but different values
\beq 
M_\Phi \ll 2 m_Q  :\  A_{1/2}^H (\tau_Q) \to 4/3 \ \ ,  \ A_{1/2}^A (\tau_Q) \to 2
\eeq
Instead, in the opposite limit, $M_\Phi \gg 2 m_Q$, chiral symmetry holds and 
the amplitudes for the CP--even and a CP--odd Higgs bosons are identical (as in 
the $b\bar b \to \Phi$ case),  
\beq 
M_\Phi \gg 2 m_Q  &:& A_{1/2}^\Phi (\tau_Q) \to   - [\log(4\tau_Q)-i\pi]^2  
/(2 \tau_Q)
\eeq
The maximal values of the amplitudes occur slightly above the kinematical threshold 
where one has for the real parts Re($A^H_{1/2}) \sim 2$ and Re($A^A_{1/2}) \sim 5$.  

At high $\tb$, the strong enhancement of the Higgs couplings to $b$--quarks and the 
suppression of the couplings to top quarks and gauge massive bosons makes that only 
these two processes are relevant, with the gluon--gluon fusion mechanisms dominantly
generated by the bottom quark loop. The cross sections $\sigma(gg \! \to\!  \Phi)$ and 
$\sigma(b\bar b \! \to\!  \Phi)$ are of the same order of magnitude and can be so large 
that  they make the process $pp\! \to \!gg\! +\! b \bar b \! \to \! \Phi \! \to \! 
\tau^+ \tau^-$, with the branching fraction for the decay $\Phi \! \to \! 
\tau^+ \tau^-$ being of order 10\% as seen previously, 
the most powerful LHC search channel for the heavier MSSM Higgs 
bosons. The $pp \to b \bar b +H/A$ mode with $A/H\to b\bar b$, which has an order of 
magnitude larger rates in principle (if no high--$p_T$ $b$--quark from production 
is required), has also been considered but the sensitivity is smaller 
as this fully hadronic  process is subject to a much larger QCD background.

At high $\tb$, the production rates are approximately the same for the $H$ and $A$ states 
in both the $b\bar b$ and $gg$ fusion cases as discussed earlier. While $\sigma(bb\to \Phi)$ 
is known up to NNLO in QCD perturbative theory \cite{bbH-NNLO}, $\sigma(gg\to \Phi)$ is instead 
known only up to NLO in the limit  $M_\Phi \gsim 2m_Q$ that we will be mainly interested in 
\cite{ggH-NLO}. For the top-quark loop, we will nevertheless also include the NNLO QCD
corrections \cite{ggH-NNLO} that are in principle only valid for $M_\Phi \lsim 2m_Q$ as 
advocated in Ref.~\cite{LHCXS}. The precise values of the cross 
section times branching fractions $\sigma (pp \to \Phi) \times {\rm BR }(\Phi \! \to \! 
\tau^+ \tau^-)$ 
for a given [$\tb, M_A]$ MSSM point have been updated in Refs.~\cite{LHCXS,BD} and the 
associated theoretical uncertainties from missing higher order perturbative contributions,
the parametrisation of the parton distribution functions and uncertainties on the 
inputs $\alpha_s$ and $b$--quark mass, have been estimated to be of the order of 25\%. 
Any effect below this level, such as the SUSY effects that we will be discussed later
in this section, should be considered as small. 

Again, at low $\tb$, the situation is very different. The cross sections for the 
$b\bar b\to \Phi$ processes are now very small as the $\Phi b\bar b$ coupling is not 
enhanced anymore. For $M_A \gsim 200$ GeV, this is also the case of the associated 
production with $t\bar t$ pairs as a result of a the small phase--space and, in the 
case of the $H$ state, of the vector boson fusion $qq \to Hqq$ and associated production 
with a gauge boson $q\bar q \to HV$ as a result of the suppressed $HVV$ coupling (the $A$
state cannot be produced in these two processes as there is no $AVV$ coupling). The 
only process which would have a reasonable production cross section would be the 
gluon-fusion process $gg \to \Phi$ with, this time, the leading contribution being 
generated by loops of top quarks that have couplings to the Higgs bosons that are only
slightly suppressed compared to the SM Higgs case. 


\begin{figure}[!h]
\mbox{
\hspace*{-7mm}
\includegraphics[scale=0.27]{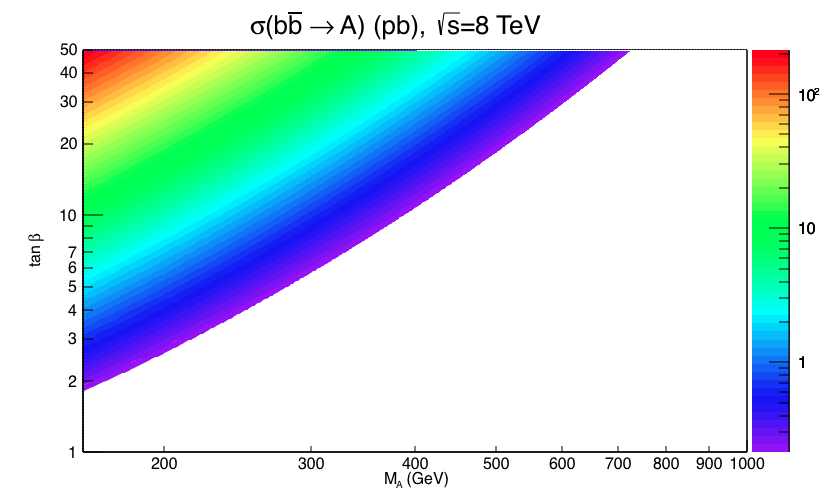} 
\hspace*{-3mm}
\includegraphics[scale=0.27]{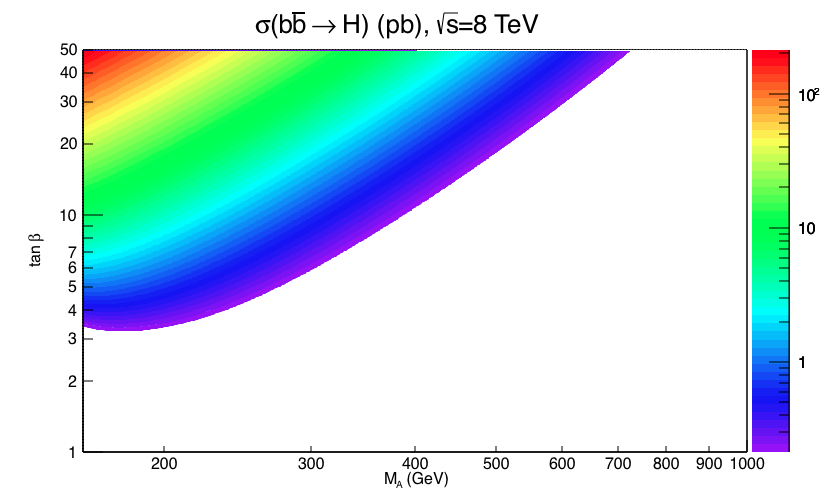}
}\\[2mm]
\mbox{
\hspace*{-7mm}
\includegraphics[scale=0.27]{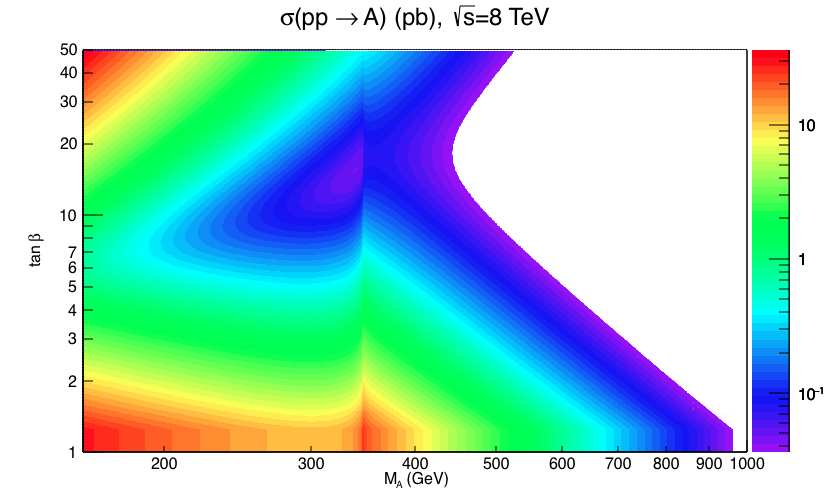}
\hspace*{-3mm}
\includegraphics[scale=0.27]{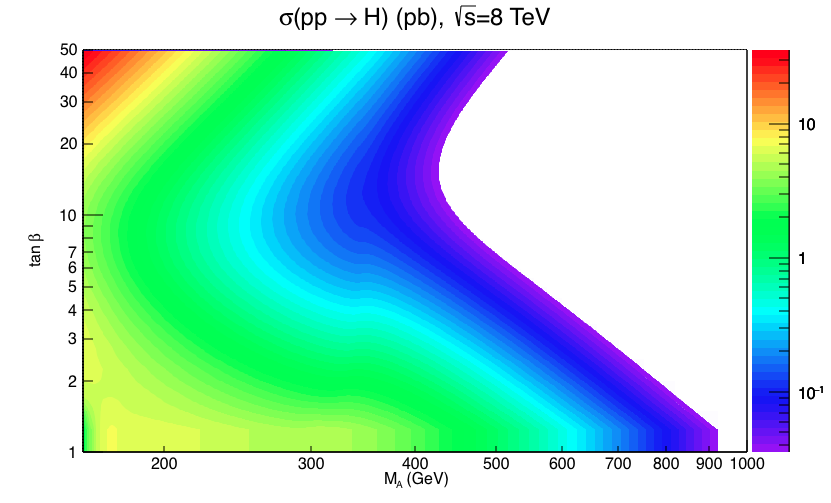}
}\\[2mm]
\mbox{
\hspace*{-7mm}
\includegraphics[scale=0.27]{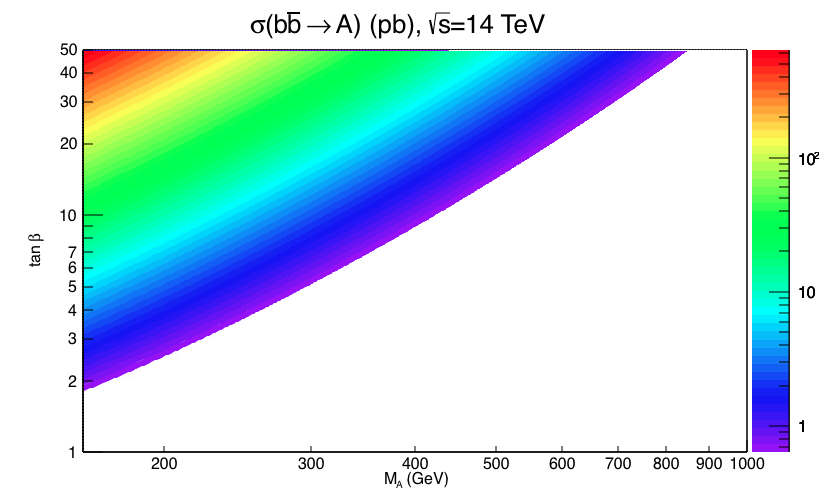}
\hspace*{-3mm}
\includegraphics[scale=0.27]{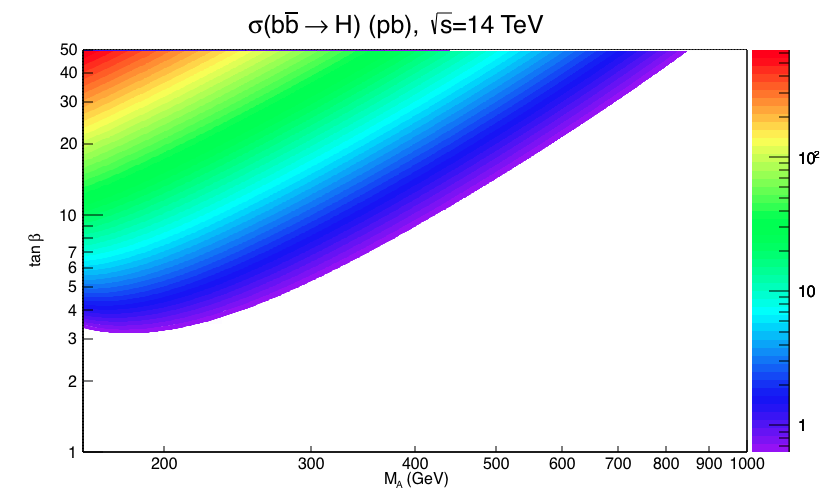}
}\\[2mm]
\mbox{
\hspace*{-7mm}
\includegraphics[scale=0.27]{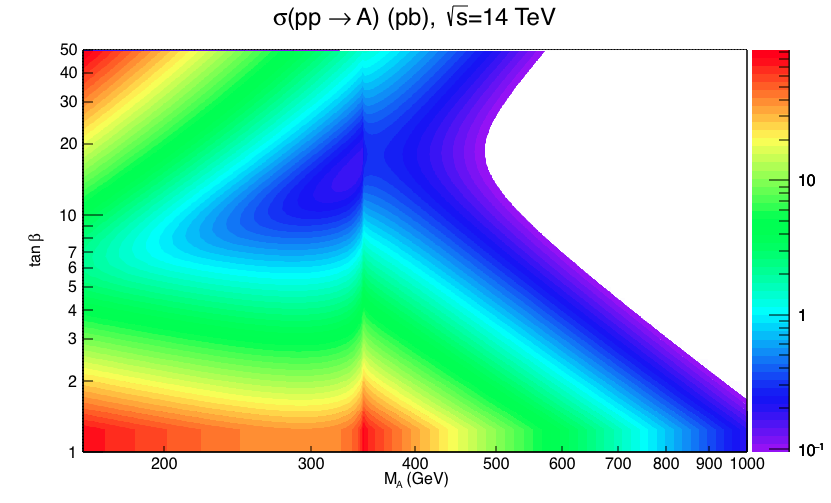}  
\hspace*{-3mm}
\includegraphics[scale=0.27]{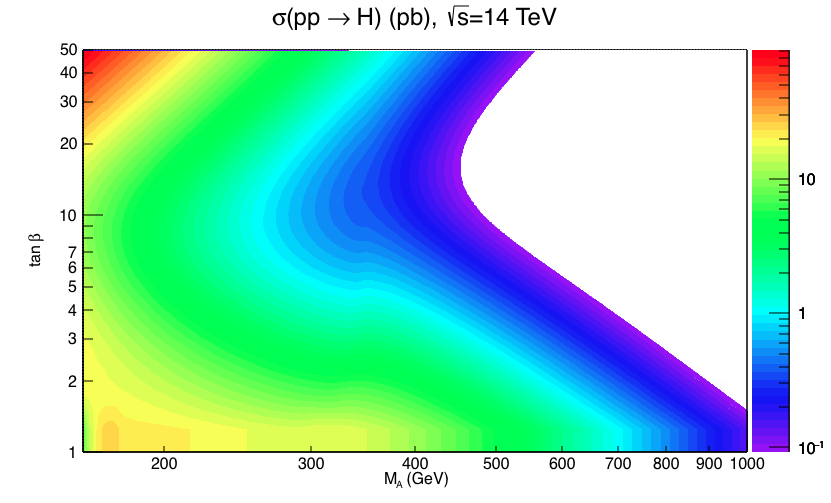}
}
\vspace*{-3mm}
\caption{The production cross sections of the MSSM heavier Higgs bosons $A$ (left) 
and $H$ (right) at the LHC with $\sqrt{s}=8$~TeV (top) and 14 TeV (bottom) in the 
$[\tan\beta,M_{A}]$ $\h$MSSM plane; the channels $gg\to H/A$ and $b\bar b \to H/A$
have been considered.}
\label{bbPhi}
\vspace*{-5mm}
\end{figure}


The production cross sections $\sigma(gg\!\to\! \Phi+X)$ and $\sigma(b\bar b \!\to\! 
\Phi+X)$ with $\Phi\!=\!A$ (left) and $\Phi\! =\! H$ (right) are displayed at the LHC in 
the  $[\tb, M_A]$ $\h$MSSM parameter space for $\sqrt s=8$ TeV (top) 
and 14 TeV (bottom); the MSTW parton distribution functions \cite{MSTW} have been used. 
The rates for $b\bar b \to \Phi$ have been obtained using the 
program {\tt SusHi} \cite{SUSHI} and an adapted version of {\tt HIGLU} \cite{HIGLU} 
has been used for $gg \to \Phi$. As can be seen, the cross sections
are rather 
large in particular at high $\tb$ and, for $gg\to \Phi$, also at low $\tb$ when the Higgs 
couplings to $b$-- or $t$--quarks are strong and at relatively low $M_A$ when  
the phase space is not too penalizing. Even for $M_A=500$ GeV (1 TeV), the production 
rates are significant  at $\sqrt s = 8$ TeV (14 TeV), if $\tb$ is sufficiently
high or low.   

\subsection{The case of the charged Higgs boson} 

\begin{figure}[!h]
\vspace*{-2.5cm}
\begin{center} 
\mbox{\hspace*{-1.cm}
\includegraphics[scale=0.38]{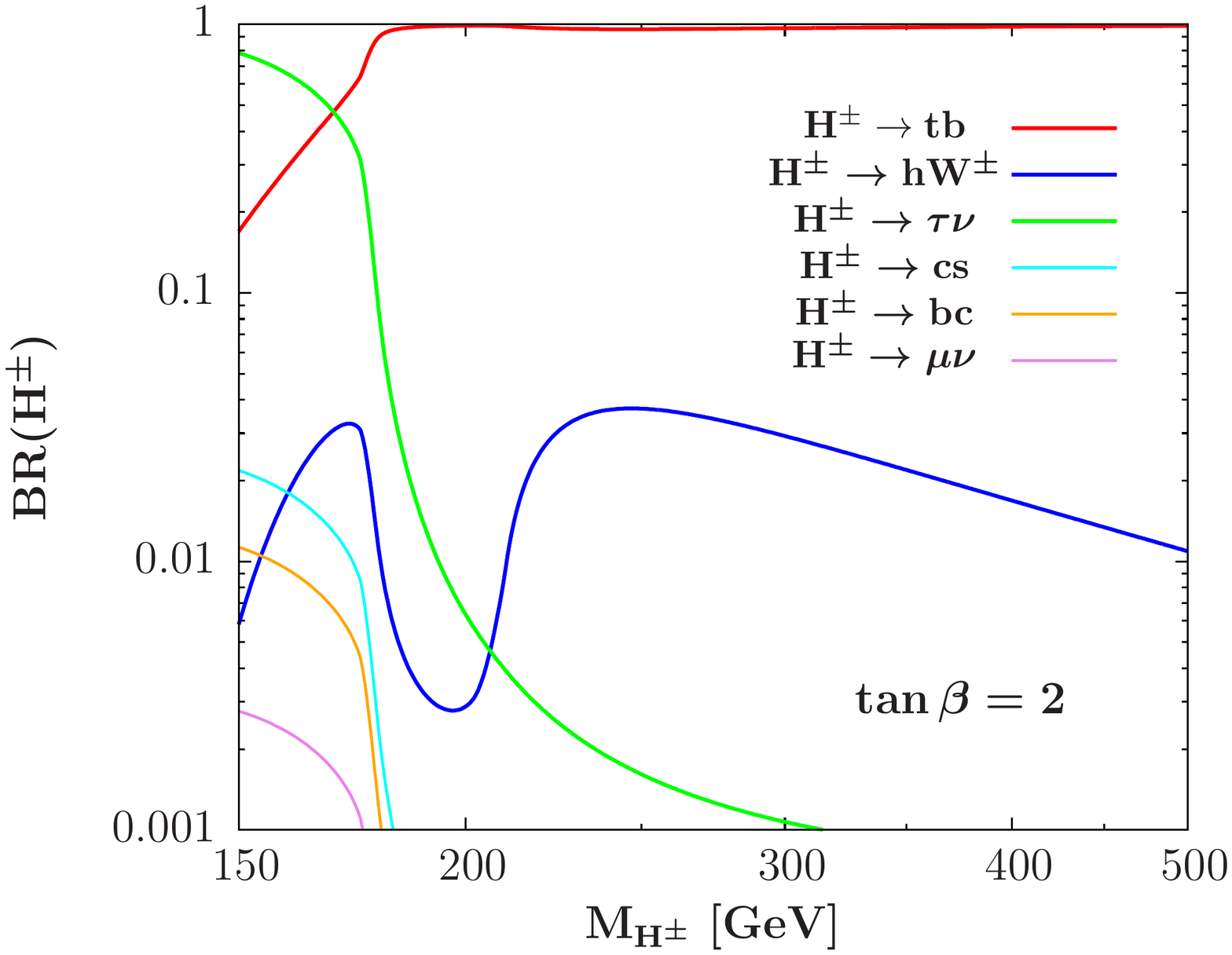}\hspace*{-0.4cm}
\includegraphics[scale=0.38]{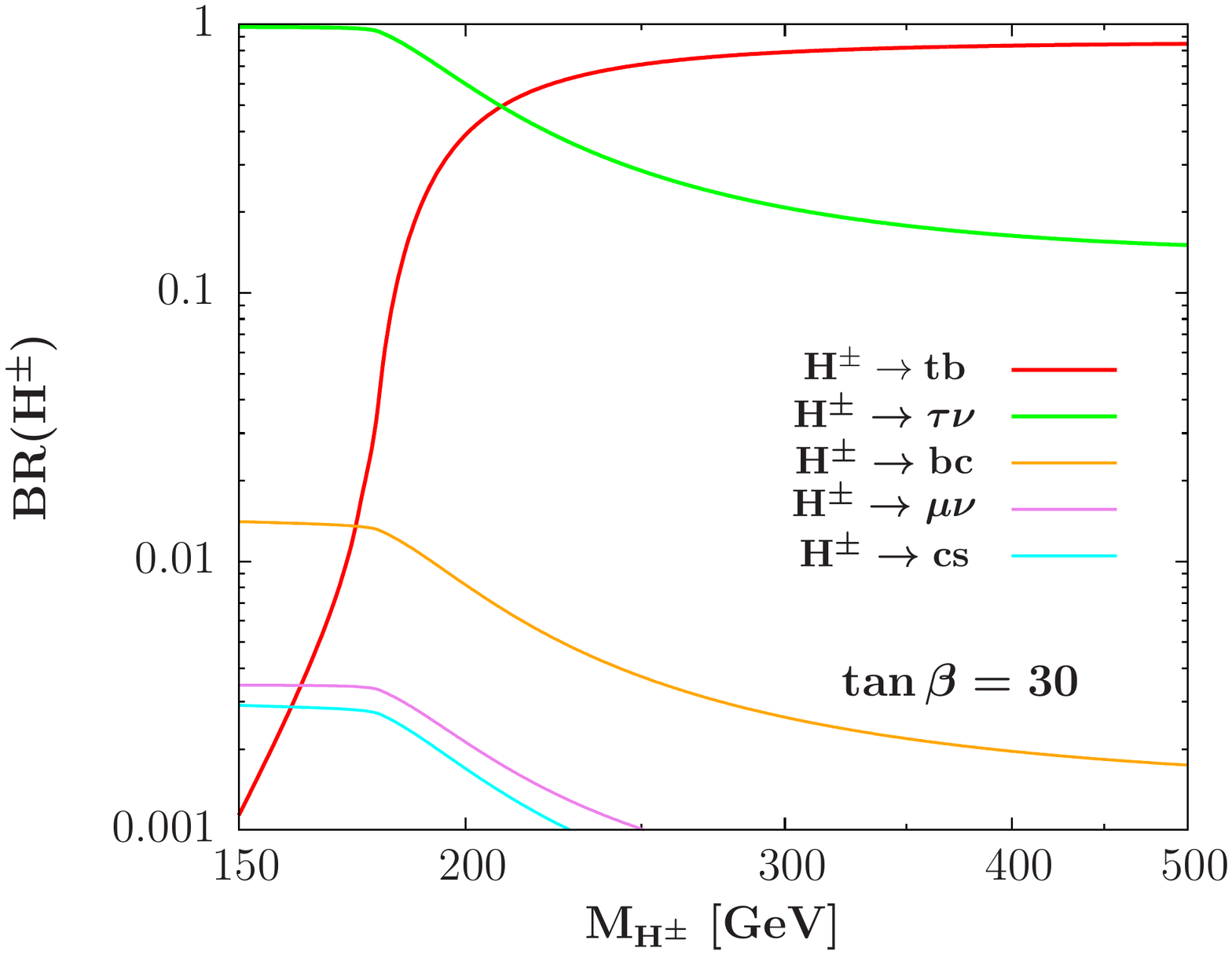} 
}
\end{center}
\vspace*{-3.1cm}
\caption{The branching ratios of the charged Higgs boson as a function of its mass
for $\tb=2$  (left) and $\tb=30$ (right); the $h$MSSM with the constraint $M_h=125$ GeV
is assumed.}
\label{fig:BRH+}
\vspace*{-.2mm}
\end{figure}

A final word should be devoted to the case of the charged Higgs boson, whose coupling 
to fermions is proportional to
\beq
g_{H^\pm \bar ud} \propto 
m_d \tb (1+\gamma_5) + m_u{\rm cot}\beta (1-\gamma_5) 
\label{eq:gH+} 
\eeq
The coupling is large at low $\tb$ when the component $m_u/\tb$ is not suppressed and 
at very large $\tb$ when the  component $m_d \tb$ is enhanced, so that many aspects 
discussed for the pseudoscalar Higgs boson hold also in this case \cite{Review}. At low
mass, $M_{H^\pm} \lsim 160$ GeV, which corresponds to $M_A \lsim 140$ GeV, the charged 
Higgs boson can be produced 
in the decay of top quarks that are copiously produced at the LHC, $gg+q\bar q
\to t \bar t$ with one top quark decaying into the dominant $t \to bW$ mode and the 
other into $t \to bH^+$. For   $M_{H^\pm} \approx 140$ GeV, the latter channel has 
a branching ratio ranging from order $\approx 10\%$ for $\tb \approx 1$ or $\tb \approx 60$
to order $\approx 1\%$ for $\tb \approx 7$--8 when the Higgs couplings are the smallest. 
In this low mass range above, the $H^\pm$ boson will decay almost
exclusively into $\tau\nu$ final states but some some competition with the hadronic
decay channel $H^+ \to c\bar s$ will occur at very low $\tb$. The $H^\pm$ branching fractions
are shown in Fig.~\ref{fig:BRH+} as a function of $M_{H^\pm}$ for two representative 
$\tb$ values, $\tb=2$ and $\tb=30$.  

\begin{figure}[!h]
\vspace*{-2mm}
\mbox{
\hspace*{-6mm}
\includegraphics[scale=0.42]{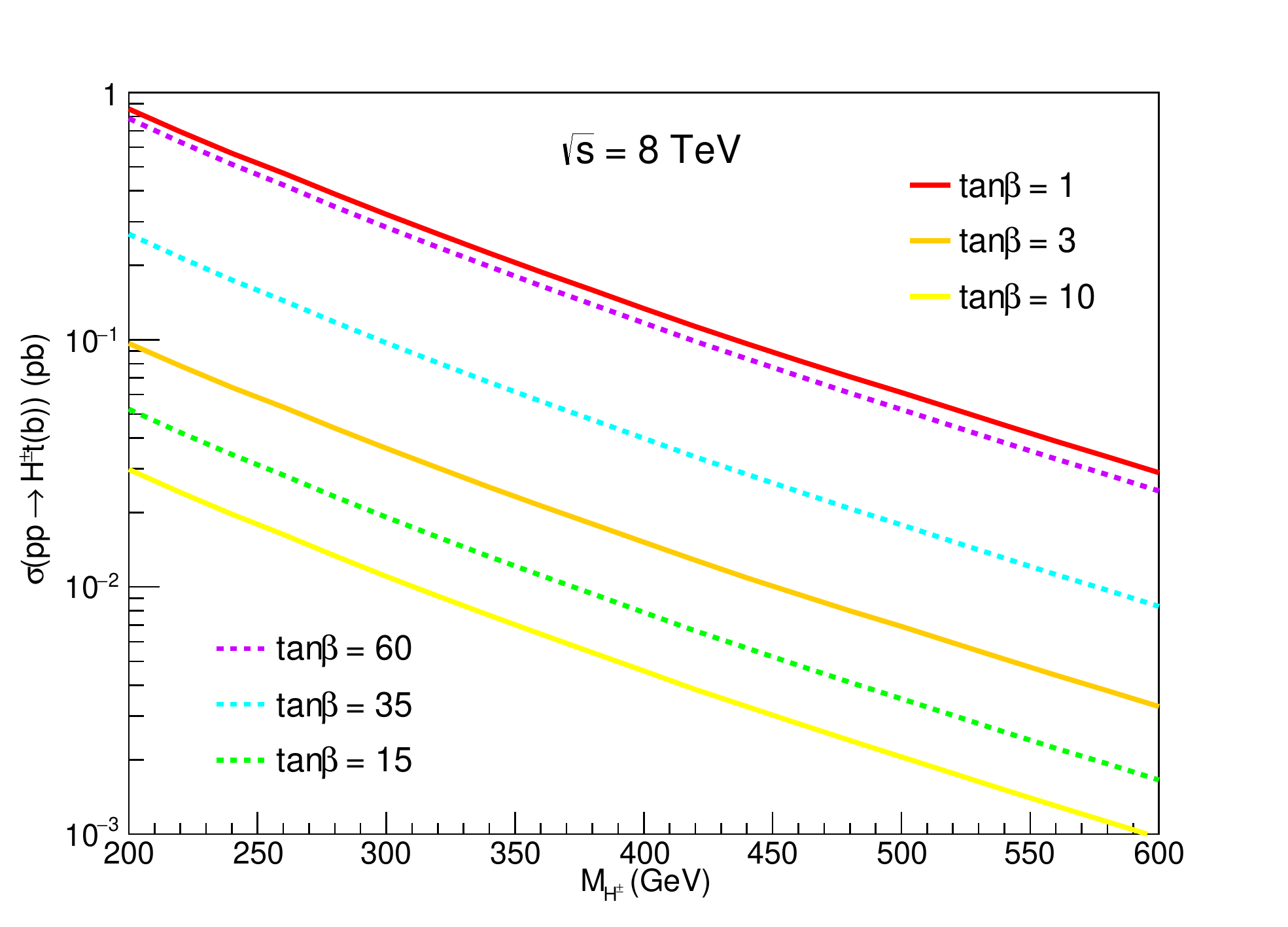} 
\hspace*{-7mm}
\includegraphics[scale=0.42]{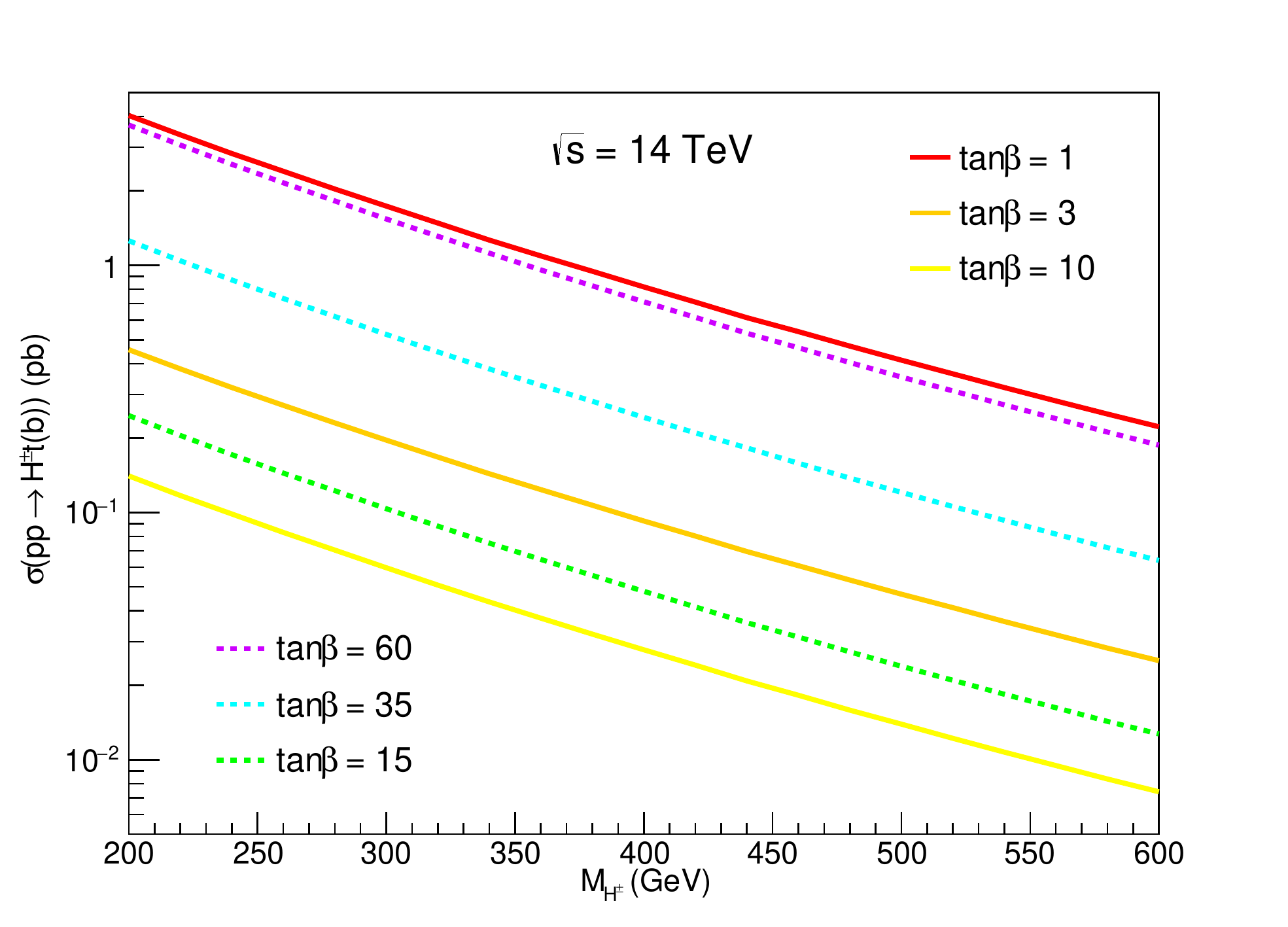}
}
\vspace*{-5mm}
\caption{The matched production cross sections $pp \to H^+ t(b)$ at the LHC with $\sqrt{s}=
8$~TeV (left) and 14 TeV (right) as a function of $M_{H^\pm}$ for several $\tan\beta$ values;
the numbers were taken from the grids given in Ref.~\cite{gridH+}.}
\label{pp:H+tb}
\vspace*{-.2mm}
\end{figure}

At higher masses, the $H^\pm$ state will be mainly produced in the three--body production  
process $pp \to tbH^\pm$ which, at high energies, is equivalent to the two--body channel 
$gb \to  H^\pm t$ if no additional final state $b$ quarks are detected \cite{ppH+}. 
Again, significant rates occur only at very low or very large values of $\tb$ when the 
$H^\pm t b$
coupling of eq.~(\ref{eq:gH+}) is large (some small additional contributions from the 
tree--level $q\bar q \to H^+ H^-$ and loop induced $gg \to H^+ H^-$ pair and associated 
$q \bar q \! \to \!  H^\pm \!+ \! A/h/H$ production modes are also possible \cite{Review}). 
The cross sections have been derived in Ref.~\cite{gridH+} where the two possibilities for 
the process,  $pp\! \to \! t \bar b H^-$ and $gb\! \to \! tH^-$, are properly matched and 
some numerical grids have been provided for the MSSM. The output of these grids for
the production rates at $\sqrt s=8$ and 14 TeV is shown 
in Fig.~\ref{pp:H+tb}  in the $[\tb, M_{H^\pm}]$ plane.

At high $\tb$, as shown in Fig.~\ref{fig:BRH+}, the $H^\pm$ decay branching fractions 
are BR($H^+ \to \tau \nu) \approx 10\%$ and BR($H^+ \to t\bar b)\approx 90\%$ exactly 
for the same reasons discussed previously for the $H/A$ particles. All the other decay
channels can be safely ignored so that the main search channel would be  $pp \to 
H^\pm t(b)$ production with $H^+\to \tau \nu$. As will be seen later, the process is 
however less powerful in probing the MSSM parameter space than the $pp \to H/A \to \tau 
\tau$ channel discussed earlier. 

At low $\tb$, the dominant decay channel will be by far the $H^+ \to t\bar b$ mode
with a branching ratio close to unity for a sufficiently heavy ${H^\pm}$ state when  
phase--space effects are irrelevant (one should note though, that slightly below the 
$m_t+m_b$ kinematical threshold, the three--body decay with an off--shell top quark, 
$H^+ \! \to \!  bt^* \! \to \! b\bar bW$, is also important in the case $\tb \approx
2$ as shown in Fig.~\ref{fig:BRH+}). 
The main search channel in this case would be the $pp \to H^\pm t(b) \to tt bb$ mode. 
This process, which is sensitive to the same area of $[\tb, M_A]$ parameter space
as the $H/A \to t\bar t$ channels discussed before i.e.
the low $\tb$ and high $M_A$ regions, has been considered in the past and found
to be of limited use as it is subject to a large QCD background \cite{Houches}. However,
a recent CMS analysis \cite{CMSH+tb} gave interesting and more optimistic results that
we will discuss in the next section. \newpage

Finally, at $H^\pm$ masses above  $M_{H^\pm} \gsim 160$ GeV, an interesting decay channel 
would occur, namely $H^\pm \to Wh$. Nevertheless, and in contrast to the similar $A \to hZ$ 
decay mode discussed previously, this channel has to compete from the beginning with the dominant $H^\pm  \to tb$ decay. Only at moderate $\tb$ and low $M_{H^\pm}$ that this decay 
has a sizeable branching ratios, of the order of a few percent, allowing for
$H^\pm$  searches in the interesting channel $pp \to tbH^\pm \to tbWh$ which,
experimentally, has not been considered so far.

\FloatBarrier

\subsection{Impact of the SUSY spectrum and dark matter}

An important question would be if the MSSM Higgs production times decay rates are not 
affected by the presence of supersymmetric particles. These could have  
two impacts: first, they could contribute virtually to the processes and modify the 
production cross sections and decay branching ratios. This issue is directly related to the 
third assumption of the $h$MSSM, namely that the couplings of the Higgs bosons are 
simply given by eqs.~(\ref{Hcoup})--(\ref{eq:Hhh}) and no direct correction is involved. 
Second, SUSY particles could appear in the decays of the Higgs particles and alter  the branching ratios for the standard channels that are searched for. This possibility 
would also invalidate the simple $h$MSSM approach as some SUSY parameters would 
be then required to describe Higgs phenomenology. Both issues have been
discussed e.g. in  Ref.~\cite{Orsay} and below, we simply summarise the main points with 
details concerning the low $\tb$ region and the decays $H/A \to t\bar t$. 

For what concerns the production processes,  besides the standard top and bottom--quark 
loops, there are also squark (and mainly stop) loops \cite{ggH-squarks} that contribute to 
the production of the CP--even $H$ boson in the gluon-fusion channel, $gg\!\to\! H$; the 
CP--odd $A$ states 
does not couple to identical sfermions and there is no--squark contribution to $gg \to A$ at lowest order. However, as the Higgs--squark couplings are not proportional to squark masses, 
the contributions are damped by powers of the squark mass squared $\propto 1/ \tilde m^2_Q$
and should be small for sufficiently heavy squarks. This is particularly the case at high 
$\tb$ values where the standard bottom--quark contributions are so strongly enhanced that
the impact of squarks becomes negligible. At low $\tb$ values, as one needs a large SUSY 
scale $M_S \gg 1$ TeV in order to accommodate an $h$ boson with a mass $M_h \approx 125$ GeV,
the impact of the too heavy squarks should also be negligible in the gluon--fusion process. 
Hence, in most cases, these SUSY loop contributions can be ignored in the production modes.  
               
SUSY particles can have a large impact also through the Higgs boson couplings. Indeed, 
besides the radiative corrections that affect the Higgs mass matrix eq.~(\ref{mass-matrix}),  
there are additional one--loop vertex corrections that modify the Higgs--fermion 
couplings and which are not described by the $\h$MSSM. These corrections are in general
only important in the case of $b$--quarks and only at high--$\tb$ and large $\mu$ values, 
since they grow as $\mu \tan\beta$. The dominant components are due to  the contributions to the Higgs--$b\bar b$ vertices from the strongly interacting sbottoms and
gluinos and the weakly interacting higgsinos with top squarks. They can be approximated by 
\cite{CR-deltab}
\beq
\Delta_b \simeq  \left[
\frac{2\alpha_s}{3\pi} m_{\tilde{g}} /{\rm max}(m_{\tilde{g}}^2, m_{\tilde{b}_i}^2) +
\frac{\lambda_t^2}{16\pi^2} A_t /{\rm max}(\mu^2, m_{\tilde{t}_i}^2) \right]
\mu \tb
\eeq
They affect mainly the heavier Higgs couplings that become in the limit 
$M_A\! \gg\! M_Z$, 
\beq
g_{Hbb} \approx g_{Abb}  \approx g_{H^\pm tb} \approx  \frac{\tb}{1+\Delta_b} ~~{\rm at~high~\tb}
\eeq
For the lighter $h$ state, the coupling $g_{hbb}$ is not affected in this limit and  
stays SM--like. 

Nevertheless, as already discussed in many places including Refs.~\cite{Orsay,BD}, this correction has only a  limited impact in the case of the full $pp\to \Phi=H/A \to \tau 
\tau$ process as the correction appears in both the production cross 
section 
$$\sigma( gg+ b\bar b \to \Phi) \propto (1+ \Delta_b)^{-2}$$ 
and in the $\tau\tau$ decay branching fraction, 
$${\rm BR}(\Phi \to \tau\tau)= \Gamma(\Phi \to \tau\tau)/[(1+ \Delta_b)^{-2}\Gamma(\Phi 
\to b\bar b)+ \Gamma(\Phi \to \tau\tau)]$$,
 and it largely cancels out in the product of the two  
\beq
\sigma \times {\rm BR} \simeq  1-  \frac15 \Delta_b 
\eeq

Hence, only when the $\Delta_b$ correction is huge and larger than  unity (a feature that
might put in danger the perturbative series) that its impact on the $pp \to \tau\tau$ 
cross section times decay rate is of the order of the theoretical uncertainty, about 25\% 
as discussed earlier.  At low $\tb$ values, and eventually also at intermediate $\tb$ values,
the $\Delta_b$ correction is not enhanced and its effects should be rather small. 
Nevertheless, this is not the case of all process and, in particular, the search channel
$pp\! \to \! H/A$ with $H/A\! \to \! b\bar b$ in which the $\Delta_b$ impact is in fact 
doubled in the production times decay rates, and the $pp \! \to \! H^\pm bt$ mode with 
$H^+ \! \to \! \tau\nu$ at low $M_{H^\pm}$ and $H^+\to  tb$ at high $M_{H^\pm}$. We will see,
however, that these processes do not play a leading role in MSSM Higgs searches at the LHC 
at high $\tb$. In summary, and to first approximation, one can thus consider that the 
$\Delta_b$ correction has a limited impact on the Higgs searches in the $\h$MSSM. 

For the second option, namely that light SUSY particles could contribute to the decays 
of the Higgs bosons, the situation is also relatively simple; see Ref.~\cite{Review} for a
review. At very high $\tb$, the partial  widths of the $H/A \to b\bar b, \tau^+\tau^-$ as 
well as $H^+ \to t\bar b, \tau^+ \nu$ decay modes are so strongly enhanced, that they leave 
no room for the SUSY decay channels. 

At low $\tb$,  high values of the SUSY scale are required, resulting in large 
squark masses  (at least in universal models in which the squark masses of the three generations are related) that make the Higgs decays into squarks kinematically closed
for reasonable $M_A$ values. If the masses of the sleptons are disconnected from the SUSY
scale and are made small enough for the decays of the heavier Higgs bosons into slepton 
pairs, $H \to \tilde \ell_i \tilde \ell_j$ and $A \to \tilde \ell_1 \tilde \ell_2$ (again, 
CP invariance forces the $A$ boson not to couple to identical sfermions), to occur. 
Nevertheless, except for the Higgs--stau couplings at sufficiently high $\tb$ values
when the competition from the standard channels is though,  the Higgs-sleptons couplings 
are in general small making these channels very rare and their impact
limited.  

Thus, only decays into charginos and neutralinos could  play a role and 
affect significantly the Higgs branching fractions in the standard 
channels\footnote{Here we will consider models in which the neutralino $\chi_1^0$ is 
the lightest SUSY particle (LSP). In gauge mediated SUSY breaking models (GMSB), a very light gravitino can be the LSP and decays of the heavier Higgs bosons into a gravitino and
a neutralino or a chargino, $H/A \to \tilde G \chi_i^0$ and $H^\pm \to \tilde G \chi_i^\pm$,
are possible (Higgs bosons do not couple to pairs of gravitinos). However, as discussed 
in Ref.~\cite{gravitino}, the partial widths for these
decays are inversely proportional to the square of the SUSY breaking scale, $ M_S^2 =
(m_{\tilde G} M_{\rm Planck})^2$, and need $M_S \lsim  {\cal O}({\rm few~100~GeV})$ to 
be substantial. 
This possibility is ruled out not only from direct SUSY particle searches but also from 
the fact that the large value of the MSSM $h$ boson mass, $M_h=125$ GeV, requires a SUSY 
scale in the multi TeV range in GMSB scenarios ($M_S \gsim 3$ TeV in minima GMSB) \cite{high-MS}.}.
Let us briefly comment on these channels. 

Three conditions must be fulfilled in order to have significant rates for Higgs decays 
into charginos and neutralinos, $H/A \to \chi^0_i \chi_j^0$ (with $i,j=1 \! \cdots \! 4$),  
$H/A \to \chi^\pm_i \chi^\mp_j$ (with $i,j=1,2$) and $H^\pm \to \chi^\pm_i \chi_j^0$ 
\cite{Review,inodecays}. 

\begin{itemize}

\item[$i)$] One needs that some of the $\chi$ states are light, $M_\Phi \gsim 2 m_{\chi}$, 
in order to allow for some decay channels to be kinematically open. 

\item[$ii)$] One needs to have significant $\Phi \chi \chi$ couplings; these couplings are maximal when the $\chi$ final states are mixtures of higgsinos and gauginos, a feature 
which requires comparable higgsino and gaugino mass parameters, $\mu \approx M_2$. 

\item[$iii)$] One needs 
that the standard Higgs decay modes are not enhanced and hence, not too low or too 
large values of $\tb$ where, respectively,  the Higgs--top and the Higgs--bottom couplings 
are enhanced. 

\end{itemize}

The maximal Higgs decay rates into charginos and neutralinos are obtained 
at moderate $\tb$ when all $\chi\chi$ channels are kinematically accessible. In this case, 
as a consequence of the unitarity of the diagonalizing chargino and neutralino mixing matrices, the sum of the partial widths do not involve  any of the elements of these 
matrices in the asymptotic regime $M_{\Phi} \gg 2m_\chi$ where phase space effects can 
be neglected. The  sum of the branching fractions of the three Higgs bosons 
$\Phi=H/A/H^\pm$ decaying into the various $\chi \chi$ final states is then simply 
given by ($\theta_W$ is the electroweak mixing angle) \cite{inodecays}
\begin{eqnarray}
{\rm BR}( \Phi \to \sum_{i,j} \chi_i \chi_j) = \frac{ \left( 1+\frac{1}{3}
\tan^2 \theta_W \right) M_W^2 }{ \left( 1+\frac{1}{3} \tan^2\theta_W \right) 
M_W^2 + \overline{m}_t^2 \cot^2 \beta + (\overline{m}_b^2 + m^2_\tau) \tan^2 \beta } 
\end{eqnarray}
when only the leading $t\bar{t}$, $b\bar{b}$ and $\tau \tau$ modes 
for the neutral and the $t\bar{b}$ and $\tau \nu$ modes for the charged Higgs bosons 
are included in the total widths. This is approximately the case as we are close to 
the decoupling limit when these SUSY channels are accessible and the other
standard decay modes such as $H\to VV,hh$ and $A\to hZ$ are suppressed.

\begin{figure}[!h]
\vspace*{-2.5cm}
\mbox{\hspace*{-3cm}
\includegraphics[scale=0.9]{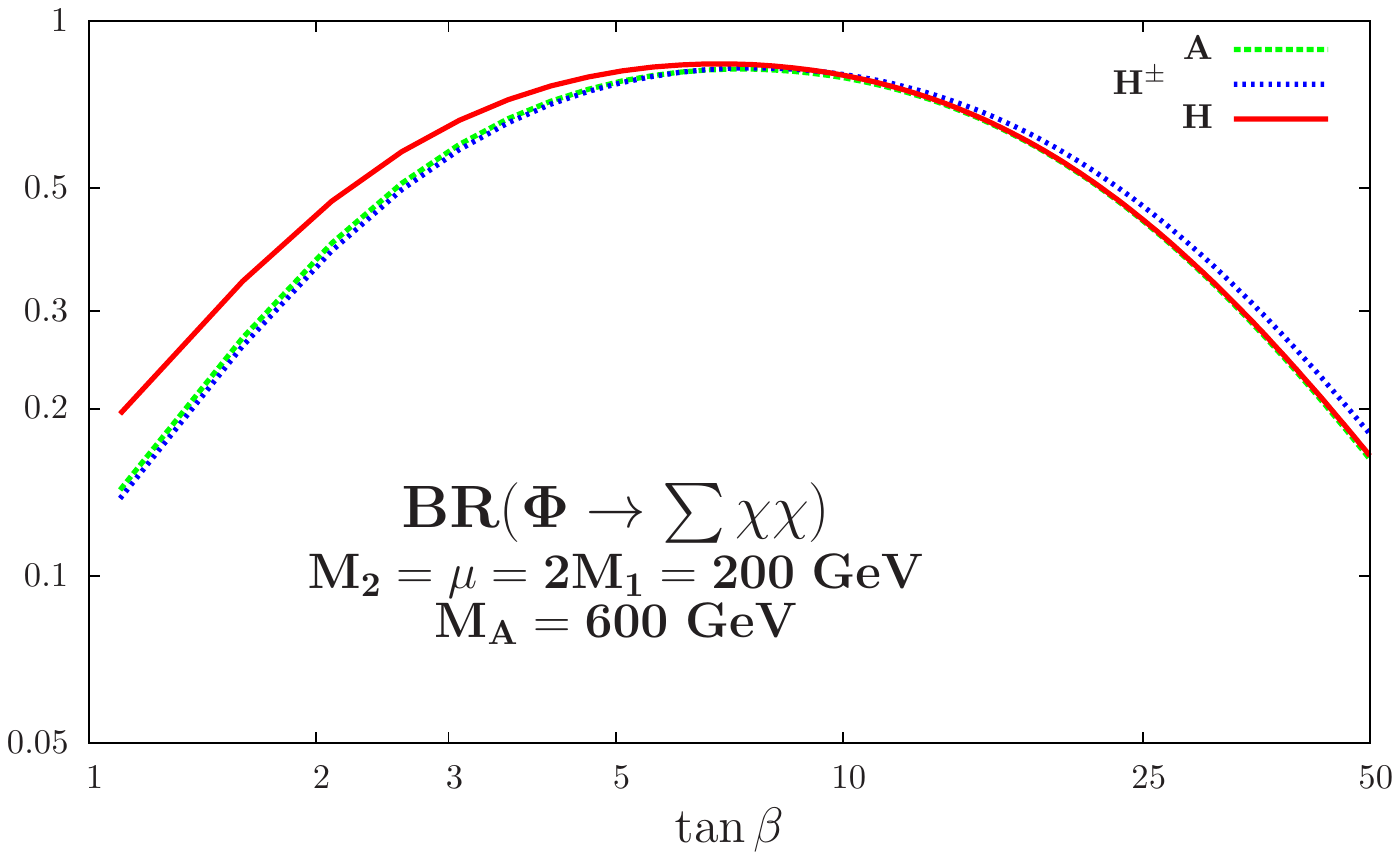} }
\vspace*{-16.7cm}
\caption{The sum of the branching ratios for the decays of the three heavier MSSM 
Higgs bosons into charginos and neutralinos as a function of $\tb$ for $M_A=600$ GeV.
}
\label{Hchi}
\vspace*{-1mm}
\end{figure}

The branching ratios when all ino states are summed up are shown 
for the three MSSM Higgs states in Fig.~\ref{Hchi} as a function of $\tb$ for 
$M_A=600$ GeV. The $h$MSSM relations for the Higgs sector have been enforced and
the other relevant SUSY parameters are fixed to  $\mu=M_2 =200$ GeV,  assuming that 
the bino and wino soft SUSY--breaking parameters are related by the unification 
condition $M_2 \approx 2M_1$. One can see that the branching ratios for the three Higgs 
particles are indeed similar and that they do not dominate at low nor at high $\tb$.  
For instance, they are less than 25\% (which is the magnitude of the theoretical uncertainty 
on the production cross sections) for $\tb \lsim 2$ and $\tb \gsim 30$ as can be seen 
from the figure. In contrast, the $\chi \chi$ branching rations can be large for 
intermediate values of $\tb$ when the Higgs couplings to top  (bottom) quarks are suppressed 
(not strongly enhanced) and, for instance, they reach the level of  $\approx 70\%$ 
at $\tb \approx 5$--10.

Nevertheless, this possibility with large Higgs decay rates into SUSY particles 
seems unlikely. First, low and comparable values of the wino and higgsino 
mass parameters $M_2 \approx \mu \lsim 300$ GeV that would lead to light charginos 
and neutralinos in the decay products of not too heavy $H/A/H^\pm$ bosons, are 
constrained by the direct searches for these particles at LEP and the LHC 
\cite{PDG,LHC-susy,trilep}. In particular, 
the associated production of the lighter chargino and the next-to-lightest neutralino 
$q \bar q  \to \chi_1^\pm  \chi_2^0$  would lead to large cross sections at $\sqrt 
s=8$ TeV and the decays $\chi_1^\pm \to W \chi_1^0$ and $\chi_2^0 \to Z \chi_1^0$,
with leptonic gauge boson decays,  
would have significant branching fractions. The search for leptons plus missing 
energy at the first run of the LHC, in particular the clean trilepton events from 
the chain $pp \to \chi_1^\pm \chi_2^0 \to WZ \chi_1^0 \chi_1^0 \to \ell \ell \ell 
E_T^{\rm mis}$, imposes severe restrictions on the parameter space. For instance, 
for $M_1 \approx 100$ GeV, the area $\mu \approx M_2 \lsim 200$ GeV that 
corresponds to the choice adopted for Fig.~\ref{Hchi} is by now excluded 
by the LHC data\footnote{Note that we have adopted the same choice of 
gaugino--higgsino parameters as in the benchmark scenario of Ref.~\cite{benchmarks} 
that has been used to interpret the experimental limits in the $pp \to \tau \tau$ 
searches made by the ATLAS and CMS collaborations \cite{tau-ATLAS,tau-CMS}. Hence, 
this choice leads to a large branching fraction for Higgs decays into $\chi \chi$
states for $M_A \gsim 300$ GeV and weakens the experimental constraints that can be 
obtained from $H/A$ searches in the $pp\to \tau\tau$ process, while it is apparently 
excluded by direct SUSY searches.} \cite{trilep}. 

To evade these experimental bounds, one needs either to increase the parameters $\mu_2\! 
\approx\! M_2$ well above  200 GeV with a consequence that the phase space for the Higgs 
decays will be limited, or to make that the light $\chi$ states are either 
pure higgsinos $(\mu \! \ll \! M_2)$ or pure gauginos ($\mu \! \gg \! M_2)$, which 
then suppresses the $Z\chi \chi$ couplings e.g. and hence the trilepton signals, leading 
to Higgs couplings to the kinematically accessible charginos and neutralinos that 
are too small.  Thus, in all these cases, the SUSY decays are suppressed and do not 
jeopardize the Higgs signals in the standard search channels.

These arguments become stronger in the case where the lightest neutralino $\chi_1^0$
is forced to be the candidate for the dark matter in the universe, with a cosmological 
relic density as measured by the WMAP/Planck teams, $0.09 \leq \Omega h^{2} \leq 
0.12$ \cite{WMAP} with $h$ being the reduced Hubble constant. Traditionally, four 
regions of the MSSM parameter space have been advocated to fulfill this condition for 
the LSP neutralino \cite{DM} and we list them below.\smallskip  

$i)$ The so--called ``well tempered neutralino" region, with a mixed gaugino--higgsino  
LSP that allows for significant  couplings to gauge and Higgs bosons; this leads to a 
good LSP annihilation rate into bosonic finals states, $\chi_1^0 \chi_1^0 
\to WW, ZZ, hZ$. For low $M_2$ and  $\mu$ values, this is the region discussed above
that is constrained by the multi-lepton plus missing energy  searches.\smallskip     

$ii)$ The bino like neutralino region, $\mu \gg M_1$,  where the main LSP annihilation
channel is into  third generation tau leptons, with the exchange of light sleptons in 
the $t$--channel, $\chi_1^0 \chi_1^0 \to  \tau^+ \tau^-$, and the $\tilde \tau$ 
co-annihilation region in which  the lightest $\tilde \tau_1$ state is almost mass 
degenerate with the LSP, $m_{\tilde \tau_1} \approx m_{\chi_1^0}$, and the correct 
relic density is provided by the process $\tilde \tau_1 \tilde \tau_1 \to$ SM particles. 
If the SUSY scale is high and a kind of universality is assumed for sfermions,  the 
$\tilde \tau_1$ state will be too heavy and both channels become inoperative.\smallskip   

$iii)$ The regions where the LSP is almost a pure higgsino or gaugino state and hence with
small couplings to the $H/A/H^\pm$ bosons. The correct cosmological relic density is provided
by the co--annihilation of the mass degenerate $\chi_1^\pm$ and $\chi_2^0$ states that need 
to be very heavy (above 1 TeV) and inaccessible in the decays of TeV Higgs bosons.
\smallskip    

$iv)$ Finally, there is the Higgs--pole region \cite{DM-Hpole} in which an efficient 
LSP annihilation into SM particles is provided by the exchange of an almost on--shell $
A$ boson in the $s$--channel; one thus needs $M_A \approx 2m_{\chi_1^0}$ (the possibility 
of $h$-boson exchange \cite{hpole} leading to $m_{\chi_1^0} \approx 60$ GeV is by 
now unlikely). In the past, the high $\tb$ region was favored and the most discussed 
annihilation channel  was  $\chi_1^0 \chi_1^0 \to A \to b\bar b$. At low $\tb$, a 
new possibility opens up, the channel $\chi_1^0 \chi_1^0 \to A \to t\bar t$ which can also 
lead to the correct relic density.\smallskip   

\begin{figure}[!h]
\vspace*{-.3cm}
\begin{center} 
\begin{tabular}{cc}
\hspace*{-6mm}
\includegraphics[scale=0.29]{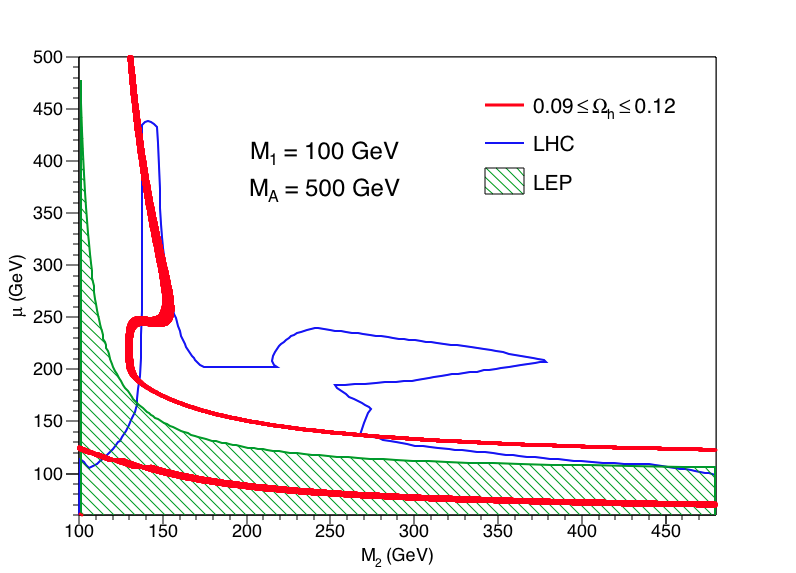}\hspace*{-4mm} 
& \hspace*{-6mm}
\includegraphics[scale=0.29]{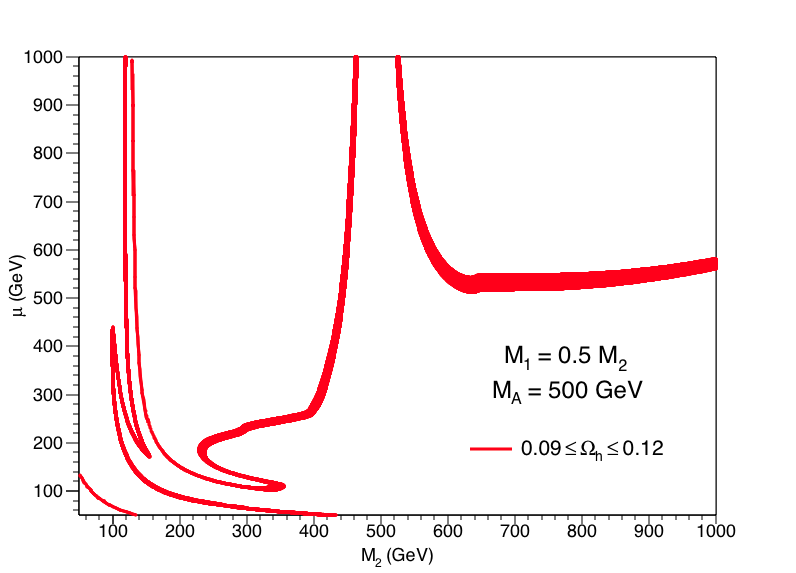} 
\end{tabular}
\vspace*{-3mm}
\caption{Points in the $[M_2,\mu]$ plane which satisfy the dark matter constraint with 
$M_{A}=500$~GeV and $\tan\beta=2$ for $M_1=100$ GeV (left) and with the unification 
condition for the gaugino masses $M_1= 0.5M_2$ (right). In the left figure, the 
parameter space excluded by the LEP and LHC direct searches of charginos and neutralinos is indicated.}
\label{Omega}
\end{center}
\vspace*{-6mm}
\end{figure}

In Fig.~\ref{Omega}, we display the areas of the [$M_2, \mu]$ parameter space  
in which the relic density of the lightest neutralino, calculated using the program 
{\tt micrOMEGAs} \cite{micromegas}, is as determined by the WMAP/Planck collaborations,
i.e. $0.09 \leq \Omega h^{2} \leq 0.12$ \cite{WMAP}. We have 
chosen $M_A\!=\!500$ GeV and $\tb\!=\! 2$ for the $\h$MSSM inputs, and assumed very heavy 
sfermions. In the left plot we have fixed the bino mass to $M_1\!=\!100$ GeV,
while in the right plot, we used the unification condition $M_2 \! = \!2M_1$. 
Outside the areas excluded by LEP and LHC ino searches at $M_1\!=\!100\;{\rm GeV}\; 
\! \approx \! m_{\chi_1^0}$, only two areas lead to a correct relic density: 
small $M_2$ or $\mu$ values ($\lsim \! 150$ GeV) that allow for $\chi_1^0 \chi_2^0$ and  
$\chi_1^0 \chi_1^\pm$ co--annihilation or a mixed bino--higgsino LSP. Another
area opens up if $m_{\chi_1^0} \! \approx \! \frac12 M_2 \! \approx \! \frac12 M_A$ as can 
be seen in the right--figure: the $A$ funnel in which the LSP efficiently annihilates through 
the channel $ \chi_1^0 \chi_1^0 \! \to \! A \! \to \! t \bar t$. In most of this area, the  
decay $A \! \to \! \chi_1^0  \chi_1^0$ is kinematically closed (or phase--space
suppressed) and, because $\chi_1^0$ is the LSP, so are all Higgs decays  into 
superparticles.  

Hence, in all cases, the requirement that the lightest neutralino is the dark matter 
in the universe with the correct relic density makes that the decays of the Higgs 
bosons into charginos and neutralinos should not occur, or at least should not 
dominate.

One concludes from all the discussions of this subsection that it is rather unlikely
that the SUSY particles make a significant impact in the phenomenology of the MSSM Higgs 
bosons, either in their virtual contributions to the production and/or decay processes 
(in particular since the SUSY effects should be larger than the $\approx 25\%$ theoretical 
uncertainty that affects the production rates) or in the direct appearance in the decays
of the heavier Higgs states. One can thus assume that the superparticles are very heavy 
and/or too weakly coupled and that they decouple from the MSSM Higgs sector, except of 
course in the radiative corrections to the CP--even mass matrix eqs.~(\ref{mass-matrix}). 

This is, in fact, another way of stating the third assumption of the $h$MSSM discussed 
in section 2 and one can thus consider this effective approach as a very good benchmark.
Ignoring the SUSY effects is a rather reasonable attitude since,  besides the tremendous
simplifications that the $h$MSSM introduces in the description of the Higgs sector, it 
leads to a straightforward interpretation of the experimental constraints, that do not 
need to be "de-convoluted" from these complicated effects when they are included. 

Nevertheless, the possibility of light charginos and neutralinos with sizable
couplings to the Higgs bosons cannot be totally excluded at the moment. On should 
therefore continue performing searches for the $H/A/H+ \chi\chi$ SUSY decays and 
other more direct searches for these particles such as the tri--lepton signal events. 
In the full MSSM with light SUSY particles, all these processes provide complementary information. 

\FloatBarrier



\section{The probing of the MSSM parameter space}

\subsection{Interpretation of the fermionic Higgs decay modes in the hMSSM}

As discussed earlier, the most efficient channels that allow to probe the MSSM parameter space 
at the LHC are the search for charged Higgs bosons coming from top quark decays, the process 
$t \to b H^+$ followed by the decay $H^+ \to \tau^+\nu$ and its charge conjugate $H^-$ process,
and the search for high mass resonances decaying into tau--lepton pairs, the processes $pp \! \to 
\! H/A \to \tau^+ \tau^-$. Both channels have been considered by the ATLAS and CMS collaborations
and we briefly summarize below the resulting constraints on the $h$MSSM.  
   
The CMS $H^\pm$ search \cite{CMSH+} was performed with the 19.7 fb$^{-1}$ 
data collected at $\sqrt s \!=\! 8$ TeV with the $\tau$-leptons decaying fully hadronically. 
95\%CL upper bounds have been set on the product of branching ratios ${\rm BR}(t\! \to\! b 
H^+)\! \times\! {\rm BR}(H^+\! \to\! \tau^+\nu)$ from 1.2\% at $M_{H^\pm}\!=\!80$ GeV 
(about the exclusion limit obtained on $M_{H^\pm}$ from LEP2 searches \cite{PDG,LEP}) to 
0.16\% at   $M_{H^\pm}\!=\!160$ GeV (beyond which phase space effects start to be too 
penalizing). The search excludes the entire range $M_{H^\pm}\lsim 140$ GeV for all values of 
$1 \leq \tb \leq 60$. For larger $H^\pm$ masses, the areas where $\tb \!\approx \! 8$ at 
$M_{H^\pm}\!= \!140$ GeV to $\tb \! \approx \! 5$--15 at $M_{H^\pm} \!= \!160$ GeV, in which 
the $m_t/\tb$ component of the $H^\pm tb$ coupling is suppressed while the $m_b \tb$ component
is not yet enhanced, remain viable at the 95\%CL. 

The ATLAS search for $H^\pm$ states \cite{ATLASH+} was also performed with the full 19.5 
fb$^{-1}$ data recorded at $\sqrt s=8$ TeV; the same channel as above, i.e. $t \to 
b H^+ \to b \tau^+\nu$ with the tau--lepton decaying hadronically, has been used. Similar 
95\%CL upper limits than CMS have been obtained on the product ${\rm BR}(t\! \to\! 
b H^+)\! \times\! {\rm BR}(H^+\! \to\! \tau^+\nu)$. Compared to the previous limits, a small additional area of the $[\tb, M_{H^\pm}$] plane,  at $\tb \! \approx \! 8$ for $M_{H^\pm} \! \approx \! 100$ GeV and to $\tb \! \approx =6$--10 at $M_{H^\pm} \! \approx \! 90$ GeV,
 remains unexcluded by the ATLAS analysis, as a result of the presence of a large $t\bar t, 
W/Z+$jet backgrounds in this mass bin. 

\begin{figure}[!h]
\vspace*{-.5cm}
\begin{center}
\begin{tabular}{c}
\includegraphics[scale=0.6]{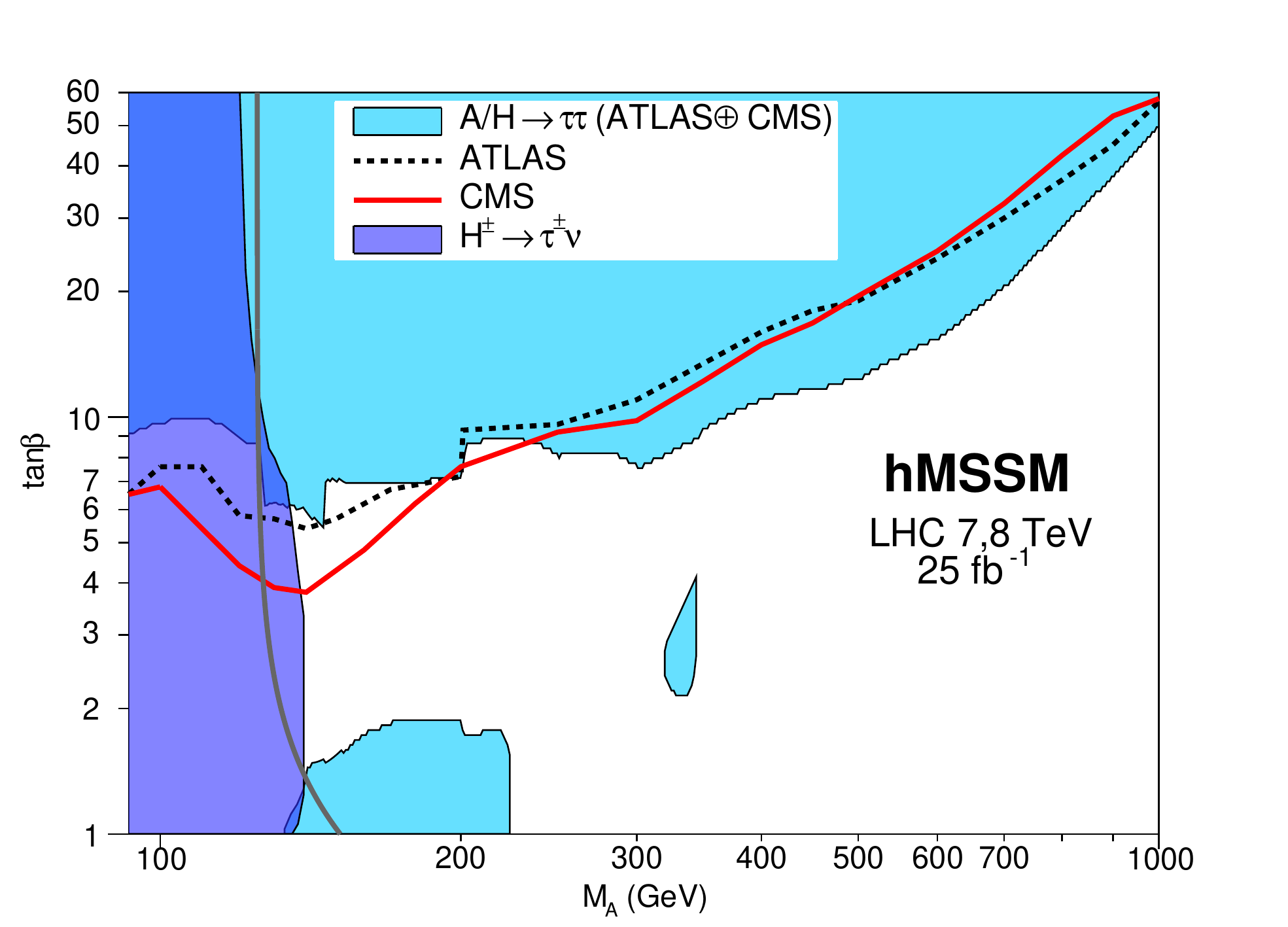} 
\end{tabular}
\vspace*{-.3cm}
\caption{The $[\tb,M_{A}]$ plane in which the $pp\! \to \! H/A \to  \tau^+\tau^-$  (light blue) 
and $t \! \to \! bH^+ \! \to b \! \tau \nu$ (dark blue) 
search limits of ATLAS and CMS are combined and interpreted within the $h$MSSM. The red-solid 
(blue-dotted) line is for the ATLAS (CMS) observed limit only in the $M_{h}^{\rm max}$ scenario
and the region at the left of the black-solid line is the one where the $h$MSSM is ill--defined. }
\label{tautau_taunu_ATLAS_CMS}
\end{center}
\vspace*{-.9cm}
\end{figure}

These limits can be turned into bounds in the [$M_A, \tb$] parameter space assuming the 
usual relation $M_{H^\pm}^2=M_A^2+M_W^2$. This is  what is illustrated by the dark blue 
area of Fig.~\ref{tautau_taunu_ATLAS_CMS} in which the constraints on the $[M_A,\tb]$ plan 
are shown: we take the limits on ${\rm BR}(t\! \to\! b H^+)\! \times\! {\rm BR}(H^+\! \to\! 
\tau^+\nu)$, that we calculate using the program {\tt SDECAY} \cite{sdecay}, and interpret
them in the $h$MSSM. We see that in our case, the entire area $M_A \lsim 140$ GeV is excluded
at low $\tb$ values, reducing to $M_A \lsim 130$ GeV at high $\tb$. In contrast to ATLAS
and CMS, which use different means to calculate the product of branching ratios, we do not
have the holes at the extreme values of $M_{H^\pm}$. 

In the previous references, both experiments performed also searches for 
heavier $H^\pm$ states with $M_{H^\pm} \!  \gsim \!m_t\!+ \! m_b \! \approx\! 180$ GeV, 
by considering the process $pp \! \to\! H^\pm t(b)$ with again $H^+ \! \to \! \tau^+\nu$. 
The areas $\tb \! \gsim \! 45~(60)$ at $M_{H^\pm} \! \approx 200~(250)$ GeV are excluded by 
the CMS collaboration (the ATLAS group assumed BR$(H^+ \! \to \! \tau^+\nu)=100\%$ in this 
area while it should be only 10\%). These limits are much less powerful than those obtained 
from the $\tau\tau$ search as will be seen shortly. However, there was in interesting CMS 
search in the channel $H^+ \! \to \! tb$ performed with 19.7 fb$^{-1}$  data at $\sqrt s=8$ 
TeV \cite{CMSH+tb}. Limits on $\sigma (pp \to tbH ^\pm)$ assuming BR$(H^+ \to tb)=1$ have 
been set and surprisingly, one is not far from being  sensitive to the very low $\tb$ area and with about a factor of two  more data, one would have probed $\tb \approx 1$  for $
M_{H^\pm}\! \approx 200$ GeV.

We note that these limits from $H^\pm$ searches exclude a substantial part of the low 
$[\tb, M_A]$ area in which the $h$MSSM is ill defined, i.e. the region at the left of the 
thick black solid line in Fig.~\ref{tautau_taunu_ATLAS_CMS}. The exclusion is valid in this
area as all the ingredients used to obtain the limits depend only on $M_{H^\pm}$ 
and $\tb$ and do not involve the CP--even Higgs parameters $M_H$ or $\alpha$ 
that were undefined. The only assumption is that the relation $M_{H^\pm}^2
=M_A^2+M_W^2$ remains valid. Hence, one problematic issue within the $h$MSSM is 
in a sense partly solved by these ``model-independent" exclusion limits 
(we do not address here the theoretical issue of the validity of the entire model 
at $\tb$ values close to unity). 
  
The most important search mode in the MSSM is certainly the $pp\to H/A \to \tau\tau$ channel. 
The ATLAS collaboration has searched for this signal using the 19.5--20.3 fb$^{-1}$ data
collected at 8 TeV \cite{tau-ATLAS} while CMS has used the full 24.6 fb$^{-1}$  data 
collected at 7+8 TeV \cite{tau-CMS}. Both collaborations consider the leptonic
($\tau_e \tau_\mu)$, semi--leptonic $(\tau_\ell \tau_{\rm had})$ and  hadronic 
($\tau_{\rm had} \tau_{\rm had}$) $\tau$ decays and CMS also considers the case
where an additional $b$--quark is present in the final state. Limits at the 95\%CL on $\sigma(pp 
\to \tau^+\tau^-)$ as a function of the invariant mass $M_{\tau \tau}$ of the tau--lepton  
pair have been given by the two experiments.

Our procedure to interpret these limits in the $h$MSSM is as follows. First, we combine 
the ATLAS and CMS 95\% CL exclusion limits on $\sigma(pp \to \tau^+\tau^-)$. 
We then  compare them  with the numbers that we obtain for the rates, namely the 
$H/A$ production cross sections calculated with the programs {\tt HIGLU} \cite{HIGLU} and 
{\tt SusHi} \cite{SUSHI} and the branching ratios BR$(H/A \to \tau\tau)$ calculated using 
{\tt HDECAY} \cite{hdecay}. These are derived assuming that the parameters $M_H$ and 
$\alpha$ given by eqs.~(\ref{hMSSM-output}) and ignoring all possible SUSY effects. 
For a given $[\tb,M_A]$ point, if in a window of $\Delta M_{\tau\tau}=40$
GeV centered on $M_A$ we have either the observed $h$ boson or the heavier $H$ state, we 
include their contribution to the signal. 

The result is shown Fig.~\ref{tautau_taunu_ATLAS_CMS} by the light blue areas of the [$M_A, \tb$] plane and is truly impressive. The largest area is the one that excludes all values above 
$\tb \approx 8$ for $M_A \lsim 300$ GeV, extending to $\tb \approx 20~(40)$ for $M_A \lsim 700~(900)$ GeV.  

The ATLAS and CMS observed limits, when interpreted in the $M_{h}^{\rm max}$ 
benchmark scenario, are also displayed in Fig.~\ref{tautau_taunu_ATLAS_CMS}.  We 
observe that below $M_A \approx 200$ GeV, our limit is less restrictive, the reason 
being that we do not make use of any refinements in order to treat the regions in which 
the three Higgs bosons have a comparable mass and, also, to deal with the observed 
signal of the lighter $h$ boson. In contrast, our limit is stronger at masses above 
$M_A \approx 300$ GeV, the reason being that while in our case there is no SUSY decays
of the Higgs bosons, the $M_{h}^{\rm max}$  scenario leads to $H/A$ decays into charginos 
and neutralinos that are significant. Indeed, for the gaugino and higgsino mass parameters 
of the $M_{h}^{\rm max}$ benchmark (and the slightly modified ones), $\mu \! =\! M_2\! 
=\! 2M_1 \! =\! 200$ GeV, the LSP has a mass of $m_{\chi_{1}^{0}} \! \approx \! 100$ GeV 
while the heavier neutralinos and the charginos have masses of the order of $\approx \! 
200$ GeV, so that many SUSY decays occur starting from $M_A \! \approx 300$ GeV and all 
of them will be present for $M_A \! \approx 500$ GeV and above. These decays will have substantial 
branching fractions, in particular in the intermediate $\tb \approx 10$ range where they 
become dominant, as can be seen from Fig.~\ref{Hchi}. Hence, the $H/A \to \tau\tau$ 
branching ratios are suppressed in this this case, resulting in a weaker exclusion 
limit (which is  unfortunate since this parameter configuration is almost certainly 
excluded by the direct searches  \cite{trilep}). 

A second step is to extrapolate the ATLAS and CMS limits to the low $\tb$ region, which 
as discussed earlier, can be described within the $h$MSSM approach in contrast to the 
benchmark scenarios used by the collaborations. Two islands were discovered during the
exploration. A first  and substantial area is at very low $\tan\beta$ and $M_A$, $\tb \lsim 2$  
and $M_A \lsim 230$ GeV.  Here, because part of the area is ill defined, we consider only
the production and the decay of the $A$ state that depend only on $\tb$ and $M_A$ as 
the decay channel $A \to hZ$, which introduces a dependence on the angle $\alpha$ through 
the $AhZ$ coupling, is not yet kinematically open. In this domain, as discussed earlier, both 
the $gg \to A$  cross section (dominated by the top--quark loop) and the branching ratio for 
the decay $A\to \tau\tau$ (which,  together with the one for $b\bar b$ and $c\bar c$, is 
the only significant one to occur) is substantial. Hence, despite of the fact that we have 
only one resonance, the cross section times branching fraction is large enough to generate 
an observable signal. The excluded area from this search removes the small residual  part of 
the ill--defined $h$MSSM region, that was left after imposing the exclusion limit from 
the $H^\pm$ searches discussed above. 

More surprising at first sight, we discovered another smaller island  at $M_A\approx 350$ GeV 
and $\tan\beta \approx 2$--4. It turns out that, around the $t\bar t$ threshold,  there is a 
very strong increase of the $gg\to A$ amplitude as the form--factor Re($A_{1/2}^A)  \approx 5$ 
is maximal at the $2m_t$ threshold\footnote{In fact the NLO QCD corrections to $\sigma(gg\! \to \! A)$
introduce a Coulomb singularity exactly at threshold \cite{ggH-NLO}. However, we have checked 
that the rate increase in this new area is not due to this unphysical feature.},  
$\tau = M_A^2/4m_t^2 \approx 1$. At the same time, BR$(A\to \tau\tau) \approx $  is substantial   
being a few percent, as the other decays except for $b\bar b$ are
slightly suppressed, $A\to hZ$ by the coupling $\approx \cos(\beta-\alpha)$ and $A\to t \bar 
t$ by phase space effects (only the three--body decay channel is kinematically open and it
is suppressed). Hence, there would have been a substantial surplus of events from the
$gg \! \to \! A \! \to \tau\tau$ process in this limited area  that is excluded by the search. 
    

\subsection{Interpretation of the bosonic Higgs decay modes in the hMSSM}

We now turn to the constraints that can be imposed on the $[\tb, M_A$] plane by considering
the  bosonic decay channels of the heavier $H$ and $A$ states. In contrast to the $H/A \! \to \! 
\tau\tau$ and $H^\pm \! \to \! \tau\nu$ searches, no interpretation of these modes has been done 
in the context of the MSSM by the ATLAS and CMS collaborations. In the following, we will
therefore adapt the constraints that have been obtained either in the context of the SM but
with a heavier Higgs state than the observed one, or in extensions of the SM other than the MSSM. We will focus on 
the experimental analyses  that provide the most stringent constraints.  

The massive gauge boson channels $H\to WW, ZZ$ have not been discussed in the context of the 
MSSM but important information can be borrowed from those performed for a heavy SM Higgs boson.
For instance, a search of the $H \! \to \! WW \! \to \! \ell \ell \nu \nu$ and $H \! \to \!
ZZ \! \to \! 4\ell$ channels have been conducted in the SM with the full event sample recorded at the LHC first phase, i.e. 4.9 fb$^{-1}$ at $\sqrt s=7$ TeV and 19.4 fb$^{-1}$ at $\sqrt s=8$ TeV for $H \to WW^*$ \cite{CMS-WW} and 5.1 fb$^{-1}$ at $\sqrt s=7$ TeV and 19.7 fb$^{-1}$ at 
$\sqrt s=8$ TeV for $H \to ZZ^*$ \cite{CMS-ZZ}. The high mass range was analyzed and the events
corresponding to the observed state with a mass of 125 GeV were considered as a background. 
In the $ZZ$ channel,  an additional CMS analysis in the $H \to ZZ^* \to 2\ell 2q$ channel 
has been made with the 19.6 fb$^{-1}$ data collected at 8 TeV \cite{CMS-2l2q}. All these analyses
exclude a significant area in the $h$MSSM parameter space at low and moderate $\tb$. 

\begin{figure}[!h]
\vspace*{-.5cm}
\begin{center}
\begin{tabular}{c}
\includegraphics[scale=0.58]{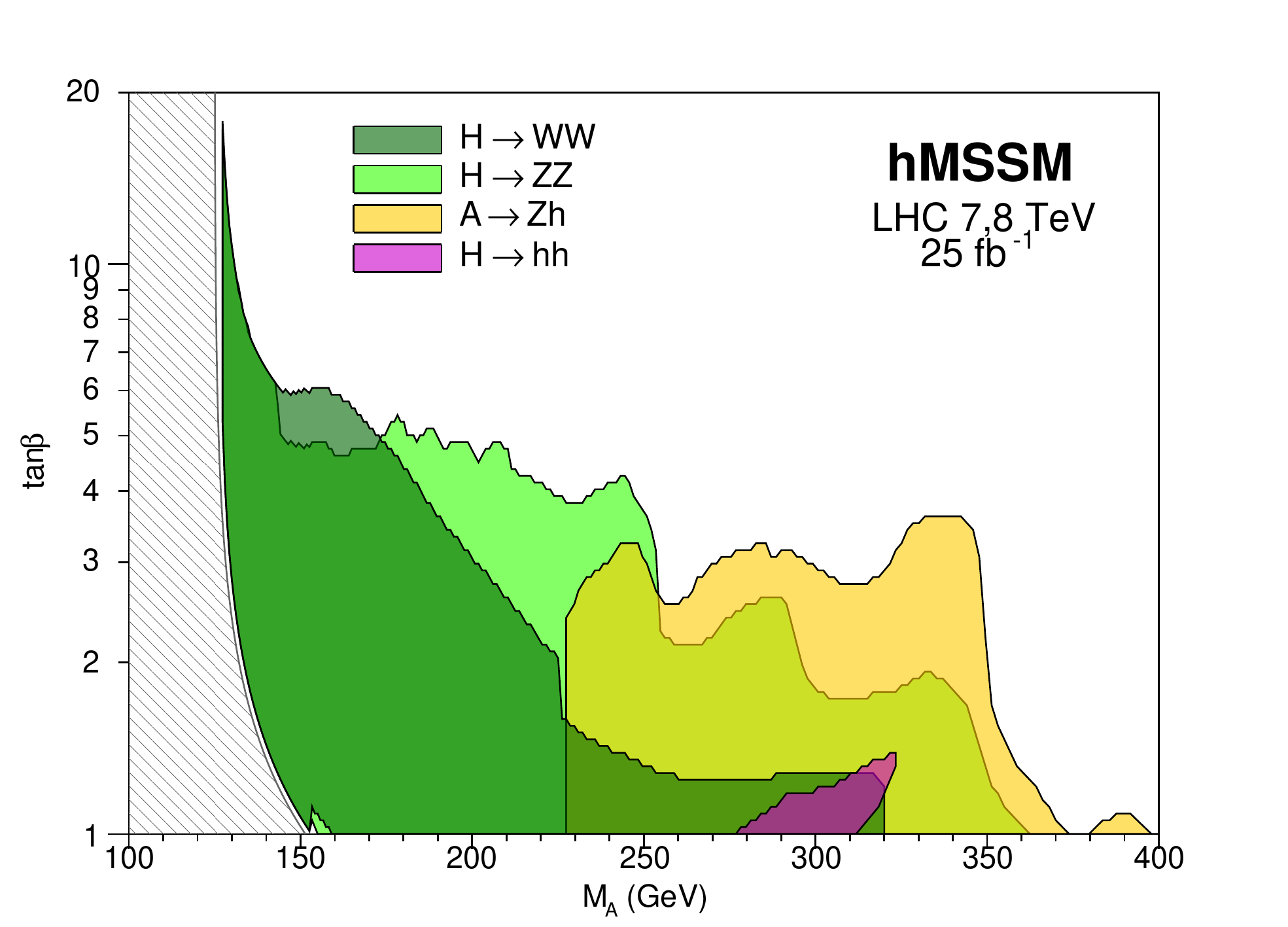} 
\end{tabular}
\vspace*{-.3cm}
\caption{Constraints in the $[\tb, M_A]$ plane of the $h$MSSM from search at the LHC for a 
heavy CP--even boson decaying into $WW,ZZ,hh$ and a heavy CP--odd boson decaying into $hZ$ 
final states. The searches are for $\sqrt{s}=7+8$~TeV c.m. energy and $25$ fb$^{-1}$ 
of accumulated data. The dashed area is the one that is ill--defined in the $h$MSSM.}
\label{AZ_hh_WW_ZZ_LHC8_25fb}
\end{center}
\vspace*{-.5cm}
\end{figure}

In the case of the $H \! \to \! WW \! \to \! \ell \ell \nu \nu$ CMS search, when all production
channels are included (there is a dominance of the $gg\to H$ mode of course) and the various
final state topologies are summed up, a Higgs particle with a SM--like coupling to gauge 
bosons is excluded from $M_H \! \approx \! 200$ GeV to $\approx \! 600$ GeV. The 95\%CL upper 
limit as a function of $M_H$ and relative to the 
SM expectation can be easily turned into an exclusion area in the $[M_A, \tb]$ plane  by 
considering the production and decay rates of the MSSM $H$ state discussed in 
section 3. The result is shown in Fig.~\ref{AZ_hh_WW_ZZ_LHC8_25fb} where the area excluded 
by this search, interpreted in the context of the $h$MSSM, is depicted in dark green. 

The exclusion area starts at relatively high $\tb$ values, $\tan\beta \! \gsim \! 10$, and
light $A$, $M_A \! \approx 140$ GeV (below this limit, we enter the domain 
in which the model is ill defined, a domain that extends to $M_A \! \approx 160$ GeV and 
$\tb \! \approx \! 1$) where one has an $H$ state with a mass $M_H \! \gsim 160$ GeV and a
coupling $g_{HVV}= \cos(\beta-\alpha)$ that is not very small as shown in the right-hand side 
of Fig.~\ref{differ}, allowing for substantial $H$ production times decay rates. For 
$\tb \! \approx \! 1$, the excluded region extends to $M_A \! \approx \! 250$ GeV, when other decay channels such as $H \! \to \! hh$ and even $H \! \to \! t \bar t$ open up and suppress 
the massive gauge boson decay modes. 

As a result of its clean final state and despite of the low statistics, the $H \! \to \! ZZ \!
\to \! 4\ell$ search turns out to be more constraining at high mass and excludes a SM--like
Higgs boson up to $M_H \! \approx$ 800 GeV (with a search domain extending to 1 TeV). While for 
low  $M_H$ values, $H \! \to \! ZZ$  is less powerful than the companion $H \! \to \! WW$ mode
as a consequence of the reduced phase space, it clearly becomes the leading channel for $M_H 
\! \gsim \! 250$ GeV. In fact, because of the higher statistics, the most  severe constraint 
is obtained in the combination of the $H \! \to \! ZZ \! \to \! 4 \ell, 2\ell 2\nu, 2\ell 2q$ 
topologies that was performed in Ref.~\cite{CMS-2l2q}. Here, the 95\%CL exclusion of 
a Higgs state with SM--like couplings extends to a mass  close to 1 TeV. 

The area excluded at 95\%CL by the non observation of these $ZZ$ final states at the LHC 
outside the $M_h \! \approx \! 125$ GeV mass window is given by the light green area of 
Fig.~\ref{AZ_hh_WW_ZZ_LHC8_25fb}. It extends from $M_A \approx 160$ GeV to $M_H \! \approx
\! 280$ GeV and concerns all values $\tb \lsim 5$. An additional small area around $M_A\! 
\approx \!300$ GeV and $\tb \lsim 2$, in which $M_H$ is close to the $2m_t$ threshold and 
the gluon--fusion amplitude $A_{1/2}^H$ is maximal thus enhancing the $gg \! \to \! 
H$ cross section, is also excluded. 

One should note that in a dedicated MSSM search, not only this $H \! \to \! ZZ$ channel but 
also the $H \! \to \! WW$ mode will lead to more effective constraints as the SM and MSSM 
Higgs particles have total decay widths that are completely different at high masses \cite{Review}. Indeed, while the SM state would have been a very wide resonance, the MSSM $H$ 
boson is a relatively narrow resonance as shown in  Fig.~\ref{fig:Gam}, allowing to select
smaller bins for the $VV$ invariant masses that lead to a more effective suppression of 
the various backgrounds.\smallskip

The resonant $H \! \to \! hh$ channel, which is important in the mass range between 250 
and slightly above 350 GeV has been considered by both the ATLAS and CMS collaborations 
with the $\approx 20$ 
fb$^{-1}$ of data collected at $\sqrt s=8$ TeV. The main focus was on the $\gamma\gamma b\bar{b}$
signature \cite{ATLAS-hh,CMS-hh} but additional searches in the 4 $b$--quark final
state have been recently reported \cite{ATLAS-4b,CMS-4b}. However, neither collaborations
has interpreted the 95\%CL exclusion limits in these channels in the context of the MSSM, 
the main reason being again that the low $\tb$ area in which these signals occur is not
theoretically accessible in the usual benchmark scenarios used for the MSSM Higgs sector.

The interpretation is however straightforward in the $h$MSSM as the trilinear self--coupling 
$\lambda_{Hhh}$ that controls the $H\to hh$ decay rate is simply given, as shown in eq.~(\ref{eq:Hhh}), in terms of the angles $\alpha,\beta$ and the radiative correction matrix element  
$\Delta M_{22}^2$ that is fixed in terms of $\tb$ and $M_A$ if the constraint $M_h\!=\!
125$ GeV is used. We have adapted the constraints from these analyses to the $h$MSSM case and 
the resulting excluded domain in the $[\tb,M_A]$ plane  is shown in purple 
in Fig.~\ref{AZ_hh_WW_ZZ_LHC8_25fb}. It covers the very low $\tb$ region, $\tb \lsim 2$, 
for the mass range between $M_A \approx 270$ GeV (which implies $M_H \gsim 250$ GeV for these low
$\tb$ values)  and $M_A \approx 330$ GeV, i.e. slightly before the $2m_t$ threshold. 

Similarly to the previous channel, the $A\to hZ$ mode has only been considered in the context of 
two Higgs doublet models \cite{Lisbon} and not in the MSSM. A CMS analysis considered the 
final state $b\bar b \ell^+\ell^-$ with 
the $\approx 20$ fb$^{-1}$ collected in 2012 at 8 TeV \cite{CMS-hZ}. A search of 
both the $A\to hZ$ and $H\to hh$ channels has been performed by CMS again in the 
multi-lepton and eventually photon finale states \cite{CMS-hZ+hh}. The impact of the 
95\%CL exclusion limits of these studies, when interpreted in the context of the 
$h$MSSM, is illustrated by the yellow area of Fig.~\ref{AZ_hh_WW_ZZ_LHC8_25fb}. The ranges $\tb \lsim 3$ and $M_A \approx 230$--350 GeV  should be 
in principle excluded with the present data.  

\subsection{Summary of the constraints at 8 TeV and projections for 14 TeV}

\begin{figure}[!h]
\vspace*{-.5cm}
\begin{center}
\begin{tabular}{c}
\includegraphics[scale=0.58]{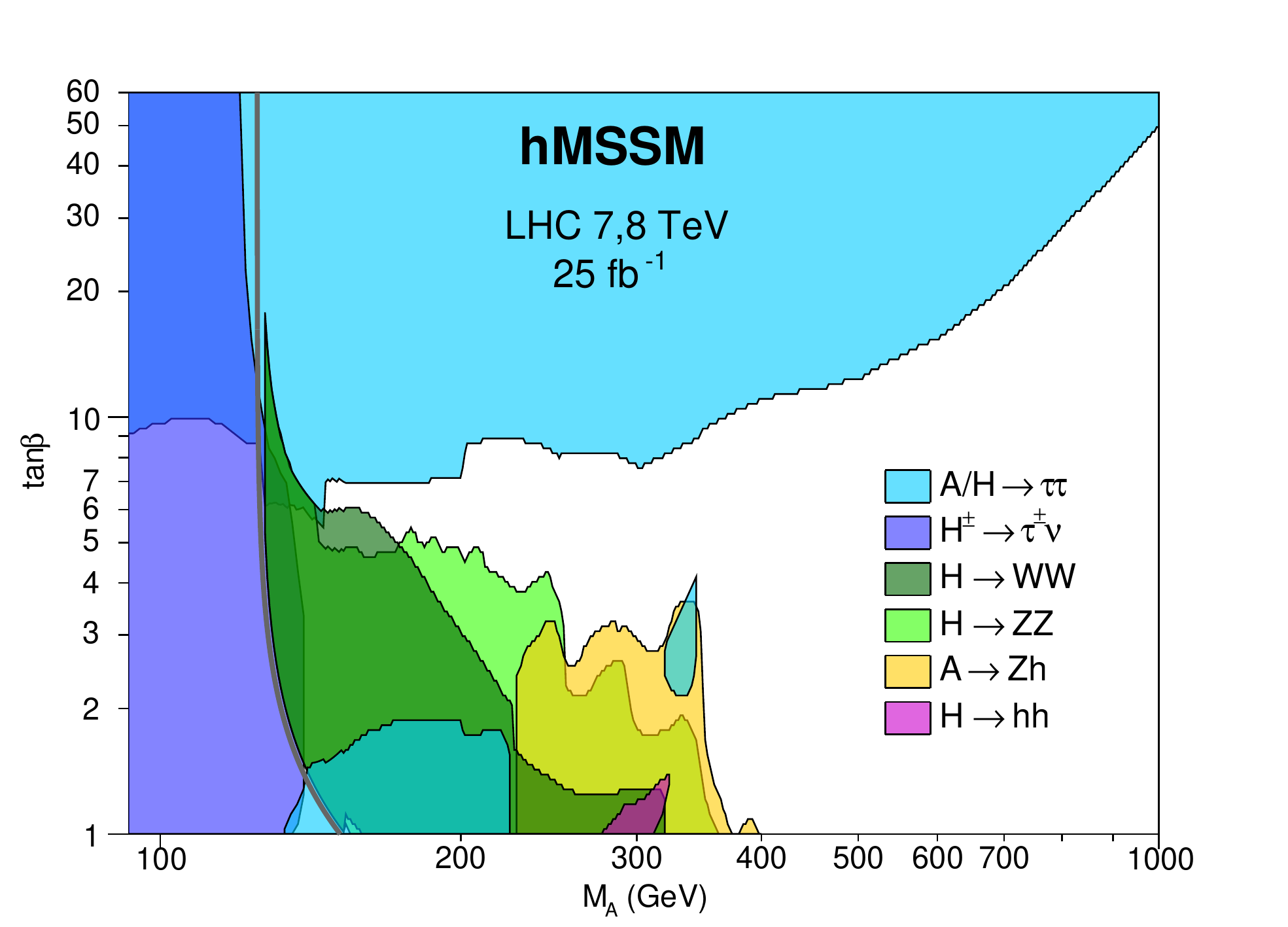} 
\end{tabular}
\vspace*{-.3cm}
\caption{The combined constraints in the $[\tb, M_A]$ plane of the $h$MSSM from searches at the LHC for the heavier $H,A$ and $H^\pm$ bosons decaying into either fermionic or bosonic final states. All the searches performed at c.m. energies up to  $\sqrt{s}=8$~TeV and $25$ fb$^{-1}$
data are included. 
}
\label{constraints_LHC8_25fb}
\end{center}
\vspace*{-10mm}
\end{figure}

Wrapping up the discussion up to this point, the impact on the $[M_A, \tb]$ plane
of the searches in the fermionic Higgs decays $H/A \to \tau\tau$ and $H^\pm \to \tau\nu$ 
and in the bosonic ones $H\to WW,ZZ,hh$ and $A\to Zh$ performed by the ATLAS and CMS
collaborations at $\sqrt s =7$+8 TeV with up to $\approx 25$ fb$^{-1}$ data are combined 
in  Fig.~\ref{constraints_LHC8_25fb}. The outcome is very impressive. The high $\tb \gsim 10$ is entirely excluded for $M_A \lsim 500$ GeV by the $\tau\tau$ searches. The range $\tb \lsim 4$ 
is excluded for $M_A \lsim 250$ GeV by the $H\to WW,ZZ$ channels.  For $\tb \lsim 2$, the 
excluded domain extends to $M_A \lsim 350$ GeV when the channels $H\to hh$ and $A\to hZ$ 
are considered. The entire low $M_A$ region, $M_A \lsim 140$ GeV, is excluded by the 
$H^+ \to \tau \nu$ search for any value of $\tb$. An additional portion of this low $M_A$ 
area is excluded by the $A\to \tau \tau$ search for low $\tb$ values.  

In fact, the entire area in which the $h$MSSM is not mathematically defined, and which is
delineated by the solid line in the figure, is excluded  by these $H^\pm$ and $A$ searches 
that do not involve the undefined CP--even $H$ boson mass $M_H$ and the mixing angle $\alpha$.  

\FloatBarrier

These constraints, if no new signal is observed, can be vastly improved at the next phase of 
the LHC with a center of mass energy up to $\sqrt s=14$ TeV and with one or two orders of magnitude accumulated data. More optimistically, this implies that the $2\sigma$ sensitivity
for a heavier MSSM Higgs boson will be drastically enhanced at the next LHC  phase.  
Starting from the expected median 95\%CL exclusion limits that have been given by the ATLAS and
CMS collaborations in the various searches performed at 8 TeV with $\approx 20$ fb$^{-1}$, we have made an extrapolation to this next LHC phase with $\sqrt s=14$ TeV and 300 fb$^{-1}$ data. 
We have naively assumed that the sensitivity will simply scale with the square root of the 
number of expected events and did not include any additional systematical effect. This comes 
from the observation that the results of the experimental analyses  are limits on the signal 
cross section at a given c.m. energy for a given resonance mass bin, $R^S_{\sqrt s} (M_A)$, 
for a channel that is subject to a given background rate $R^B_{\sqrt s} (M_A)$ at this mass 
bin, when the integrated luminosity is fixed at a value ${\cal L}_{\sqrt s}$. 
Knowing the sensitivity limit $R^{S}_{8}(M_A)$ at $\sqrt s=8$ TeV, one derives the 
associated limit at $\sqrt s= 14$ TeV using
\beq
R^{S}_{14}({M_A}) = 
\sqrt{ {\cal L}_{8} / {\cal L}_{14} }  \times
\sqrt{ R^{B}_{14}({M_A}) / R^{B}_{8}({M_A}) } \times  
R^{S}_{8}({M_A}) 
\eeq
Having the knowledge of only the signal cross sections $\sigma^S_{\sqrt s} (M_A)$ for the 
various points and not the corresponding background rates, we assume that the latter simply and
very naively scale like the signal cross sections. This is the case of some channels 
of interest, such as $gg\to H/A \to t\bar t$ whose main background is $gg\to t\bar t$ and 
as both are $gg$ initiated processes, they roughly scale with the $gg$ luminosity at 
higher energies. However, for many other channels such as $H\to WW,ZZ$ or $A/H \to \tau \tau$, the irreducible background is mostly due to $q\bar q$ annihilation which increases more
slowly with energy than the initiated $gg$ signal processes. This makes our approach rather conservative.

With this assumption, one obtains for the sensitivity at $\sqrt s\!=\!14$ TeV, $R_{14}^S(M_A)$,  needed to set the exclusion limit, that we turn into a 95\%CL sensitivity, for a given $M_A$ 
\beq 
R^{S}_{14}({M_A}) \approx \sqrt{ {\cal L}_{8} / {\cal L}_{14} }  \times 
\sqrt{ \sigma^{S}_{14}({M_A}) / \sigma^{S}_{8}({M_A}) } \times R^{S}_{8}({M_A}) 
\eeq
  
\begin{figure}[!h]
\vspace*{-3mm}
\begin{center}
\begin{tabular}{c}
\mbox{\hspace*{-7mm}
\includegraphics[scale=0.42]{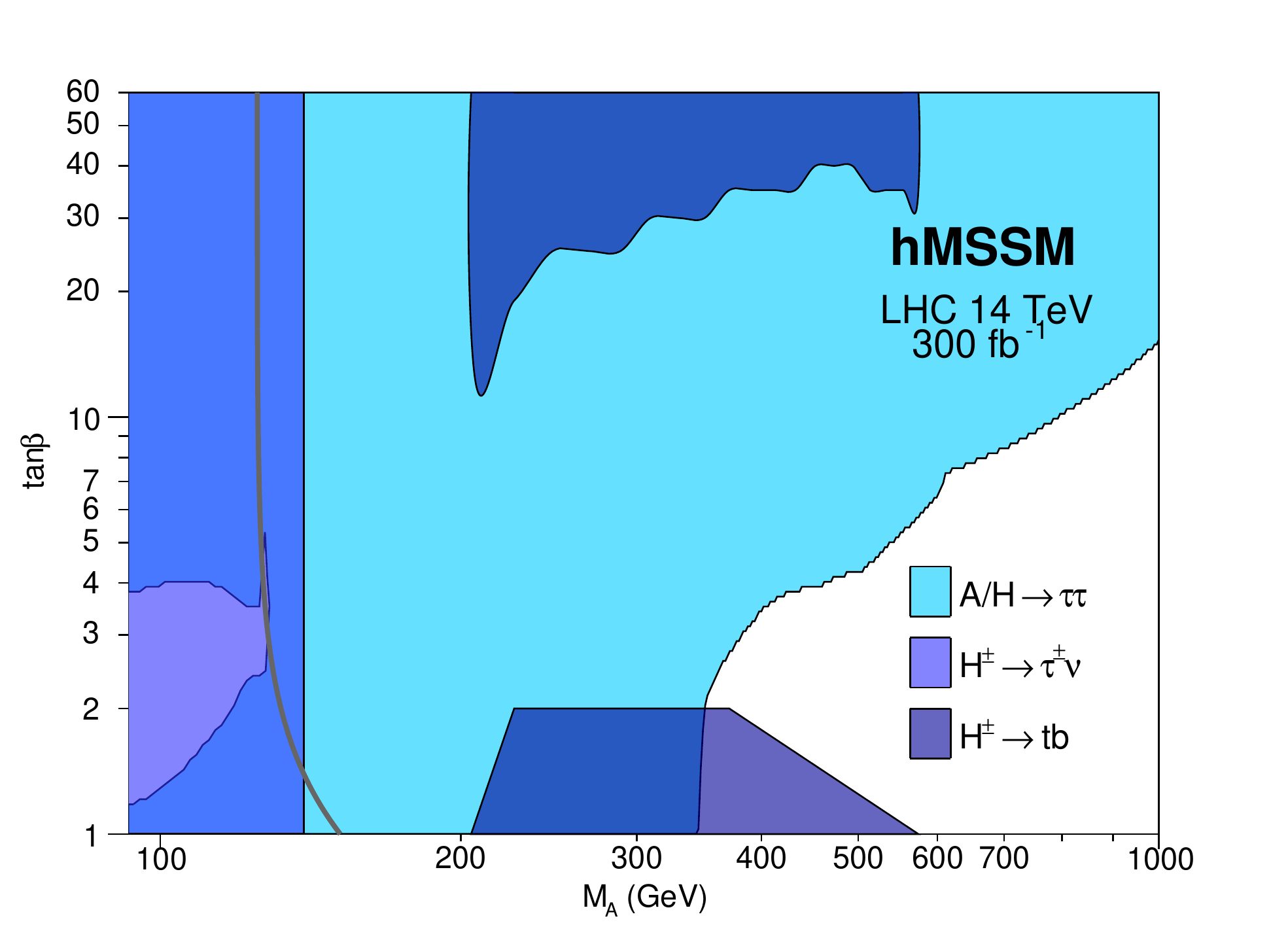}\hspace*{-7mm} 
\includegraphics[scale=0.42]{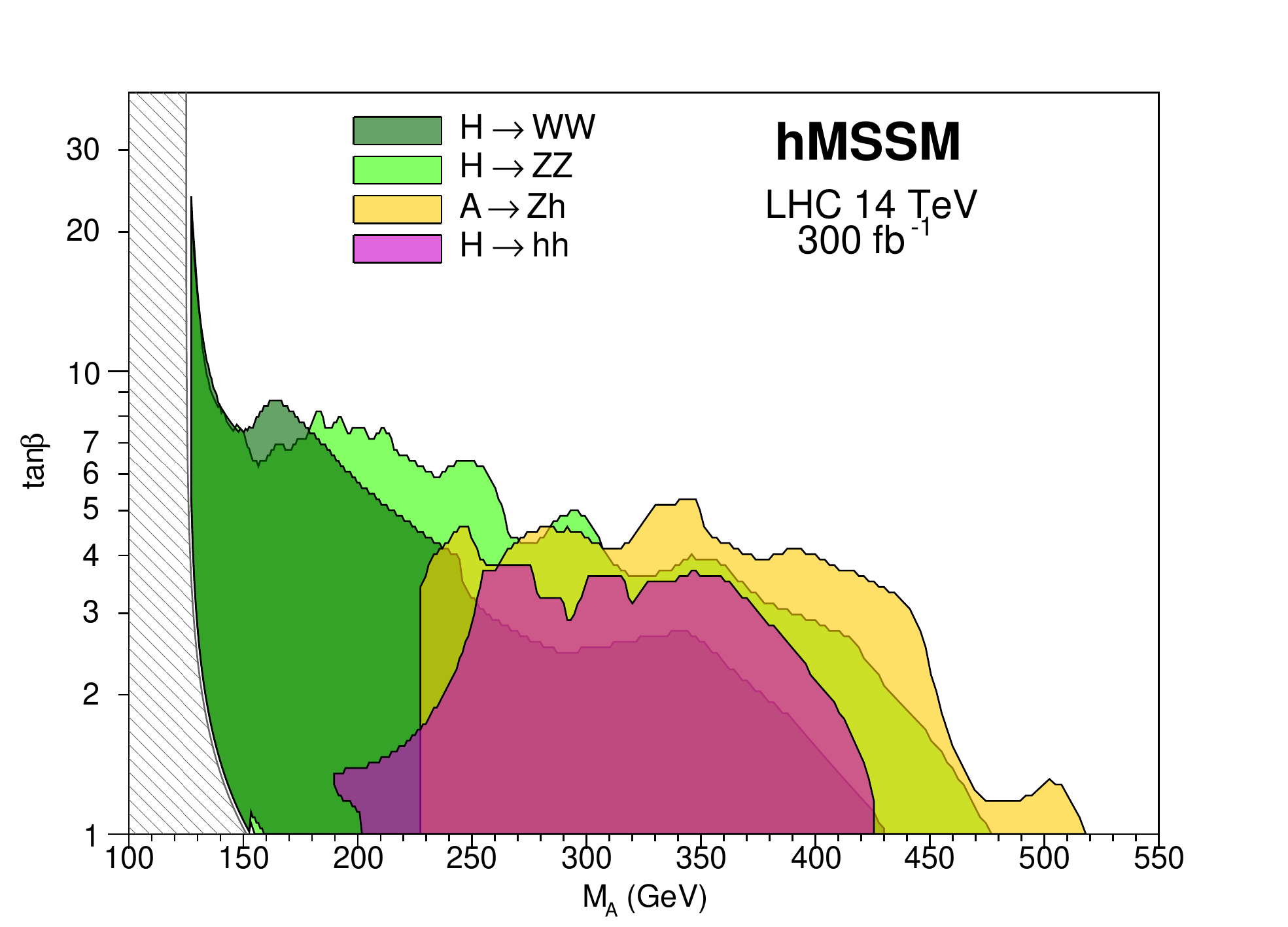} 
}
\end{tabular}
\vspace*{-.5cm}
\caption{Projections for the LHC with $\sqrt{s}=14$~TeV and $300$ fb$^{-1}$ data for
the $2\sigma$ sensitivity in the $h$MSSM $[\tb, M_A]$ plane from the search for $A/H^\pm$ states 
in their fermionic decays (left) and $A/H$ states in their bosonic decays (right). The
same color code as at $\sqrt s\!=\!8$ TeV has been used and, for the fermionic channels, we 
add a constraint from the $H^+ \to tb$ mode depicted in dark blue.}
\label{constraints_LHC14_300fb-two}
\end{center}
\vspace*{-.7cm}
\end{figure}

\begin{figure}[!h]
\vspace*{-3mm}
\begin{center}
\begin{tabular}{c}
\includegraphics[scale=0.6]{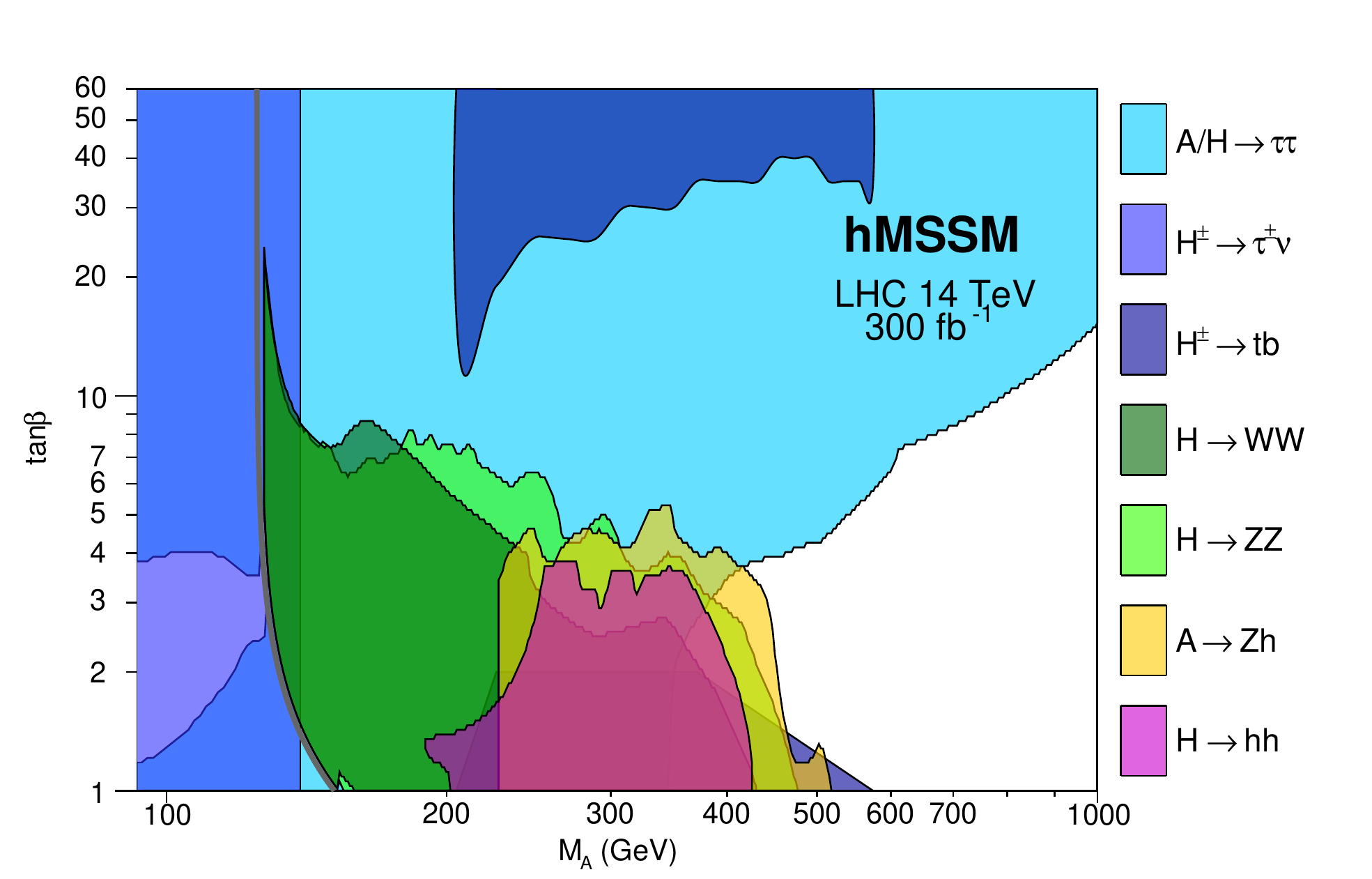} 
\end{tabular}
\vspace*{-.3cm}
\caption{Projections for the LHC with $\sqrt{s}=14$~TeV and $300$ fb$^{-1}$ data for
the $2\sigma$ sensitivity in the $h$MSSM $[\tb, M_A]$ plane when the searches for the
$A/H/H^\pm$ states in their fermionic and bosonic decays are combined.}
\label{constraints_LHC14_300fb}
\end{center}
\vspace*{-.7cm}
\end{figure}

The output of this procedure is presented in the $[\tb, M_A]$ $h$MSSM plane in 
Figs.~\ref{constraints_LHC14_300fb-two}  for the fermionic (left) and bosonic (right)
Higgs search channels. In the former case, we have included 
in addition the channel $pp \to tbH^+ \to tb tb$ which now shows some sensitivity a low 
$\tb$ and not too high $M_{H^\pm}$ values. The combined expected 95\%CL sensitivities are 
shown in Fig.~\ref{constraints_LHC14_300fb} and, as can be seen, a vast improvement of 
the current sensitivity to the MSSM parameter space is foreseen in all channels. 
This is particularly the case of the $A/H\to \tau\tau$ channels which alone, closes the 
entire region below $M_A\lsim 350$ GeV for any $\tb$ value, while the $H\to WW,ZZ$ modes 
which show sensitivity up to $M_A \approx 600$ GeV at very low $\tb$. In the Higgs mass range 
in which they are relevant, i.e. below the $t\bar t$ threshold, the channels $H\to hh$ and 
$A\to hZ$ start to probe rather high $\tb$ values, $\tb \approx 4$ and $\tb \approx 6$, respectively.

Nevertheless, there will remain an area of the $h$MSSM parameter space, at $\tb \lsim 4$ and masses above $M_A\approx 400$ GeV to name it, which will not be accessible by
the channels that have been considered so far in the search of the heavier $H/A$ and $H^\pm$ 
states. To probe this area, the high luminosity option of the LHC with ${\cal L}=3$ ab$^{-1}$ 
data or a higher energy $pp$ collider, such as the presently discussed Fcc--pp at $\sqrt s \approx 100$ TeV will be necessary. However, as it was discussed in many instances in this paper, this virgin area is the ideal territory to perform searches in the $gg \to H/A \to t\bar 
t$ channel to which we turn our attention now.

\subsection{Including the $\mathbf{pp\to H/A\to t\bar t}$ channel}

As it was discussed at length in the previous section, for low $\tb$ and high $M_A$ values, 
the decay modes $H/A \to t\bar t$ of the heavier MSSM scalar and pseudoscalar Higgs states
will largely become the dominant ones while the $gg\to H/A$ cross sections are 
still substantial thanks to the large Higgs coupling to the top quarks that mediate
the production  process.  Hence, the search for resonances decaying into $t\bar t$  final states 
will be mandatory in order to probe these areas of the $[M_A, \tb]$ parameter space 
at the LHC. However,  a peak in the invariant mass distribution of the $t\bar t$ system, 
that one generally expects in the narrow--width approximation, is not the only signature of a 
Higgs resonance in this case. Indeed, the  $gg\to H/A$ signal will interfere with the QCD  
$t\bar t$ background  which, at LHC energies, is mainly generated by the gluon--fusion channel, 
$gg\to t\bar t$. The signal--background interference will depend on the CP nature of 
the $\Phi=H/A$ boson and on its mass and total decay width; it can be either constructive or destructive, leading to a rather complex signature with a peak--dip structure of the 
$t\bar t$ invariant mass distribution. 

These aspects are known since already some time and have especially been discussed in the context of a heavy SM Higgs state \cite{Htt0} and, hence, for the CP--even Higgs case. The
slightly more involved MSSM situation, as there are one CP--even and one CP--odd 
resonances that are close in mass, has been addressed  only in a very few places; see 
for instance Refs.~\cite{Htt1,Htt}. Dedicated analyses have been performed at the parton--level 
only and do not make use of recent developments like boosted heavy quark techniques 
\cite{boost} that could allow to enhance the observability of the Higgs signal. The ATLAS and 
CMS collaborations have performed searches for heavy states decaying into $t\bar t$ pairs
\cite{ATLAS-tt,CMS-tt} but did not specifically address the complicated Higgs situation as  
only electroweak spin--one resonances, like new neutral gauge bosons or electroweak 
Kaluza--Klein excitations, were considered. In these two cases, the main production channel is
$q\bar q$ annihilation and there is no interference with the (colored) QCD $q\bar q \to t\bar 
t$ background and the resonance signal simply appears as a peak in the invariant mass
distribution  of the $t\bar t$ pair.  
 
A full and realistic Monte--Carlo simulation  of the $gg\to H/A\to t\bar t$ process including
the effects of the interference and taking into account reconstruction and detector aspects 
is beyond the scope of this paper, and will be postponed to a future publication 
\cite{preparation}. Here, we will simply make a very crude estimate of the sensitivity that 
can be achieved in this channel, relying on previous ATLAS \cite{ATLAS-tt} and CMS 
\cite{CMS-tt} analyses performed at $\sqrt s=8$ 
TeV c.m. energy in the spin--one resonance context mentioned above. We will naively consider 
the number of signal and background events, applying very simple kinematical cuts and 
ignoring the complicated interference effects, and delineate the area in the [$\tan\beta,M_A$] 
$h$MSSM parameter space in which one has $N_{\rm signal}/ \sqrt{N_{\rm bkg}} \geq s$. The    significance $s=5$ would correspond to a $5\sigma$ observation of the Higgs signal 
while $s=2$ would be a first hint of the new effect; in the absence of any effect,
$s=2$ would correspond to the 95\%CL exclusion limit of the phenomenon. To further simplify 
our analysis, we will assume that the two heavy $A$ and $H$ states are mass degenerate so that 
the signal rate is simply the sum of the $A$ and $H$ production cross section times the respective branching ratios in their decays into $t\bar{t}$ pairs (which, as we have already
seen, is a  good approximation). 

The main ingredients of the analysis are as follows. The normalization of the Higgs signal 
has been 
obtained using the programs  {\tt HIGLU} for the production cross sections and {\tt HDECAY} 
for the decay branching ratios. The total cross section of the SM background (which will serve 
as a normalization) has been obtained using the program  {\tt Top++} \cite{Czakon:2011xx}. 
For the input $m_t=173.2$ GeV one obtains for the background rate at the first stage of the LHC 
with $\sqrt{s}=8$ TeV
\begin{eqnarray}
\sigma_{\rm tot}^{\rm QCD} (pp \to t\bar t) =247.7 ~ ^{+ 6.3}_{- 8.5} ~ ^{+ 11.5}_{- 11.5} 
~{\rm pb ~at~} \sqrt{s}=8~{\rm TeV} 
\end{eqnarray}
when the renormalisation and factorization scales are fixed to  $\mu_R=\mu_F=m_t$.  
In this equation, the first error is the one due to the scale variation within a factor 
of two from the central scale, and the second one the PDF+$\alpha_s$ uncertainty. 
This value for the cross section is obtained at NNLO in QCD including the resummation 
of next-to-next-to-leading logarithmic (NNLL) soft gluon terms and it turns out 
that it is only 3\% larger than the value of the cross section when evaluated at NNLO 
\cite{tt-NNLO}. Note that at $\sqrt s=14$ TeV, using the same approximation and 
ingredients, one would obtain for the cross section
 \begin{eqnarray}
\sigma_{\rm tot}^{\rm QCD} (pp \to t\bar t)  =966.0 ~ ^{+ 22.7}_{- 33.9}~ ^{+ 40.5}_{- 40.5} 
~{\rm pb ~at~} \sqrt{s}=14~{\rm TeV} 
\end{eqnarray}

\begin{figure}[!t]
\vspace*{-4mm}
\begin{center}
\includegraphics[scale=0.4]{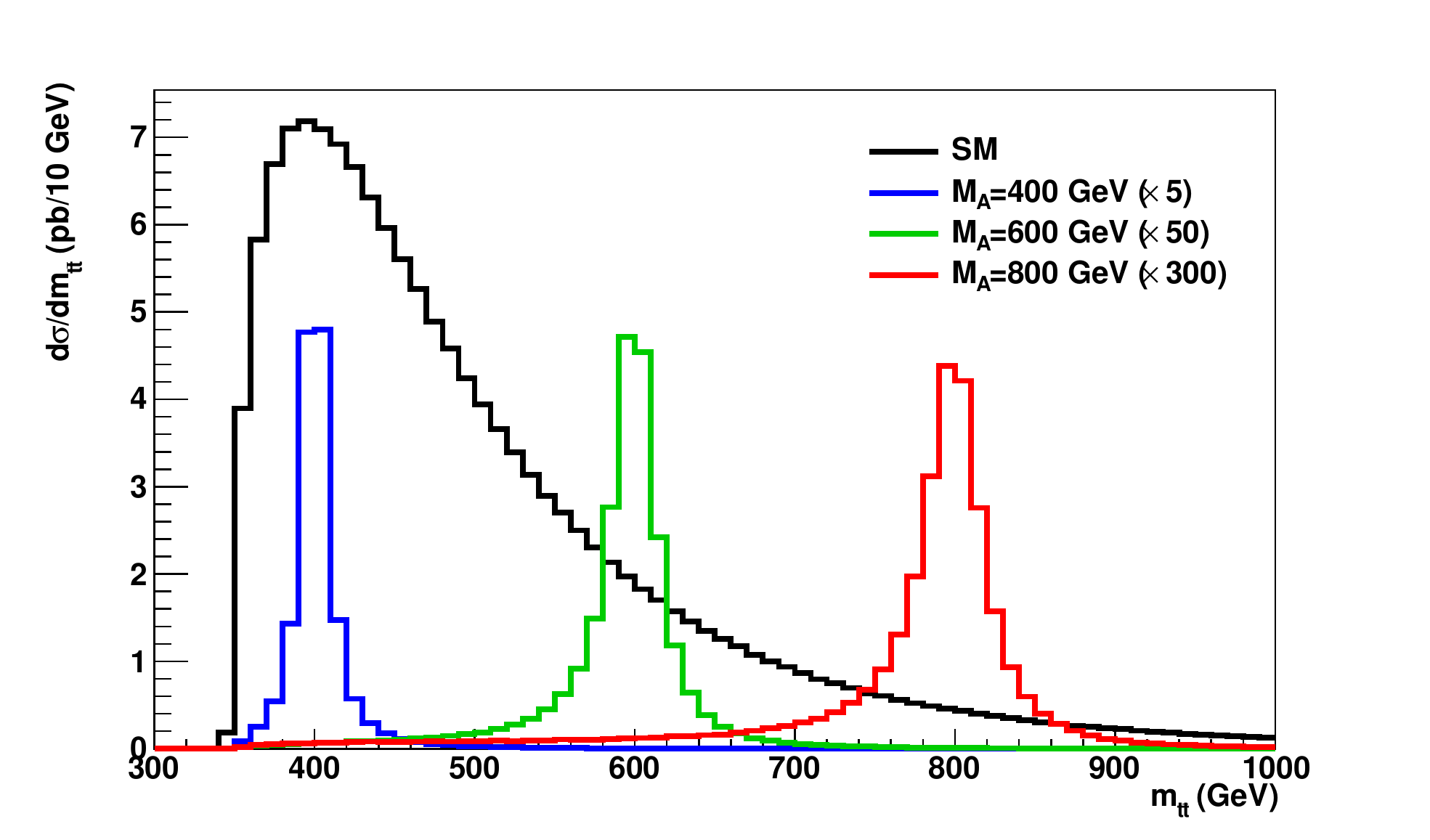}\\[-2mm] 
\mbox{\hspace*{-3mm}
\includegraphics[scale=0.4]{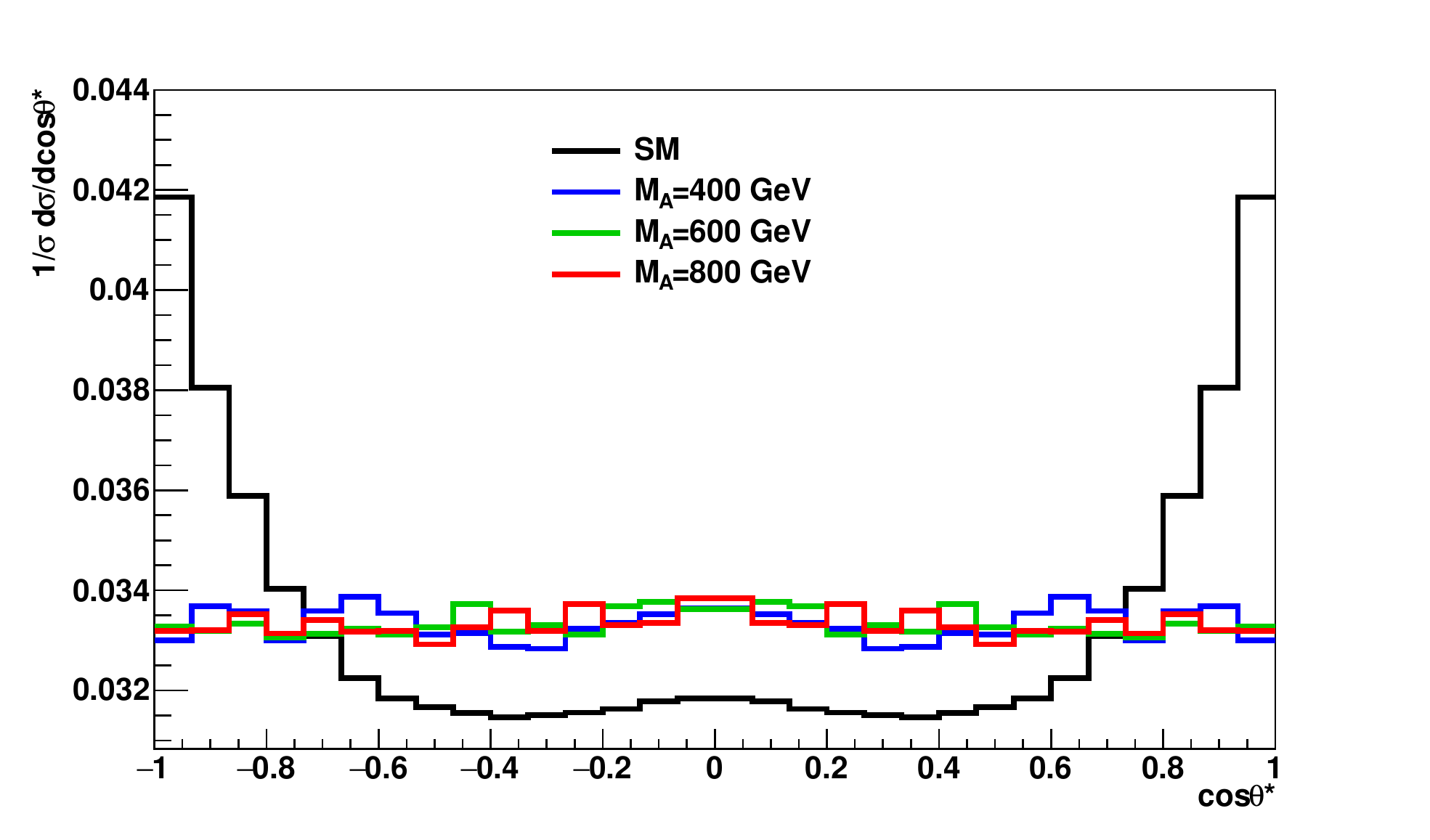}\hspace*{-3mm}
\includegraphics[scale=0.4]{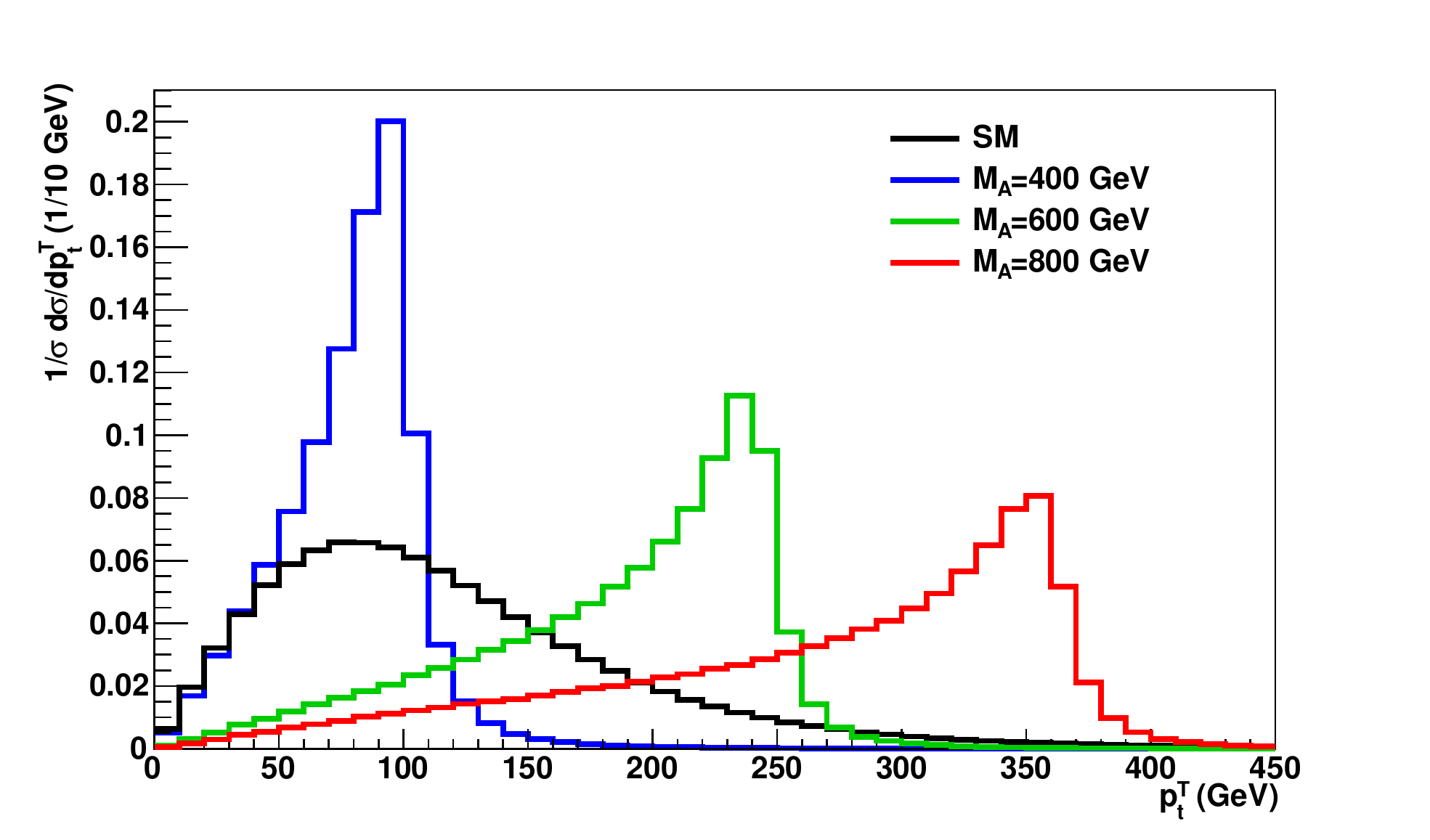} 
}
\vspace*{-6mm}
\caption{The $pp\to t\bar t$ signal and background at $\sqrt s=14$ TeV
for $M_A=400,600$ and 800 GeV and $\tb=1$: the invariant $m_{tt}$ (top), 
$\cos\theta^*$ (bottom left) and $p_T^t$ (bottom right) distributions. }
\label{pptott-cuts}
\end{center}
\vspace*{-8mm}
\end{figure}

Using the program MadGraph5~\cite{madgraph}, we have generated the signal and
background cross sections for the process $pp\to t\bar t$.   
The differential cross section as a function of the invariant mass of the $t\bar t$
system, $d\sigma/dm_{tt}$,  is shown at $\sqrt s=8$ TeV in the upper 
part of Fig.~\ref{pptott-cuts} where mass bins of 10 GeV have been assumed. 
We overlay on the continuum QCD background distribution (in black solid line), the 
distributions for the $A$ signal only (the colored lines) with $\tb=1$ and three possible 
mass values, $M_A=400,600$ and $800$ GeV. In order to see the signals in the figure,  we 
have multiplied the distributions by a factor of 5, 50 and 300, respectively.

In order to enhance the significance $s$, one could apply very basics kinematical
cuts that suppress the background while leaving the signal almost unaffected. In 
the left and right--hand sides of Fig.~\ref{pptott-cuts}, we show two distributions
(as we are interested in the shapes only, the distributions have not been 
re-weighted with the correct $K$--factors etc.. and the integrated areas thus 
correspond to the Monte--Carlo cross sections). The
first one is $1/\sigma \times d\sigma/d\cos \theta*$ where $\theta*$ the helicity 
angle between the off--shell Higgs boson boosted back into the top quark pair rest 
frame and the top quark pair direction (left). As can be seen, while the  signal 
distribution is  almost flat, the background is peaked in the forward and backward 
directions; a cut  $|\cos \theta*| \leq 0.8$ for instance would remove a large sample 
of background events. A second distribution per 10 GeV bin is in terms of the transverse 
momentum of the top quarks,  $1/\sigma \times d\sigma/dp_T$ (right).   They show a 
characteristic behavior for the signal events, with a pronounced peak and then a sharp
drop. One grossly estimates that, for the mass value $M_A=800$ GeV for instance, a cut 
on the $p_T^t$ distribution could allow to suppress the background by a factor of 
$\approx 6$.

\begin{figure}[!h]
\vspace*{-4.5mm}
\begin{center}
\begin{tabular}{cc}
\hspace*{-6mm}
\includegraphics[scale=0.4]{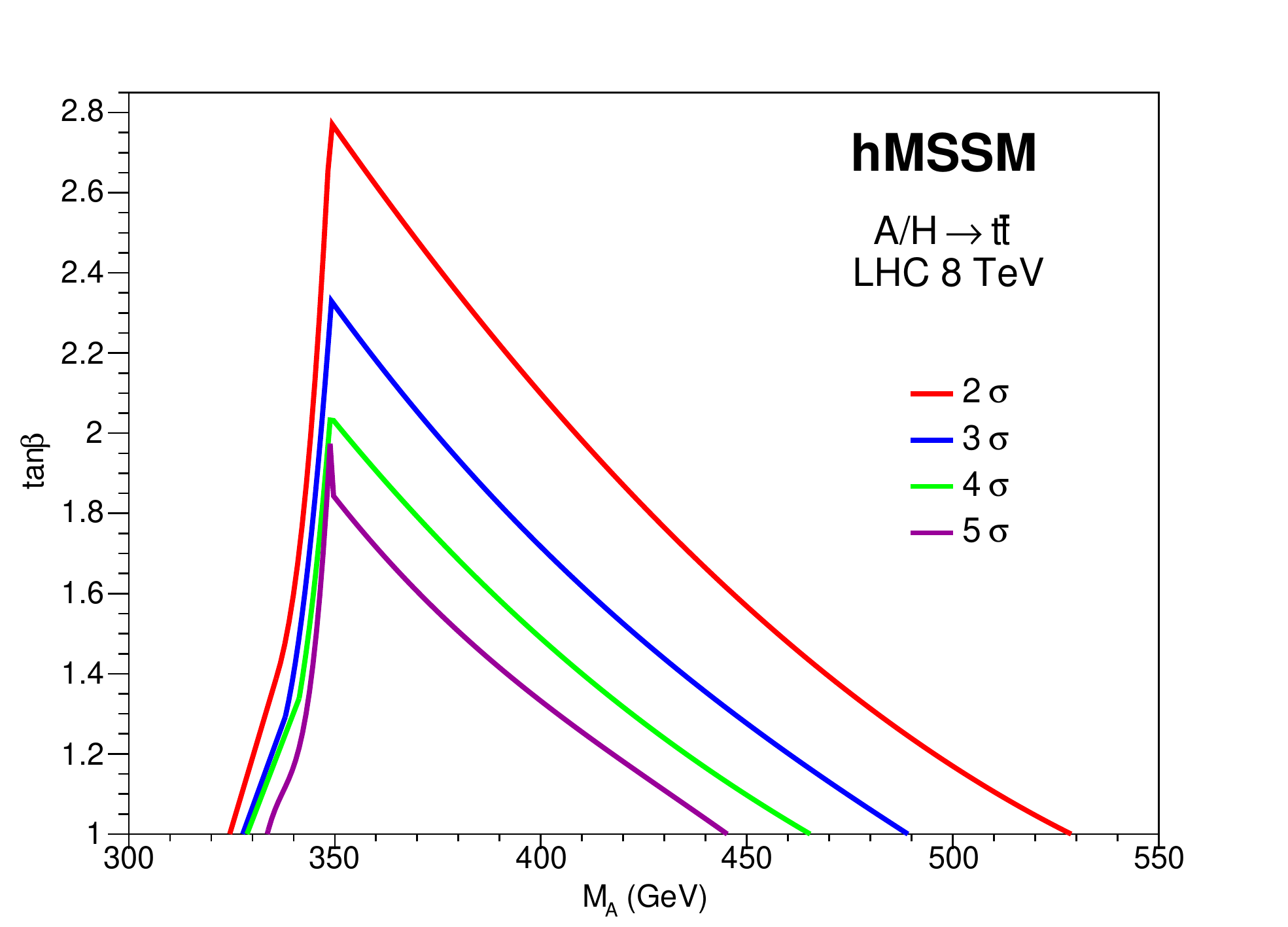} 
&\hspace*{-8mm}
\includegraphics[scale=0.4]{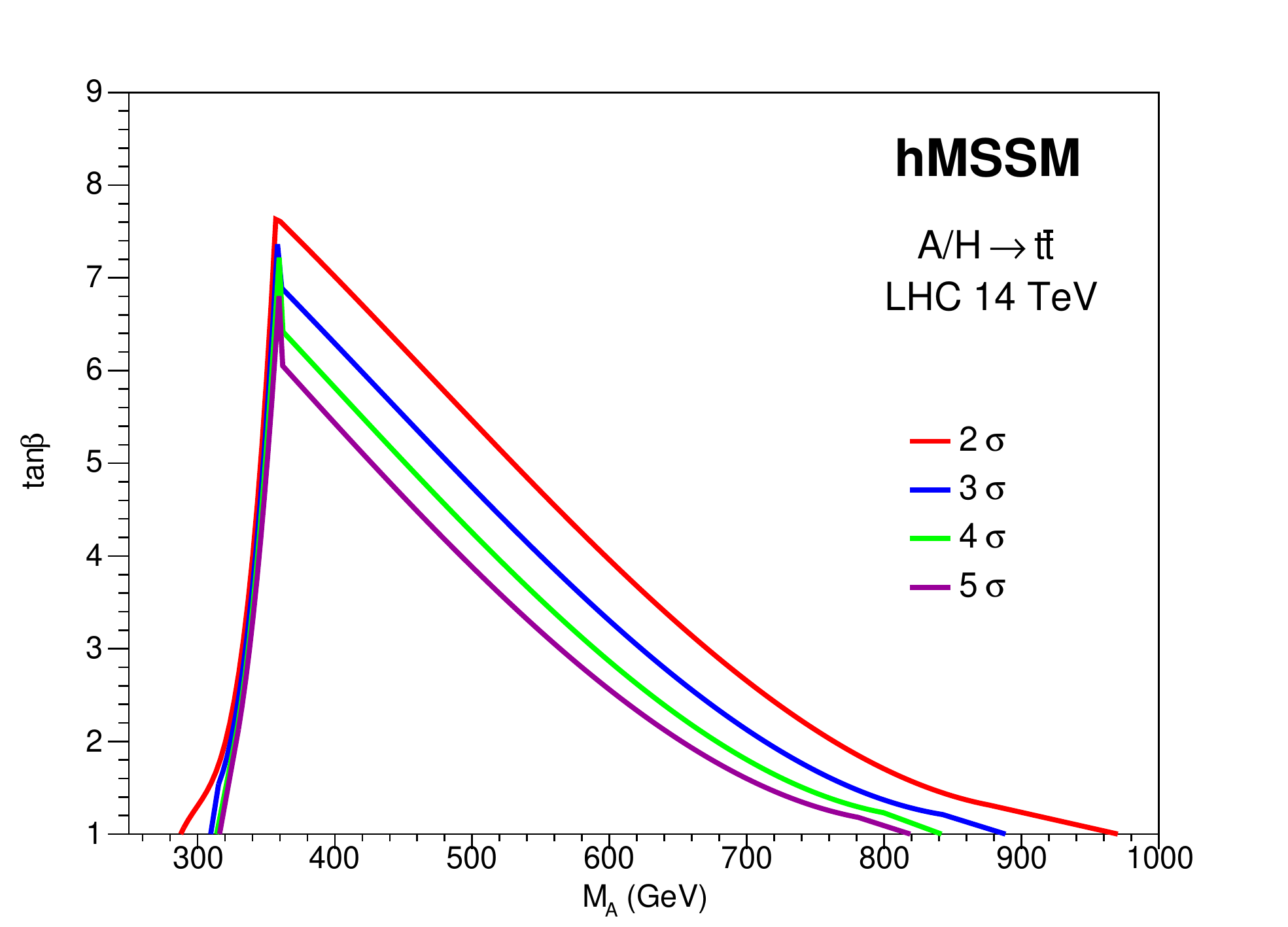} 
\end{tabular}
\vspace*{-4mm}
\caption{Sensitivity in the $gg\to H/A\to tt$ channel in the $[\tb, M_A]$ plane of the 
$h$MSSM at the $2,3,4,5\sigma$ level  $\sqrt{s}=8$~TeV and $25$ fb$^{-1}$ (left) and 
$\sqrt{s}=14$~TeV and $300$ fb$^{-1}$ (right). }
\label{expectations-tt}
\end{center}
\vspace*{-6mm}
\end{figure}

Assuming that when applying all kinematical cuts, one could suppress the $t\bar t$ 
QCD background by an order of magnitude without significantly altering the Higgs signal, we delineate
in Fig.~\ref{expectations-tt} the regions of the $[\tb, M_{A}]$ plane in which one 
would expect $N_{\rm signal}/ \sqrt{N_{\rm bkg}} \geq 2,3,4,5$. For the previous 
LHC run with $\sqrt s=8$ TeV and 25 fb$^{-1}$ data (left), one observes that a two 
$2\sigma$ ``evidence" for a new resonance, or a 95\%CL exclusion of the relevant point 
of the MSSM parameter space in the absence of any effect,  can be achieved for $\tan 
\beta \approx 2.5$ if $M_A \approx 350$ GeV and $\tan \beta \approx 1$ when $M_A
 \approx 550$ GeV.  A $5\sigma$ discovery could be achieved in this case in a much 
 smaller area of the parameter space, $\tan\beta\approx 1.5$--1 and  $M_A \approx 
350$--450 GeV.  The situation could be vastly improved at the next LHC run with 
$\sqrt s=14$ TeV and 300 fb$^{-1}$ data (right) as one could be sensitive to  
$\tan\beta$ values $\tb \approx 7$ for $M_A\approx 350$ GeV and mass values $M_A 
\approx 1$ TeV for $\tb \approx 1$.

\begin{figure}[!h]
\begin{center}
\vspace*{-3mm}
\begin{tabular}{c}
\includegraphics[scale=0.6]{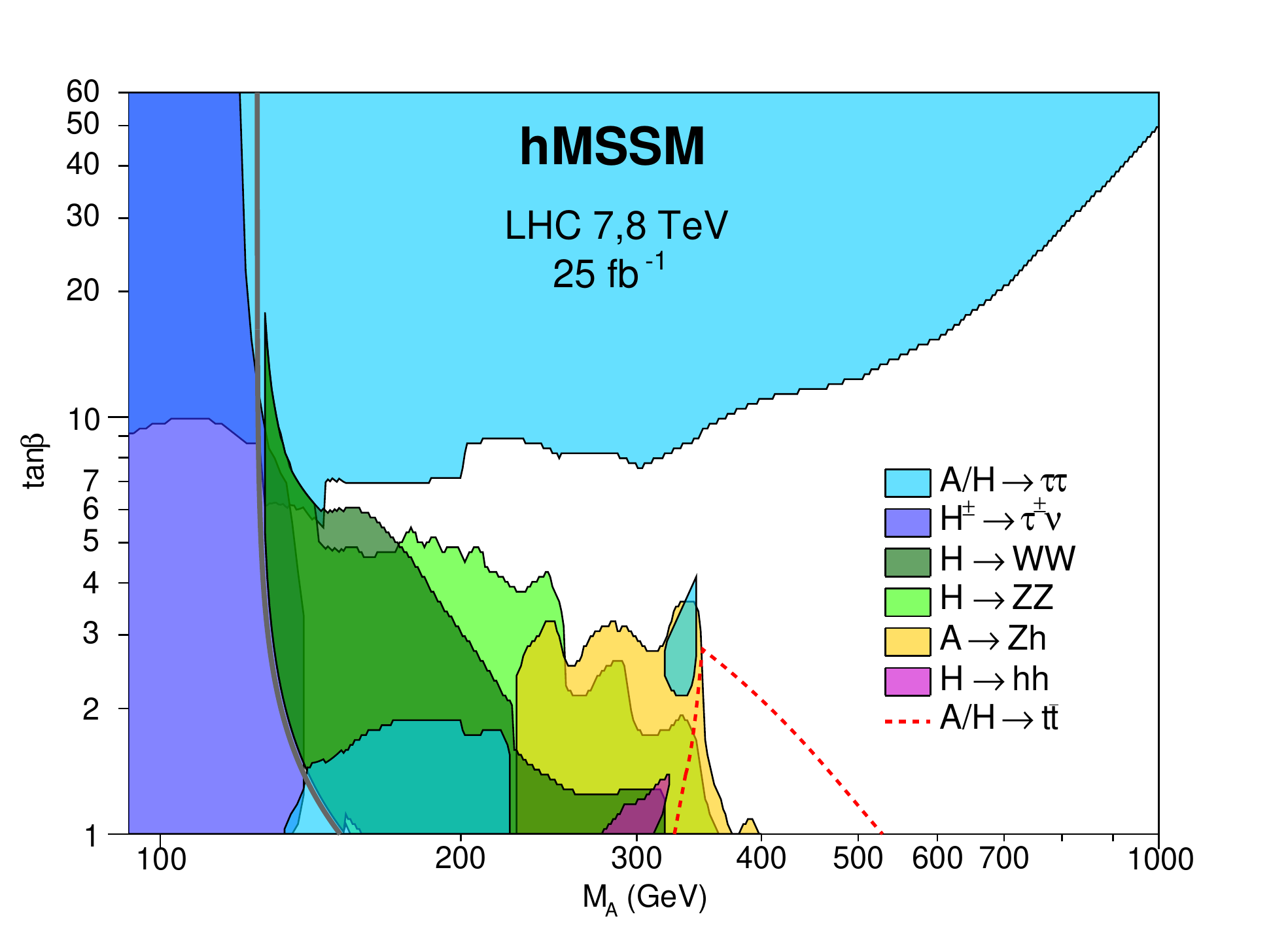} 
\end{tabular}
\vspace*{-3mm}
\caption{Expectations for the $2\sigma$ sensitivity in the $h$MSSM $[\tb, M_A]$ plane when 
the searches for the $A/H/H^\pm$ states in all channels, including the $gg\to H/A\to tt$ process, are combined at the LHC with $\sqrt{s}=8$~TeV and $25$ fb$^{-1}$ data.} 
\label{all-constraints_LHC8}
\end{center}
\vspace*{-6mm}
\end{figure}

\begin{figure}[!h]
\begin{center}
\vspace*{-3mm}
\begin{tabular}{c}
\includegraphics[scale=0.65]{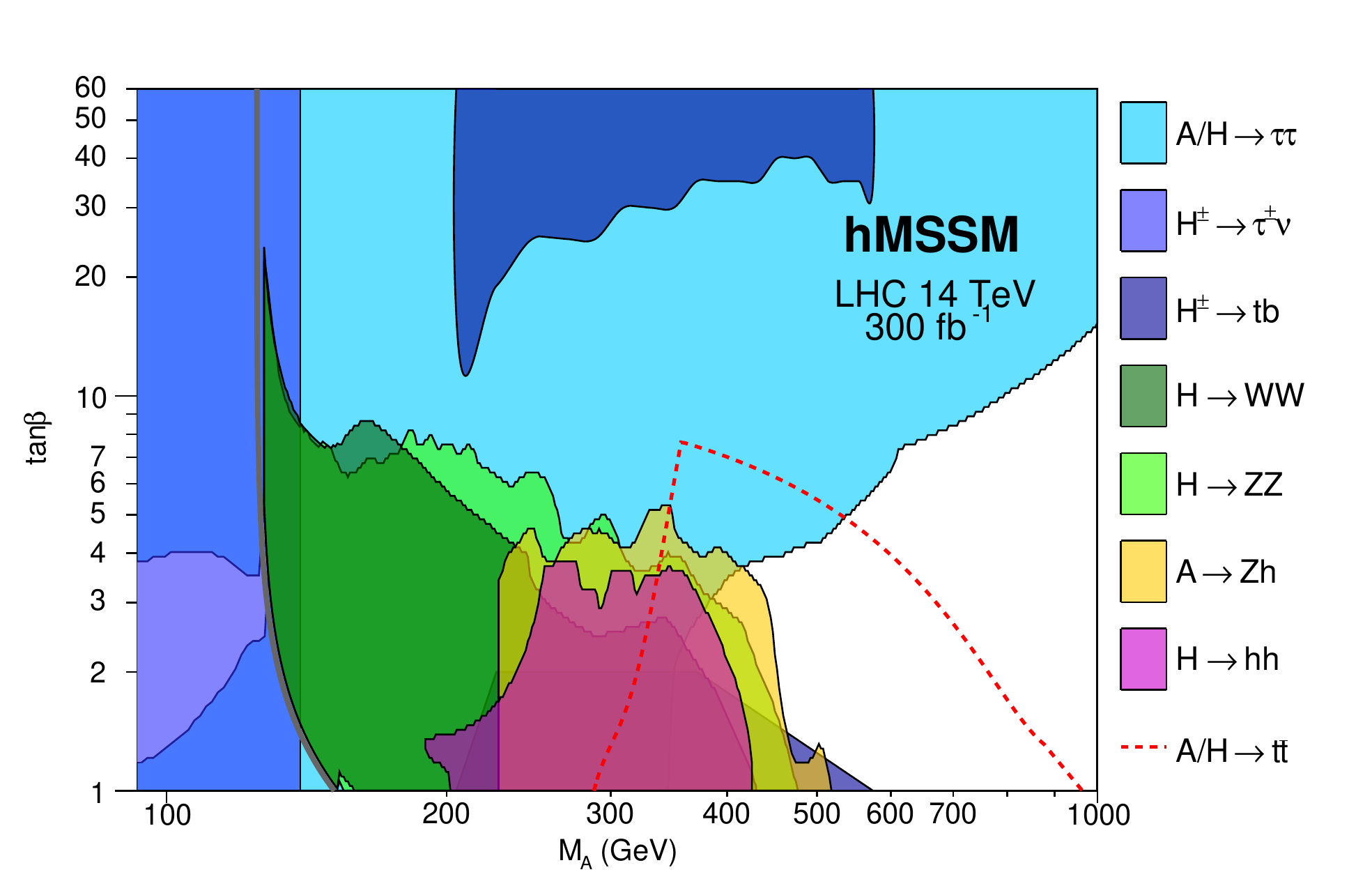} 
\end{tabular}
\vspace*{-3mm}
\caption{The same as in Fig.~19 but at the LHC with $\sqrt{s}=14$~TeV and $300$ fb$^{-1}$ data.}
\label{all-constraints_LHC14}

\end{center}
\vspace*{-19mm}
\end{figure}

\begin{figure}[!h]
\begin{center}
\vspace*{-3mm}
\begin{tabular}{c}
\includegraphics[scale=0.65]{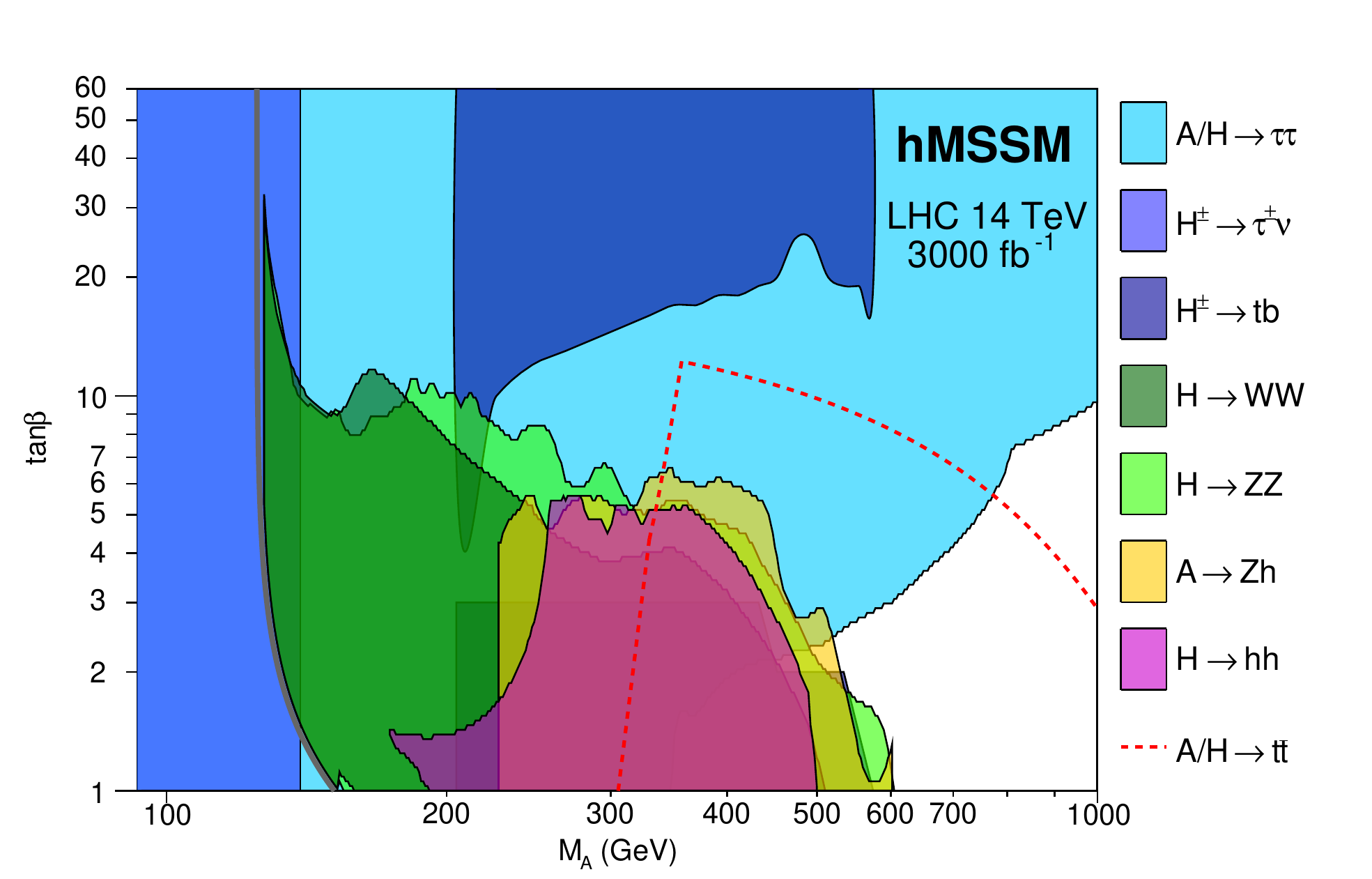} 
\end{tabular}
\vspace*{-3mm}
\caption{The same as Fig.~20 but at the high--luminosity LHC option 
with  3000 fb$^{-1}$ data. }
\label{all-constraints_LHC14-HL}
\end{center}
\vspace*{-9mm}
\end{figure}

The $2\sigma$ sensitivity in the $[\tb, M_{A}]$ plane when the $H/A \to t\bar t$
reach (using the the assumptions above) is superimposed to the sensitivity
in all the fermionic and bosonic channels discussed previously is displayed in
Figs.~\ref{all-constraints_LHC8} and \ref{all-constraints_LHC14} for, respectively, 
the previous and the next LHC phases. As can be seen, a vast improvement in the
sensitivity is expected if the $H/A\! \to \! t\bar t$ channel is included, in particular 
at the forthcoming LHC run with $\sqrt s=14$ TeV and 300 fb$^{-1}$ data. 
The improvement is even more impressive at the high--luminosity LHC option, when 
the luminosity is increased to 3000 fb$^{-1}$; see Fig.~\ref{all-constraints_LHC14-HL}.
In this case, almost the entire $h$MSSM  parameter space, up to $M_A$ values close to 
$\approx 1$ TeV, can be probed.

\newpage

\FloatBarrier

\newpage

\section{Conclusions}

In this paper, we have addressed  the issue of covering the entire parameter space of the MSSM 
Higgs sector at the LHC by considering the search  of the heavier $H,A$ and $H^\pm$ states that are predicted in the  model, in addition to the already observed lightest $h$ boson. These searches should not only be restricted to the channels that have been considered so far by the ATLAS and CMS collaborations,  namely those with a surplus of $\tau \nu$ events and those with high mass 
resonances decaying into $\tau$ lepton pairs, which would signal the presence of new
contributions  from the $t \to bH^+ \to b \tau\nu$ and $pp\to H/A \to \tau\tau$ processes,  
which are mainly relevant for the high $\tb$ region of the MSSM Higgs sector. Search for heavier 
Higgs bosons should also be conducted in channels that are more appropriate for the probing of 
the low $\tb$ region and which, until now,  have been overlooked. 

We have first discussed and refined the $h$MSSM approach introduced in Ref.~\cite{habemus} in 
which the dominant radiative corrections to the MSSM Higgs sector, that introduce a dependence
on numerous SUSY parameters, are traded against the measured mass $M_h=125$ GeV
of the Higgs boson which was observed at the LHC, thus allowing to describe again the entire
Higgs sector of the model with only two input parameters. This simple,
economical and ``model independent" approach permits to reopen the low $\tb$ region, at 
the expense of considering the possibility that the scale of SUSY--breaking is extremely high,
$M_S \gg 1$ TeV, and that the model is severely fine-tuned. The $h$MSSM  is expected 
to be viable down to values $\tan\beta \approx 2$ and, for higher $\tb$ values, reproduces to 
a very good approximation the standard results of the MSSM Higgs sector. This is particularly
true if the higgsino mass parameter is much smaller than the SUSY--breaking scale, $\mu \ll 
M_S$, an assumption that is natural at low $\tb$ values which imply a very high SUSY--breaking
scale. Thus, searches for new signals in the MSSM Higgs sector can be performed in the  
entire $[\tb, M_A]$ parameter space, in a reliable way for $\tb \gsim 2$. Nevertheless, in 
an effective approach, one can eventually extrapolate to values of $\tb$ very close to unity, 
despite of the fact that the scale $M_S$ required to reach this value is so high that its renders the model not only too fine-tuned but also potentially inconsistent.  

We have then analyzed the production and decay modes of the $H,A$ and $H^\pm$ particles at 
the LHC, with a special attention to  the low $\tb$ region in which the top quark plays a
prominent role, as its couplings to the Higgs bosons are not strongly suppressed compared
to the SM case. We have first shown that the searches that are presently conducted by
ATLAS and CMS  can also be relevant at low $\tb$. This is for instance the case 
of the $pp\to A \to \tau\tau$ and $pp \to tbH^+ \to tbtb$ processes at low to moderate $M_A$
values. We have then shown that search channels such as  $H\to WW,ZZ, hh$ and $A\to hZ$, 
when interpreted in the context of the $h$MSSM, can also probe the low $\tb$ and not too 
high $M_A$ regions. In fact, already with the 25 fb$^{-1}$ data collected at energies up to 
$\sqrt s=8$ TeV, the ATLAS and CMS collaborations exclude  the possibility $\tb \lsim  4\; (2)$ 
up to pseudoscalar Higgs masses of $M_A \approx 250 \; (350)$ GeV. At the upcoming stage 
of the LHC, with an expected energy and luminosity of $\sqrt s=14$ TeV and ${\cal L}=300$ 
fb$^{-1}$, the entire parameter space i.e. for any value of $\tb$ could be probed up to $M_A
\approx 400$ GeV,  when combining the searches in the usual fermionic channels and in the 
additional bosonic channels discussed here. 

An important message conveyed by the present paper is that, in order to fill or close the 
gap in the MSSM $[\tb, M_A]$ plane left by the fermionic and bosonic searches mentioned above,  
one should definitely consider the $pp\to H/A \to t\bar t$ process. Indeed, at low $\tb$ 
and for Higgs masses above the $t\bar t$ kinematical threshold, the decays 
$H/A\to t\bar t$ become the dominant ones, suppressing the rates for the other decay channels 
to a very low if not negligible level. On the other hand, the $gg\to H/A$ production mode 
has a still significant cross section as the top quark that generates this loop process 
has substantial couplings to the $H/A$ states at sufficiently low $\tb$ values. This is not 
a very easy
search channel in view of the formidable $pp\to t\bar t$ QCD background. Nevertheless, it
exhibits very special and interesting features such as an interference with the QCD background
that leads to a rather involved peak--dip structure of the signal. 

We have not performed a detailed and realistic study of this process but attempted to roughly
quantify the observation of a signal at the LHC, relying on present ATLAS and CMS analyses
in searches for heavy (non Higgs) resonances decaying into top quark pairs at  8 TeV center 
of mass energies, and discussed its possible implications. It appears that the channel 
$gg\to H/A\to t\bar t$, would be capable of covering partly the area at low $\tb$ and high 
$M_A$, hence allowing for a full coverage of the $[\tb, M_A]$ plane of the MSSM up to 
Higgs masses $M_A \approx 600$ GeV with 300 fb$^{-1}$ data at $\sqrt s=14$ TeV. At the high luminosity option of the LHC with 3000 fb$^{-1}$ data, one could reach a full coverage 
of the MSSM parameter space for pseudoscalar masses closer to $M_A \approx$ 1 TeV. 

More refined analyses are required in order to firmly establish the viability of the 
various processes discussed here, in particular the $H/A \to t\bar t$ channel. In view 
of the important role that it could  play in the probing of the MSSM parameter space, 
the latter process is worth investigating in a more realistic way, including the 
interference between the Higgs signal and the QCD background.  This is what we plan 
to do in a forthcoming publication \cite{preparation}. \bigskip 

\noindent {\bf Acknowledgements}:
Discussions with the LHC Higgs cross section working group on the $h$MSSM, in particular, 
S. Heinemeyer, G. Lee, M. Muhlleitner, S. Nikitenko, N. Rompotis,  P. Slavich, M. Spira,   
C. Wagner and G. Weiglein, are greatfully acknowledged. 
AD is supported by the ERC advanced grant Higgs@LHC and JQ by the STFC Grant ST/J002798/1.

\end{document}